\documentclass[twocolumn]{aastex63}
\usepackage{multirow}
\usepackage[figuresright]{rotating}
\usepackage{subfigure}
\usepackage{epic,eepic}
\usepackage{graphicx}
\usepackage{longtable}
\usepackage{float}
\usepackage{lineno}
\usepackage{pifont}
\usepackage{gensymb}
\usepackage{multirow}
\usepackage{amsmath}
\usepackage{color}

\shorttitle{}
\shortauthors{Zhao et al.}

\begin{document}

\title{Characterizing Orbital Parameters of Hot Subdwarf Binaries with Multiple Spectroscopic Surveys}

\correspondingauthor{Song Wang}
\email{songw@bao.ac.cn}

\author{Xinlin Zhao}
\affiliation{Department of Physics and Chongqing Key Laboratory for Strongly Coupled Physics, Chongqing University, Chongqing 401331, China}

\author{Song Wang}
\affiliation{Key Laboratory of Optical Astronomy, National Astronomical Observatories, Chinese Academy of Sciences, Beijing 100101, China}
\affiliation{Institute for Frontiers in Astronomy and Astrophysics, Beijing Normal University, Beijing 102206, China}

\author{Zhenxin Lei}
\affiliation{Key Laboratory of Stars and Interstellar Medium, Xiangtan University, Xiangtan 411105, China}

\author{Yangyang Dong}
\affiliation{Key Laboratory of Stars and Interstellar Medium, Xiangtan University, Xiangtan 411105, China}

\author{Buhui Lv}
\affiliation{Key Laboratory of Optical Astronomy, National Astronomical Observatories, Chinese Academy of Sciences, Beijing 100101, China}
\affiliation{School of Astronomy and Space Sciences, University of Chinese Academy of Sciences, Beijing 100049, China}

\author{Chuanjie Zheng}
\affiliation{Key Laboratory of Optical Astronomy, National Astronomical Observatories, Chinese Academy of Sciences, Beijing 100101, China}
\affiliation{School of Astronomy and Space Sciences, University of Chinese Academy of Sciences, Beijing 100049, China}

\author{Xiaohong Yang}
\affiliation{Department of Physics and Chongqing Key Laboratory for Strongly Coupled Physics, Chongqing University, Chongqing 401331, China}

\author{Jifeng Liu}
\affiliation{Key Laboratory of Optical Astronomy, National Astronomical Observatories, Chinese Academy of Sciences, Beijing 100101, China}
\affiliation{Institute for Frontiers in Astronomy and Astrophysics, Beijing Normal University, Beijing 102206, China}
\affiliation{College of Astronomy and Space Sciences, University of Chinese Academy of Sciences, Beijing 100049, China}
\affiliation{New Cornerstone Science Laboratory, National Astronomical Observatories, Chinese Academy of Sciences, Beijing 100101, China}

\begin{abstract}


Hot subdwarfs (HSDs) provide critical insights into the physical mechanisms governing binary evolution. 
In this work, we conduct a systematic analysis of 157 HSDs, selected from Gaia EDR3 and characterized using multi-survey spectroscopic data.
Atmospheric parameters of these HSDs are derived via a convolutional neural network (CNN) method and template-matching method.
Based on the atmospheric parameters from CNN method, these HSDs exhibit a median mass of $0.45^{+0.19}_{-0.17} M_{\odot}$ and radius of $0.18^{+0.04}_{-0.05} R_{\odot}$, consistent with earlier work.
Orbital parameters of 23 systems are determined through the fitting of radial velocity data and light curves, with 11 of them being new solutions.
We find that reflection-dominated binaries typically have periods longer than 0.1 d and host low-mass main-sequence companions ($\sim$ 0.2 $M_{\odot}$) with rotation-inflated radii. In contrast, binaries including an HSD and a white dwarf show very short periods ($P < 0.2$ d), with the closest systems hosting more massive white dwarfs.
Most of these systems share a similar mass--period distribution with that of post-common-envelope binaries, supporting a common-envelope origin.

\end{abstract}


\section{Introduction}
\label{intro.sec}

Hot subdwarfs (HSDs) are located on the extreme horizontal branch (EHB) of the Hertzsprung-Russell (HR) diagram and exhibit spectral types O (sdOs) or B (sdBs). 
They display high effective temperatures, ranging from 20000 K to 70000 K, and have a typical mass of approximately 0.46 $M_{\odot}$ \citep{2023ApJ...953..122L,2026A&A...707A...6D}. 
Most HSDs are believed to be core helium-burning objects that lost the majority of their hydrogen envelopes near the tip of the red giant branch (RGB). 
HSDs have significantly advanced our understanding of the formation and evolution pathways of binaries \citep{2002MNRAS.336..449H,2003MNRAS.341..669H,2020RAA....20..161H}.
A comprehensive review of this topic can be found in \cite{2026enap....2..488H}.

The details of the formation channels of HSDs remain poorly constrained.
Observations indicate that approximately one-third of HSDs reside in composite sdB binaries with main-sequence (MS) companions of spectral types F, G, or K \citep{2013A&A...559A..54V,2018MNRAS.473..693V,2026arXiv260205008M}. 
These systems typically exhibit long orbital periods, often ranging from several hundred to over a thousand days.
Another third reside in single-lined systems with compact companions such as WDs or low-mass main sequence stars (e.g., M-type stars and brown dwarfs) \citep{2002MNRAS.333..231M,2015A&A...576A..44K,2019A&A...630A..80S}. 
In contrast to the former case, these compact systems are predominantly short-period binaries, with orbital periods mostly less than 10 days \citep{2015A&A...576A..44K,2022A&A...666A.182S}.
The remaining HSDs are apparently single, isolated stars, although some of them may reside in long-period binaries or exhibit amplitudes of radial velocity (RV) below the detection threshold.
For HSDs in binaries, two main mass-loss scenarios are invoked during the Roche-lobe overflow (RLOF) phase of their progenitor. 
When the mass ratio is relatively modest (e.g., $q<1.2-1.5$) \citep{2008ASPC..392...15P}, stable RLOF can efficiently strip the hydrogen envelope of the progenitor, leading to the formation of an HSD.
In contrast, if the system has a large mass ratio or a short orbital period, the mass transfer becomes dynamically unstable, triggering a common envelope (CE) phase; the subsequent ejection of the envelope then produces the observed HSDs.
For single HSDs, their formation has mostly been attributed to merger channels from short-period binary systems with double helium white dwarfs \citep{1984ApJ...277..355W,2012MNRAS.419..452Z,2016MNRAS.463.2756H,2018MNRAS.476.5303S,2026A&A...710A..99P}.
Hence, detailed studies of HSD binaries provide crucial insights into their formation pathways and offer stringent tests for CE evolution models.
Moreover, HSD binaries hosting massive white dwarfs represent a potential progenitor population for Type Ia supernovae \citep{2004MNRAS.350.1301H,2021NatAs...5.1052P,2023RAA....23h2001L} and also serve as promising gravitational sources for Laser Interferometer Space Antenna (LISA) \citep{2020ApJ...891...45K,2020ApJ...898L..25K,2025arXiv251025653T}.

Several previous studies \citep{2011A&A...526A..39G,2011A&A...530A..28G,2012ASPC..452..129G,2015A&A...577A..26G,2015A&A...576A..44K,2023A&A...673A..90S} have conducted detailed analyses of a number of HSD binaries using RV measurements and light curves (LCs), making significant contributions to our understanding of the formation and evolution of HSD binaries.
However, these studies assumed the typical values for the mass and radius of HSDs, leading to non-negligible systematic biases in the inferred orbital parameters and the properties of the companion.
For example, in the case of LAN 11 \citep{2025SCPMA..6869511L}, the spectroscopic mass of the HSD ($\sim 0.6 M_{\odot}$) is clearly larger than the typical mass ($\sim 0.46 M_{\odot}$). Using the typical mass would lead to an incorrect companion mass estimate.

In this work, we selected a sample of HSDs from the catalogue of Gaia EDR3 \citep{2022A&A...662A..40C} and then measured the systematic parameters using multiple sets of spectroscopic data.
We derived atmospheric parameters for 157 HSDs using both template matching and a machine‑learning approach. Based on these parameters, we then computed the masses and radii of the targets. 
Finally, for a group of systems with reliable RV measurements and LCs, we solved for their orbital parameters.
In Section \ref{sample_data.sec}, we detail the sample selection and data reduction.
Section \ref{atmo.sec} presents the stellar information of the HSDs, including the distance, atmospheric parameters, and the mass.
In Section \ref{orbit_fitting.sec}, we investigate the systematic information for 23 binaries through the RV fitting and the LC fitting, including the orbital period, eccentricity, and the inclination angle.
%
%
Section \ref{dis.sec} discusses the nature of the companion and possible formation mechanisms.
Finally, a summary is given in Section \ref{summary.sec}.

\begin{figure}
    \center
    \includegraphics[width=0.5\textwidth]{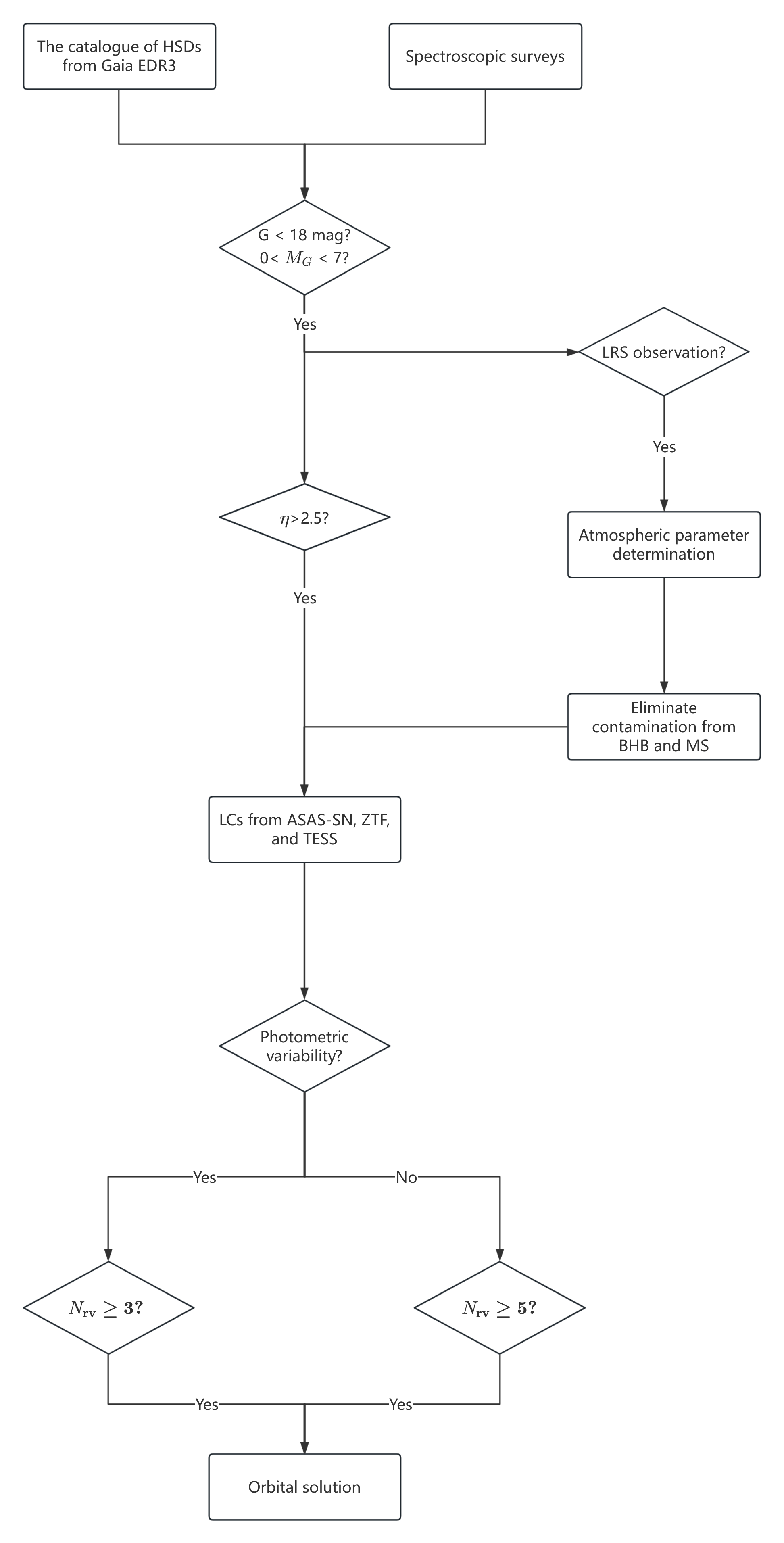}
    \caption{A summary flowchart illustrating the workflow used in this paper.}
    \label{flowchart.fig}
\end{figure}

\section{Sample selection and data reduction}
\label{sample_data.sec}

\subsection{Initial sample}
\label{sample.sec}

In this study, we used the HSD catalogue (including 6616 known HSDs and 61585 candidates) based on Gaia EDR3 \citep{2021A&A...649A...1G,2022A&A...662A..40C} and cross-matched it with spectroscopic data from LAMOST DR11, BOSS (SDSS-IV/V), SEGUE, APOGEE (SDSS-IV/V), and GALAH for subsequent analysis.

The Large Sky Area Multi-Object Fiber Spectroscopic Telescope (LAMOST) has been conducting both low-resolution (R$\sim$1800) and medium-resolution (R$\sim$7500) optical spectroscopic surveys. 
The wavelength coverage for low-resolution spectra (LRS) spans from 3690 \AA\ to 9100 \AA\  \citep{2015RAA....15.1095L}. 
For medium-resolution spectra (MRS), the observed wavelength range is split into two arms: the blue arm covering 4950--5350 \AA\ and the red arm covering 6300--6800 \AA\  \citep{2020arXiv200507210L,2021RAA....21..292W}. 
LAMOST DR11 provides approximately 12 million LRS spectra and 13 million MRS spectra, marking a significant contribution to large-scale spectroscopic studies.

The Sloan Digital Sky Survey (SDSS) is a major multi-epoch, multi-wavelength imaging and spectroscopic redshift survey conducted using the 2.5-meter Sloan Foundation Telescope at Apache Point Observatory and the 2.5-meter du Pont Telescope at Las Campanas Observatory. 
In the optical regime, SDSS employs the Baryon Oscillation Spectroscopic Survey (BOSS) \citep{2013AJ....146...32S} to observe targets over a wavelength range of 3700--9800 \AA, and the Sloan Extension for Galactic Understanding and Exploration (SEGUE) \citep{2009AJ....137.4377Y} to cover 3850--9200 \AA. 
In the infrared band, specifically from 15000 \AA \ to 17000 \AA, observations are carried out using the Apache Point Observatory Galactic Evolution Experiment (APOGEE) survey \citep{2019PASP..131e5001W}.

The Galactic Archaeology with HERMES (GALAH) survey is a large observing program conducted with the High Efficiency and Resolution Multi-Element Spectrograph on the Anglo-Australian Telescope, designed to investigate the formation and chemical evolution of the Milky Way \citep{2012ASPC..458..421Z}. 
HERMES simultaneously obtains high-resolution spectra (R$\approx$28000) for up to 400 stars at a time. 
The fourth data release of the GALAH survey provides spectra for 1085520 observations of 917588 stars within the Milky Way \citep{2025PASA...42...51B}, enabling detailed studies of Galactic archaeology.


\subsection{Sample cleaning}

Visual inspection revealed that the spectra of sources with $G$-band magnitudes greater than 18 mag are too poor in quality to yield a reliable RV measurement. 
Thus, we first excluded these faint sources ($G>18$ mag) from the cross-matched sample.

Second, we built the Hertzsprung–Russell (HR) diagram to check the distribution of the initial sample.
We obtained the geometric distances provided by Gaia Early Data Release 3 (EDR3) \citep{2020yCat.1350....0G,2021AJ....161..147B}, which employed a probabilistic method to infer distances using Gaia parallaxes and a prior based on a three-dimensional model of the Milky Way.
Assuming $R_{V}=3.1$, the extinction ($A_{V}$) for each system was estimated using the three-dimensional dust reddening map from Pan-STARRS DR1, specifically the {\it Bayestar2019} model \citep{2015ApJ...810...25G}.
Based on the location of HSDs on the HR diagram from \cite{2024A&A...686A..25D}, we further excluded systems with absolute magnitude $M_G<0$ mag or $M_G>6$ mag to minimize potential contamination.

Third, we identified Blue Horizontal Branch (BHB) stars in our sample using the $H_{\delta}$ line, following the method proposed by \cite{2008ApJ...684.1143X}. 
Specifically, we fitted the $H_{\delta}$ profiles from our LRS spectra using a S\'ersic profile \citep{1968adga.book.....S}.
By calculating the full width of the line at 20\% below the local continuum ($D_{0.2}$) and the flux relative to the continuum at the line core ($f_m$), we applied the criteria of $17\,\text{\AA} \le D_{0.2} \le 28.5\,\text{\AA}$ and $0.1 \le f_m \le 0.3$ \citep{2008ApJ...684.1143X}. 
As a result, four objects (Gaia DR3 3741871891637318656, 3962948461251171200, 4003621011270550528, 797745271351197952) were identified as BHB stars and excluded from our sample.

\begin{figure}
    \center
    \includegraphics[width=0.49\textwidth]{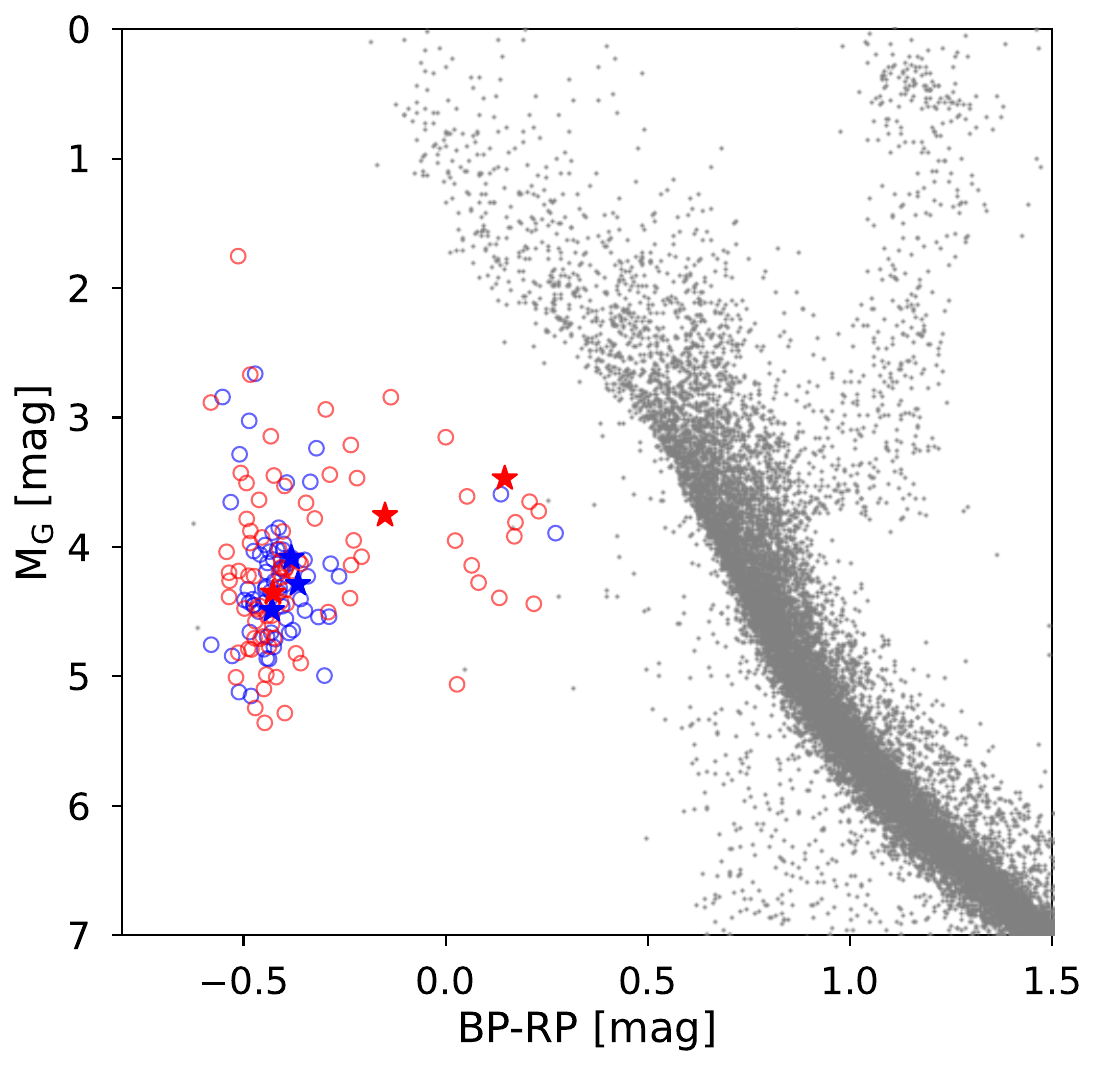}
    \caption{Position of our sample on the HR diagram. Red and blue markers correspond to systems with $\eta<2.5$ and $\eta>2.5$, respectively (see Section \ref{orbit.sec}). Circles represent known HSDs, whereas stars indicate HSD candidates from \cite{2022A&A...662A..40C}. The gray points are plotted for a comparison, which are from {\it Gaia} EDR3 with distances $d <$ 100 pc, $G_{\rm mag}$ between 4--16 mag, and galactic latitudes $|b|$ $>$ 40 deg.} 
    \label{HR.fig}
\end{figure}

After these steps, we obtained a sample of 196 HSD candidates.
However, contamination from BHB and main sequence (MS) stars remains possible. 
According to the atmospheric parameters (see Section \ref{atmo_pars.sec}), we classified candidates with $T_{\rm eff} < 20000$ K and log$g < 5.0$ as BHB stars, and those with log$g < 4.5$ as B-type MS stars \citep{2012MNRAS.427.2180N}. 
After excluding these contaminants, 157 HSDs (151 known HSDs and 6 candidates) were left in the final sample.
It should be noted that nearly all spectra of our samples exhibit a single-lined spectral morphology.
Figure \ref{flowchart.fig} presents a flowchart illustrating the sample selection procedure.
Figure \ref{HR.fig} shows the position of the sample in the HR diagram.
The $\eta$ in the HR diagram denotes the RV variability of each systems (see Section \ref{orbit.sec}).
%

Finally, we collected the LCs from the All-Sky Automated Survey for Supernovae (ASAS-SN)\footnote{https://asas-sn.osu.edu/}, the Zwicky Transient Facility (ZTF)\footnote{https://irsa.ipac.caltech.edu/Missions/ztf.html}, and the Transiting Exoplanet Survey Satellite (TESS). 
TESS LCs for the sample were downloaded using the {\it lightkurve} Python package \citep{2018ascl.soft12013L}. 
We favored the 20s-cadence data and used the 120s-cadence data as a complement.
Potential contamination from nearby sources can be excluded by inspecting the DSS2 images.
We also examined the field-of-view of TESS and the CROWDSAP parameter for all our targets. 
We found that the majority of the sources exhibit CROWDSAP values close to 1, indicating that no stars are blending into the target pixel.
%
%
Using the Lomb-Scargle (LS) periodogram method \citep{1989ApJ...338..277P}, we estimated the orbital periods from these LCs.
Periods from TESS LCs are prioritized due to the higher precision of the data.
For sources exhibiting clear photometric variability signatures (e.g., ellipsoidal modulation, reflection effects), we require at least three spectroscopic observations; 
for those without such photometric signatures (and thus photometry provides no constraint on the orbit), 
we require at least five spectroscopic observations.
The systems lacking measured atmospheric parameters were also excluded (Section \ref{atmo_pars.sec}).
%
%

\subsection{RV measurements for LAMOST spectra}
\label{data.sec}

In the binary sample (see Section \ref{orbit_fitting.sec}), approximately 80\% of the RV measurements are derived from LRS and MRS in LAMOST DR11. 
Therefore, to obtain reliable Keplerian orbital solutions, we re-estimated the RVs for LRS and MRS.
For other surveys, we preferred to use the RVs from catalogs.

Due to the limited spectral quality of LAMOST data, we determined the RV for each spectrum from the Doppler shift of the $H_{\alpha}$ line.
We first selected the spectral region around the $H_{\alpha}$ line (6550--6580 \AA) and fitted its profile with a Gaussian function to estimate the initial wavelength of the line centre.
To improve the precision of the RV measurement, a second Gaussian fit was performed on a narrower window spanning $\pm$2 \AA \ around the preliminary line center. 
All fitting results were visually inspected.
Spectra exhibiting poor fits were refitted with an appropriately adjusted wavelength range.
As a result, RVs were calculated using the $H_{\alpha}$ line center, and the uncertainties were estimated from the fitting errors.
Table \ref{rvdata.tab} lists the RV data of our sample.
As an example, Figure \ref{Ha_fitting_1914.fig} presents a schematic of the $H_{\alpha}$ line fitting for Gaia DR3 1914301803258669440 based on the MRS spectra.

Most of RV data provided by BOSS (SDSS-IV/V), SEGUE, APOGEE (SDSS-IV/V), and GALAH exhibit large uncertainties. 
To derive more accurate orbital parameters in our subsequent analysis, we selected only data points with RV uncertainties smaller than 10 km/s for RV fitting and those smaller than 15 km/s for LC and RV joint fitting. 
Consequently, the number of RV data meeting the $<10$ km/s criterion from LAMOST, BOSS (SDSS-IV/V), SEGUE, APOGEE (SDSS-IV/V), and GALAH are 1120, 1, 51, 13, and 1, respectively. 
For the $<15$ km/s criterion, the corresponding numbers are 1339, 3, 51, 13, and 1.

\section{Stellar parameters of hot subdwarfs}
\label{atmo.sec}

\subsection{Atmospheric parameters}
\label{atmo_pars.sec}

\begin{figure}
    \center
    \includegraphics[width=0.48\textwidth]{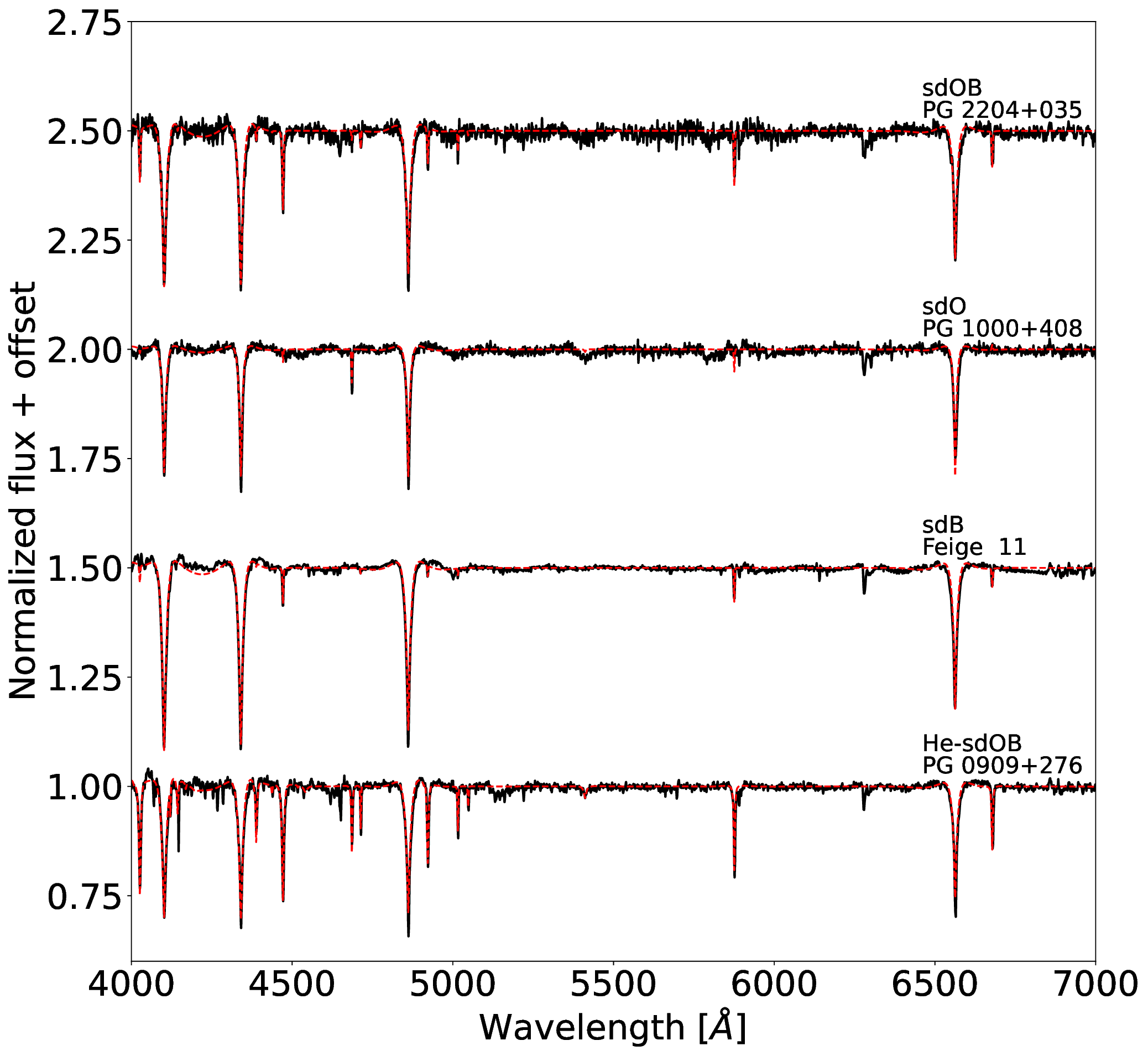}
    \caption{Example schematic of LRS spectral fitting based on the CNN method. The black lines indicate the observed LRS, while the red lines are the best-fit templates.} 
    \label{spectra_fitting.fig}
\end{figure}

For all systems with available LAMOST LRS, we adopted two methods to estimate the atmospheric parameters of HSDs, using the spectrum with the highest signal-to-noise ratio (S/N) for each system.
In the first method, we employed a template-matching approach to estimate the atmospheric parameters.
Theoretical templates were generated using synthetic spectra from the Synspec code \citep{2007ApJS..169...83L}, based on non-LTE model atmospheres computed with Tlusty \citep{2017arXiv170601859H}.
%
We adopted a grid spanning effective temperatures ($T_{\rm eff}$) from 20000 to 56000 K,  surface gravities (log$g$) from 5.0 to 6.3 and helium abundance from -4.3 to 2.0  ($\log(n{\rm He}/n{\rm H})$).
The best-fit atmospheric parameters were determined by minimizing the $\chi^{2}$ difference between the observed LRS and the theoretical templates \citep{2012MNRAS.427.2180N, 2018ApJ...868...70L}.
In the second method, we employed a deep learning model to estimate atmospheric parameters of HSDs. 
The model was built using the convolutional neural network (CNN) in TensorFlow framework, incorporating both ‌channel attention and spatial attention mechanisms‌ \citep{woo2018cbamconvolutionalblockattention, hu2019squeezeandexcitationnetworks} to enhance the efficiency and accuracy of extracting spectral features from HSDs. 
The training dataset comprised ‌11396 synthetic spectra with added noise‌ calculated by \citet{2014ASPC..481...95N} and ‌945 observed HSD spectra‌ from the LAMOST spectral library indentified in \citet{2018ApJ...868...70L, 2019ApJ...881..135L, 2020ApJ...889..117L, 2023ApJ...942..109L} . 
As an example, Figure \ref{spectra_fitting.fig} illustrates the schematic of LRS spectral fitting using the CNN method.

\begin{figure*}
    \center
    \includegraphics[width=0.98\textwidth]{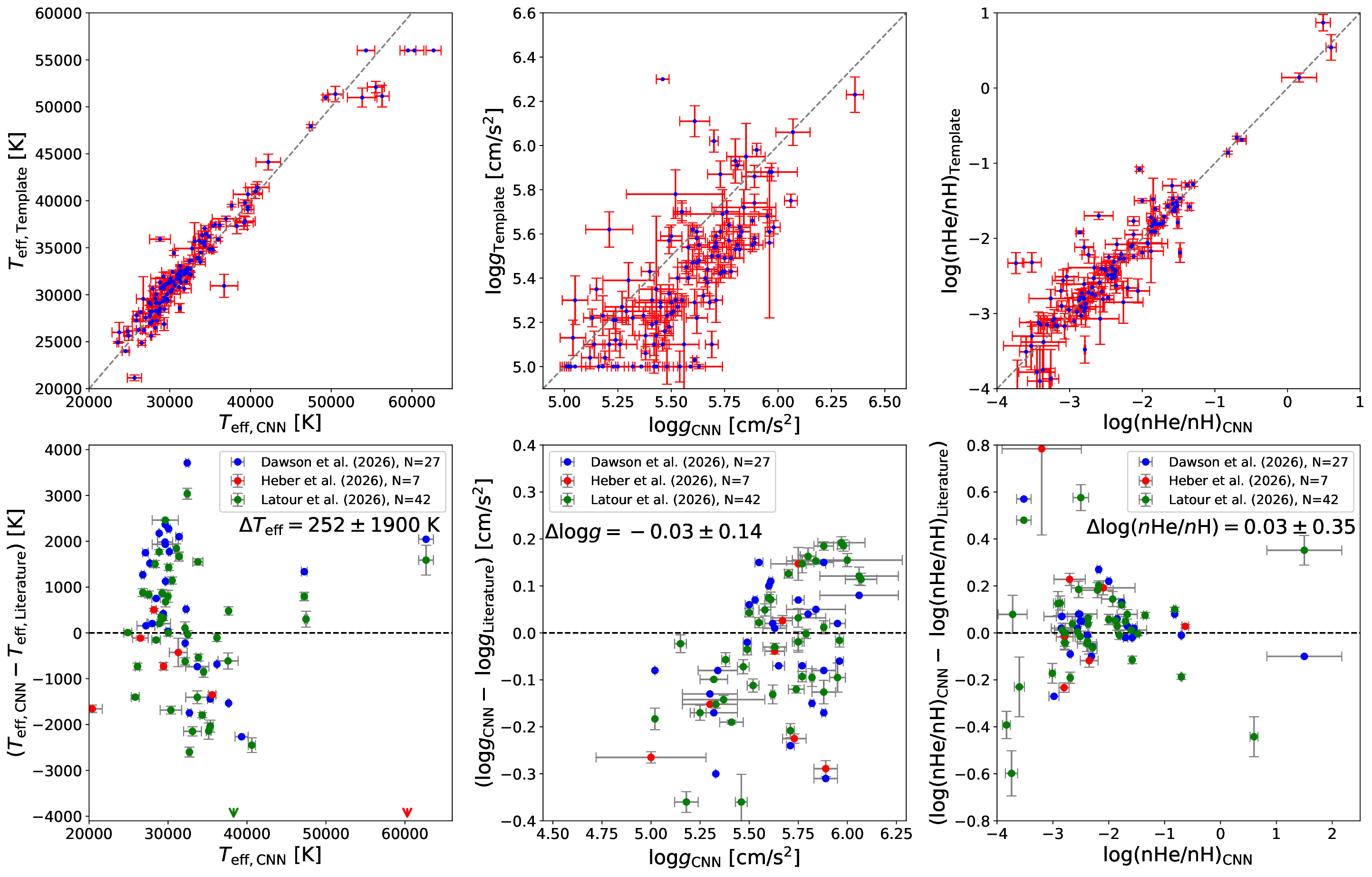}
    \caption{Top row: Comparison of atmospheric parameters derived from the CNN method and the template-matching method. Bottom row:Comparison of the derived atmospheric parameters between this work and previous studies.} 
    \label{comparison_pars.fig}
\end{figure*}

The model spectra used in template-matching method do not include metal lines, which means that metal lines were excluded from both the TLUSTY model atmospheres and subsequent spectral synthesis.
This may cause slight deviations in the derived effective temperature and surface gravity.
Generally, the direction and magnitude of these shifts depend on the specific stellar type and its metallicity \citep{2011ApJ...733..100L}.
For example, for HSDs with $T_{\rm eff}<40000$ K, neglecting metal lines leads to an overestimation of $T_{\rm eff}$ and an underestimation of log$g$; conversely, for hotter HSDs, this will yield an underestimated $T_{\rm eff}$ and an overestimated log$g$.
In additon, the template-matching method does not extrapolate beyond the boundaries of the atmospheric parameter grid.
Consequently, when the atmospheric parameters approached the boundaries of the template grid (e.g., log$g=5.0$ or $T_{\rm eff}=$56000 K), the fitting results from template-matching method becomes unreliable (Figure \ref{comparison_pars.fig}).

Although the templates used in the CNN method also do not include metal lines, the inclusion of observed spectra in the training set can significantly reduce systematic offsets, yielding more robust atmospheric parameter estimates. 
For a detailed description of the model architecture and training sample, see Sections 2 and 3 of \citet{lei2025deeplearningdrivenatmosphericparameter}. 
Comparison of the CNN results with previous studies \citep{2026A&A...705A.248L,2026A&A...707A...6D,2026A&A...708A.115H} show offsets of $\Delta T_{\rm eff}=252\pm1900$ K, $\Delta$log$g=-0.03\pm0.14$, and $\Delta \log(n{\rm He}/n{\rm H})=0.03\pm0.35$ in effective temperature, surface gravity, and helium abundance (Figure \ref{comparison_pars.fig}), respectively.
These slight discrepancies indicate that the results derived from the CNN method are reliable.
Therefore, in subsequent analyses, we preferred to adopt the atmospheric parameters provided by the CNN method.

For systems without LRS observations, we adopted atmospheric parameters from the available catalogs \citep{2018ApJ...868...70L,2021ApJS..256...28L,2022A&A...662A..40C}.
The average of atmospheric parameters from various catalogs is used as the final estimate for systems with multiple measurements.
The uncertainty corresponds to the root sum of squares of the errors from multiple measurements.
Table \ref{stellar_pars.tab} and \ref{stellar_pars_catalogs.tab} list the parameters of our sample.

\begin{figure*}
    \center
    \includegraphics[width=0.98\textwidth]{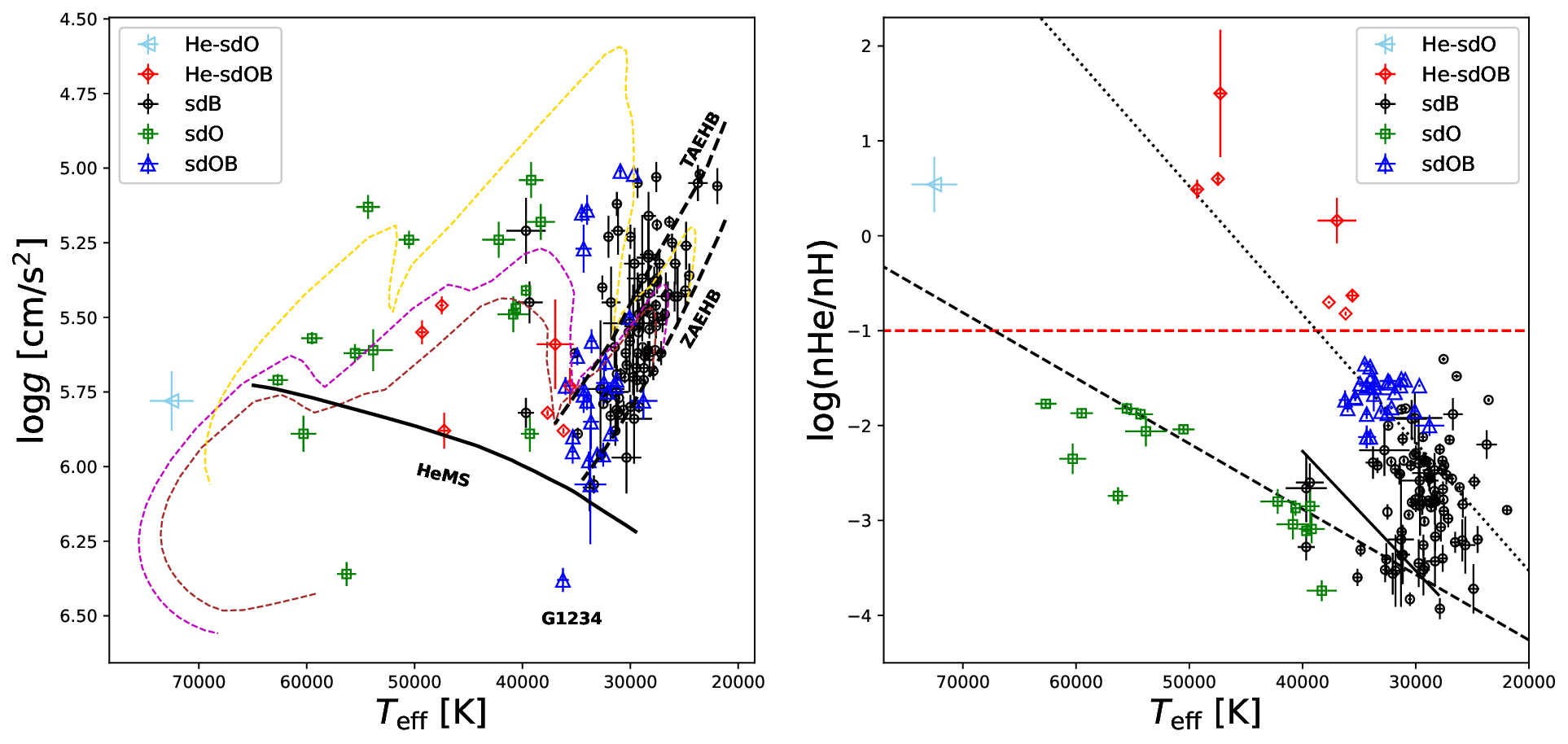}
    \caption{Left panel: The Kiel diagram for our sample. The dashed black lines show the ZAEHB and TAEHB sequences for [Fe/H]=-1.48 from \cite{1993ApJ...419..596D}, while the black solid line indicates the helium main sequence (HeMS) from \cite{1971AcA....21....1P}. Right panel: $T_{\rm eff} - \log(n{\rm He}/n{\rm H})$ diagram. The red dashed line denotes the solar He abundance (e.g., $\log(n{\rm He}/n{\rm H})=-1$). The dotted and solid lines represent the linear regression line fitted by \cite{2003A&A...400..939E}, while the dash line denotes the best-fit relation for the He-weak sequence reported by \cite{2012MNRAS.427.2180N}.}
    \label{teff_vs_logg_and_loghe.fig}
\end{figure*}

We classified the HSDs in our sample, based on LAMOST LRS spectra following \cite{2018ApJ...868...70L} and previous catalogs \citep{2018ApJ...868...70L,2021ApJS..256...28L,2022A&A...662A..40C}, into 93 sdB, 26 sdOB, 7 He-sdOB, 15 sdO, 1 He-sdO stars.
Figure \ref{teff_vs_logg_and_loghe.fig} (left panel) shows the Kiel diagram (i.e., $T_{\rm eff}$ versus log$g$) for our sample. 
The yellow, purple, and brown dashed curves correspond to evolutionary tracks \citep{1993ApJ...419..596D} of hot EHB stars with masses of 0.495 $M_{\odot}$, 0.490 $M_{\odot}$ and 0.488 $M_{\odot}$, respectively.
Most sdB stars lie between the zero-age extreme horizontal branch (ZAEHB) and the terminal-age extreme horizontal branch (TAEHB), indicating they are undergoing core helium burning. 
%
%
The sdOB stars occupy similar region but exhibit higher temperatures. 
%
%
Other types (including He-sdOB, sdO and He-sdO stars) show a much more dispersed distribution, characterized by significantly higher temperatures than other hot subdwarfs.
One should {\bf note} that the evolutionary tracks from \cite{1993ApJ...419..596D} represent a sequence of varying envelope masses for a fixed core mass. 
Incorporating tracks with different core masses would expand the covered parameter space, potentially providing a better match for these systems.
G1234 deviated significantly from these evolutionary tracks, further indicating they may have already evolved into white dwarfs.

In the $T_{\rm eff} - \log(n{\rm He}/n{\rm H})$ diagram (Figure \ref{teff_vs_logg_and_loghe.fig}, right panel), the black dotted and solid lines represent the linear regression fits for the He-rich and He-weak sequences from \cite{2003A&A...400..939E}. 
The black dashed line indicates the best-fit for the He-weak sequence from \cite{2012MNRAS.427.2180N}. 
Our sample clearly follows these trends, with helium abundance increasing alongside effective temperature. 
Specifically, the sdB, sdOB, and He-sdOB stars follow the helium-rich sequence, but helium abundance regimes: approximately $-4<\log(n{\rm He}/n{\rm H})<-2$ for sdBs, $-2<\log(n{\rm He}/n{\rm H})<-1$ for sdOBs, and $-1<\log(n{\rm He}/n{\rm H})<2$ for He-sdOBs, respectively.
In contrast, the sdO stars follow the helium-week sequence.
Notably, there are some helium-poor sdBs with $\log(n{\rm He}/n{\rm H})<-3$. 
The majority of these sdBs have masses below $0.41 M_{\odot}$, implying longer evolutionary timescales \citep{2003MNRAS.341..669H}. 
Consequently, they may undergo more extended diffusion, which could explain the extremely low helium abundances observed in these sdBs.



\begin{table*}
\caption{Atmospheric parameters derived from CNN method for our samples. \label{stellar_pars.tab}}
\centering
\setlength{\tabcolsep}{1mm}
\begin{center}
\begin{tabular}{llcccccc}
\hline\noalign{\smallskip}
Gaia ID & SIMBAD ID & R.A. & Decl. & $T_{\rm eff, CNN}$ & log$g_{\rm CNN}$ & $\log(n{\rm He}/n{\rm H})_{\rm CNN}$ & Type$^a$ \\
 & & ($^{\circ}$) & ($^{\circ}$) & (K) &  &  &   \\
\hline\noalign{\smallskip}
2767874292175410560 & PG 0004+133 & 1.89071 & 13.59924 & $30081\pm213$ & $5.50\pm0.02$ & $-1.85\pm0.03$ &  1\\
2850670743266825600 & GALEX J001052.4+263000 & 2.71848 & 26.5002 & $28341\pm1060$ & $5.30\pm0.12$ & $-2.47\pm0.18$ &  1 \\
2581810261598516096 & PG 0048+091 & 12.8622 & 9.35915 & $31244\pm389$ & $5.12\pm0.04$ & $-3.12\pm0.08$ &  1 \\
312628749626419328 & SDSS J010113.32+312555.3 & 15.30553 & 31.43203 & $28654\pm201$ & $5.63\pm0.03$ & $-2.86\pm0.05$ &  1 \\
2551900379931546240 & Feige 11 & 16.09036 & 4.22683 & $28877\pm43$ & $5.60\pm0.01$ & $-2.79\pm0.01$ &  1 \\
369576820516013824 & FBS 0102+362 & 16.20348 & 36.46185 & $32284\pm169$ & $5.65\pm0.03$ & $-1.55\pm0.03$ &  1 \\
369573831218501376 & FBS 0106+374 & 17.4405 & 37.76062 & $29324\pm557$ & $5.43\pm0.08$ & $-3.52\pm0.27$ &  1 \\
401413450281523584 & UCAC4 686-007194 & 18.32048 & 47.19162 & $56304\pm880$ & $6.36\pm0.04$ & $-2.74\pm0.09$ &  1 \\
371711041304831616 & FBS 0117+396 & 20.09559 & 39.84984 & $27519\pm352$ & $5.37\pm0.05$ & $-2.90\pm0.10$ &  1 \\
397395014455786624 & UCAC4 679-007631 & 21.53859 & 45.78795 & $36732\pm1709$ & $5.56\pm0.07$ & $-2.39\pm0.25$ &  1 \\
2592234628262141312 & PG 0123+159 & 21.65452 & 16.1365 & $31056\pm114$ & $5.77\pm0.02$ & $-2.37\pm0.03$ &  1 \\
344155939885410816 & ATO J029.0051+40.0561 & 29.00523 & 40.05595 & $29640\pm84$ & $5.53\pm0.01$ & $-2.69\pm0.02$ & 2 \\
2514278566658235776 & PG 0209+017 & 33.11088 & 1.92149 & $29427\pm359$ & $5.67\pm0.05$ & $-2.79\pm0.15$ &  1 \\
341195058150383104 & FBS 0229+439 & 38.2164 & 44.19087 & $35329\pm140$ & $5.90\pm0.02$ & $-1.71\pm0.02$ &  1 \\
123108816565129344 & GALEX J030128.0+301537 & 45.36673 & 30.26018 & $60301\pm1187$ & $5.89\pm0.06$ & $-2.35\pm0.16$ &  1 \\
3266182479530739328 & PG 0313+005 & 49.08389 & 0.70634 & $32032\pm443$ & $5.23\pm0.07$ & $-3.56\pm0.22$ &  1 \\
9132341717393792 & PG 0319+055 & 50.41116 & 5.64446 & $31828\pm257$ & $5.83\pm0.04$ & $-2.46\pm0.07$ &  1 \\
64368713521485568 & UCAC4 561-007652 & 56.02222 & 22.07297 & $28231\pm215$ & $5.54\pm0.03$ & $-2.74\pm0.05$ & 1 \\
3299455110137611776 & HZ 12 & 66.04043 & 8.55467 & $27541\pm162$ & $5.19\pm0.02$ & $-1.30\pm0.02$ & 1 \\
\noalign{\smallskip}\hline
\end{tabular}
\end{center}
$^a$Type 1 and 2 represent known HSDs and candidates from \cite{2022A&A...662A..40C}, respectively. A portion is shown here for guidance regarding its form and content.
\end{table*}

\begin{table*}
\caption{Atmospheric parameters derived from previous studies. \label{stellar_pars_catalogs.tab}}
\centering
\setlength{\tabcolsep}{1mm}
\begin{center}
\begin{tabular}{llccccccc}
\hline\noalign{\smallskip}
Gaia ID & SIMBAD ID & R.A. & Decl. & $T_{\rm eff}$ & log$g$ & $\log(n{\rm He}/n{\rm H})$ & Ref.$^a$ & Type$^b$ \\
 & & ($^{\circ}$) & ($^{\circ}$) & (K) &  &   & & \\
\hline\noalign{\smallskip}
435211617384833536 & Cl Melotte 20 488 & 50.41551 & 47.45519 & $27990\pm651$ & $5.34\pm0.10$ & $-2.52\pm0.31$ & 1 & 1 \\
3131204469305148288 & GSC 00141-01628 &  96.59278 & 4.07328 & $27138\pm349$ & $5.62\pm0.04$ & $-2.98\pm0.09$ &  1 & 1 \\
895907607894131840 & HS 0740+3734 &  115.99267 & 37.45781 & $20400\pm1273$ & $5.00\pm0.28$ & $-2.10\pm0.57$ & 1 & 1 \\
706760168755893888 & Ton 358 &  133.22757 & 31.72682 & $28900\pm1414$ & $5.37\pm0.21$ & $-2.50\pm0.14$ & 1 & 1 \\
825628714431805056 & BD+48 1777 &  142.69547 & 48.27289 & $47257\pm366$ & $5.88\pm0.06$ & $1.50\pm0.67$ &  1,2 & 1 \\
3839919084402034688 & TYC 4895-599-1 &  145.27204 & -0.79881 & $75065\pm2873$ & $5.37\pm0.14$ & $0.27\pm0.11$ &  1,2 & 1 \\
821412838957446400 & US 1027 &  147.72904 & 46.06812 & $29736\pm810$ & $5.53\pm0.14$ & $-2.40\pm0.50$ &  1,2 & 1 \\
723000539612586752 & Ton 1281 &  160.91386 & 23.15164 & 34800 & $5.26\pm0.01$ & $-1.50\pm0.01$ &  1 & 1 \\
3808164540051860096 & GALEX J104725.1+010847 &  161.85461 & 1.14642 & $21935\pm352$ & $5.06\pm0.06$ & $-2.89\pm0.04$ &  2 & 2 \\
3959631234670040704 & BD+25 2534 &  189.34799 & 25.06651 & $33707\pm1462$ & $6.06\pm0.20$ & $-1.67\pm0.18$ &  1,3 & 1 \\
3675067076961979264 & BD-07 3477b &  191.08437 & -8.67142 & $28494\pm542$ & $5.63\pm0.06$ & $-2.18\pm0.03$ &  1 & 1 \\
3637481302758519040 & NAME NY Vir c &  204.70059 & -2.03035 & $32500\pm50$ & $5.79\pm0.01$ & $-2.91\pm0.08$ &  3 & 1 \\
1172548341214173952 & PG 1431+081 &  218.44825 & 7.90471 & $36600\pm849$ & $6.16\pm0.18$ & $-0.50\pm0.01$ &  1 & 1 \\
1180527806334711936 & PG 1502+113 &  226.30628 & 11.14342 & $34049\pm771$ & $5.84\pm0.16$ & $-2.41\pm0.17$ &  1,2 & 1 \\
1639999937328089088 & PG 1547+632 &  237.04883 & 63.05226 & $31300\pm1131$ & $5.75\pm0.14$ & $-3.20\pm0.71$ &  1 & 1 \\
2151301534625728768 & V* LS Dra &  276.21838 & 57.78987 & $33100\pm1131$ & $6.00\pm0.28$ & $-1.60\pm0.28$ & 1 & 1 \\
2160160432952810496 & FBS 1832+631 &  278.2043 & 63.15294 & $26800\pm990$ & $5.29\pm0.13$ & $-2.60\pm0.14$ &  1 & 1 \\
4077650785664863360 & HD 171858 &  279.48604 & -23.19321 & $27200\pm1131$ & $5.30\pm0.14$ & $-2.84\pm0.10$ &  1 & 1 \\
4299431347569705216 & GALEX J201337.6+092801 &  303.40667 & 9.46707 & $35185\pm267$ & $5.35\pm0.04$ & $-2.68\pm0.04$ &  1 & 1 \\
1876904938896587008 & HS 2233+2332 &  338.94613 & 23.79422 & $26500\pm990$ & $5.30\pm0.14$ & $-2.70\pm0.28$ &  1 & 1 \\
2846162921688127360 & Balloon 90100001 &  348.83956 & 29.08368 & 29400 & $5.33\pm0.01$ & $-2.54\pm0.01$ & 1 & 1 \\
1920513288042722432 & HS 2333+3927 &  353.92711 & 39.74084 & $37600\pm1273$ & $5.75\pm0.14$ & $-2.20\pm0.28$ &  1 & 1 \\
\noalign{\smallskip}\hline
\end{tabular}
\end{center}
$^a$Reference 1 is \citet{2022A&A...662A..40C}, 2 is \citet{2021ApJS..256...28L}, and 3 is \citet{2018ApJ...868...70L}. \\
$^b$Type 1 and 2 represent known HSDs and candidates from \cite{2022A&A...662A..40C}, respectively.
\end{table*}

\subsection{Estimation of radius and mass}
\label{est_mass_rad.sec}

Two methods were adopted to estimate the radius of HSDs in this work.
In the first method, we determined the radii of HSDs following \citep{2023ApJ...953..122L}:
\begin{equation}
\label{rad.eq}
\begin{split}
    \frac{R^{2}}{d^{2}} = \frac{f({\lambda})}{F({\lambda})},
\end{split}
\end{equation}
where $F(\lambda)$ is the model flux at the stellar surface and $f(\lambda)$ is the observed flux. $R$ and $d$ represent the radius and distance, respectively.
Considering that our templates for atmospheric parameter estimation cover 3200--7000 \AA, only the $Gaia$ $BP$-band flux was used for the calculation. 
$F(\lambda)$ was estimated by convolving the best-fit template spectrum (obtained from the CNN method) with the $BP$ filter transmission curve.
Meanwhile, the observed $BP$-band flux $f(\lambda)$ was derived from the observed magnitude after correcting for extinction.

In the second method, we calculated the radius ($R$) for each HSD using Gaia magnitudes according to the following formulae:
\begin{equation}
\label{mass_rad.eq}
\begin{split}
    R &= \frac{L_{\odot}}{4\pi \sigma T^{4}} \\
      &\quad \times 10^{-\frac{\displaystyle m_{\mathrm{\lambda}} - 5\log d - A_{\mathrm{\lambda}} - M_{\odot} + BC_{\mathrm{\lambda}} + 5}{\displaystyle 2.5}} ,
\end{split}
\end{equation}
where $\sigma$ is the the Stefan-Boltzmann constant. $L_{\odot}$ is solar bolometric luminosity (3.828$\times10^{33}$ erg/s) and $M_{\odot}$ is solar bolometric magnitude (4.74 mag). Here, $m_{\rm \lambda}$ is the apparent magnitude of three bands ($G$, $BP$, and $RP$), $d$ is the distance, and $A_{\rm \lambda}$ is the extinction of each band. The bolometric correction $BC_{\rm \lambda}$ is computed using the {\it isochrones} package \citep{2015ascl.soft03010M}, based on the stellar atmospheric parameters as input.
The final radius was determined by averaging the results derived from the three bands.
The uncertainties are the root sum of squares of the errors.

\begin{figure}[t]
    \center
    \includegraphics[width=0.48\textwidth]{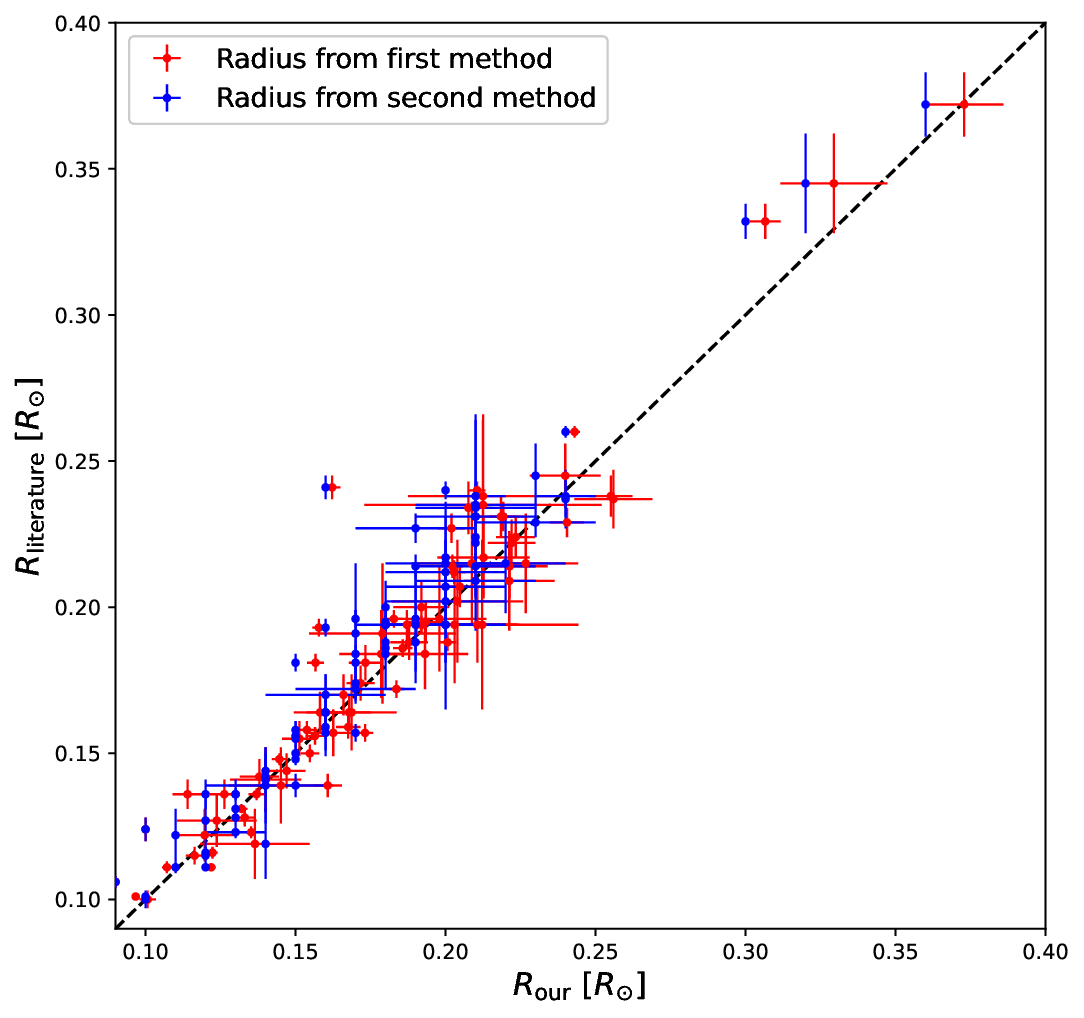}
    \caption{Comparison of the radii derived from the two methods with the results from \cite{2023ApJ...953..122L}. The red and blue dots represent the radii from the first and second methods, respectively.}
    \label{radius_compare.fig}
\end{figure}

Figure \ref{radius_compare.fig} presents a comparison of the radii derived from the two methods with the results from \cite{2023ApJ...953..122L}, which determined stellar radius using the spectral energy distribution (SED) fitting method.
The radii derived from the first method are consistent with those reported by \cite{2023ApJ...953..122L}, while the second method yields radii that are systematically slightly underestimated. This offset is likely attributable to uncertainties in $BC_{\rm \lambda}$, which is based on single-star evolutionary models. Therefore, we chose to use the radius estimates from the first method in the following analysis.

\begin{figure*}[t]
    \center
    \includegraphics[width=0.95\textwidth]{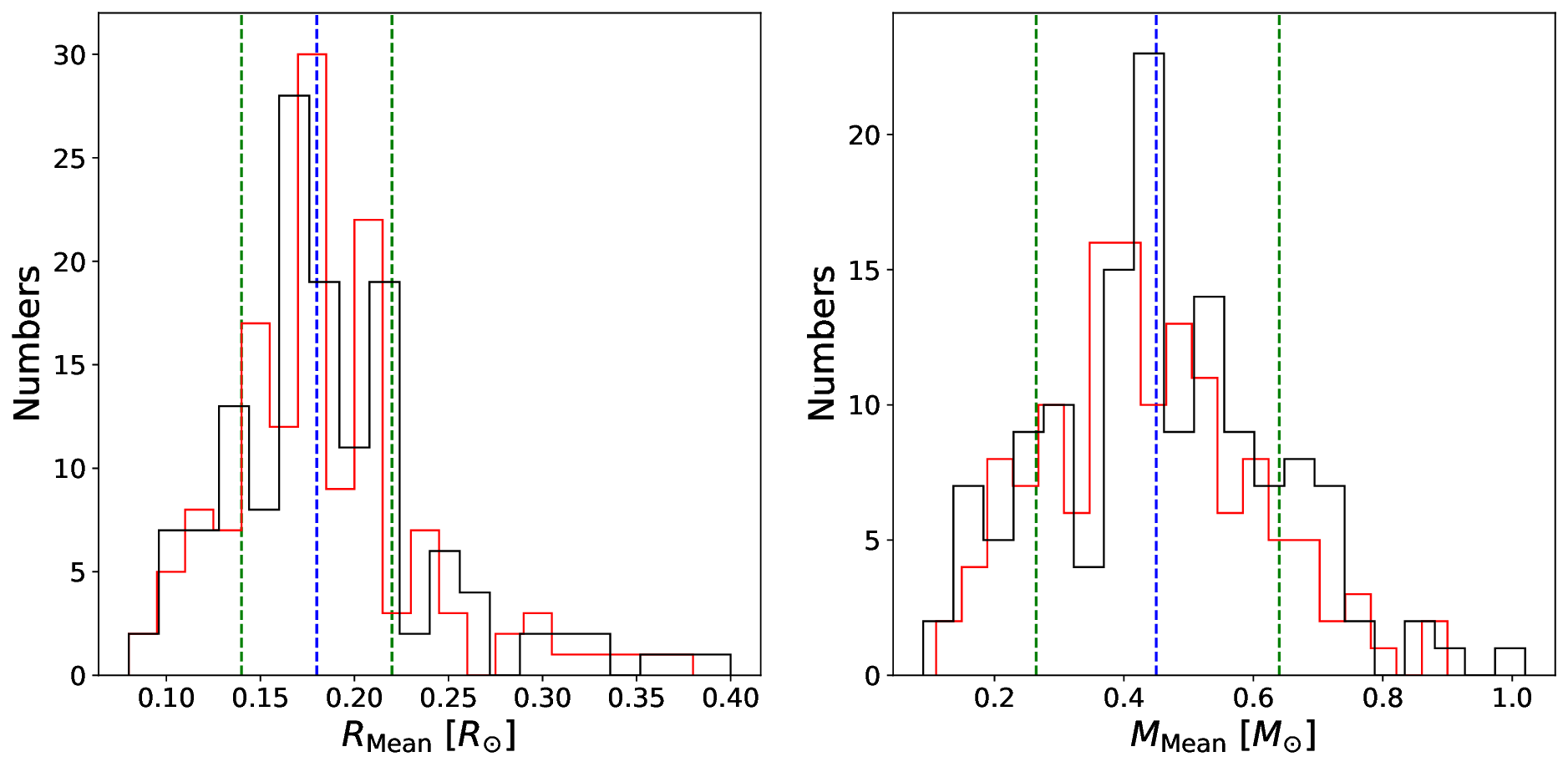}
    \caption{Distribution of the radius and mass of HSDs from atmospheric parameters derived using the CNN method. The black and red histograms correspond to the radius and mass estimates derived from the first method and the second method, respectively. The blue and green lines represent the median, along with the 16th and 84th percentiles, of the radius and mass estimates derived from the first method.} 
    \label{mass_radius_distribution.fig}
\end{figure*}

The mass of the HSDs was derived as follows:
\begin{equation}
\label{mass.eq}
    M = \frac{g R^{2}}{G},
\end{equation}
where $G$ denotes the gravitational constant.
Figure \ref{mass_radius_distribution.fig} shows the distribution of the radius and mass, peaking at $0.18^{+0.04}_{-0.05} R_{\odot}$ and $0.45^{+0.19}_{-0.17}\ M_{\odot}$ (estimated from the first method), respectively.
The peak mass is consistent with that reported in previous studies \citep[e.g., {\bf $\sim0.46\ M_{\odot}$};][]{2023ApJ...953..122L,2026A&A...707A...6D}.
%
%
%
The radius and mass  estimates for our sample are listed in Table \ref{mass.tab}.

\begin{table*}
\caption{Mass and radius estimates for our sample. The superscripts "1" and "2" denote the estimates from the first and second methods, respectively. \label{mass.tab}}
\centering
\setlength{\tabcolsep}{1.5mm}
\begin{center}
\begin{tabular}{llcccc}
\hline\noalign{\smallskip}
Gaia ID & SIMBAD ID & $M^{1}$ & $R^{1}$ & $M^{2}$ & $R^{2}$ \\
 &  & ($M_{\odot}$) & ($R_{\odot}$) & ($M_{\odot}$) & ($R_{\odot}$) \\
\hline\noalign{\smallskip}
2767874292175410560 & PG 0004+133 & $0.46\pm0.03$ & $0.20\pm0.01$ & $0.42\pm0.03$ & $0.19\pm0.01$ \\
2850670743266825600 & GALEX J001052.4+263000 & $0.25\pm0.12$ & $0.19\pm0.04$ & $0.23\pm0.12$ & $0.18\pm0.02$ \\
2581810261598516096 & PG 0048+091 & $0.24\pm0.03$ & $0.22\pm0.01$ & $0.29\pm0.05$ & $0.24\pm0.02$ \\
312628749626419328 & SDSS J010113.32+312555.3 & $0.64\pm0.05$ & $0.20\pm0.01$ & $0.58\pm0.07$ & $0.19\pm0.01$ \\
2551900379931546240 & Feige 11 & $0.53\pm0.02$ & $0.19\pm0.01$ & $0.47\pm0.02$ & $0.18\pm0.01$ \\
369576820516013824 & FBS 0102+362 & $0.40\pm0.03$ & $0.16\pm0.01$ & $0.37\pm0.05$ & $0.15\pm0.01$ \\
369573831218501376 & FBS 0106+374 & $0.41\pm0.08$ & $0.20\pm0.01$ & $0.37\pm0.13$ & $0.20\pm0.02$ \\
401413450281523584 & UCAC4 686-007194 & $0.90\pm0.17$ & $0.10\pm0.01$ & $0.90\pm0.17$ & $0.10\pm0.01$ \\
371711041304831616 & FBS 0117+396 & $0.42\pm0.07$ & $0.22\pm0.02$ & $0.37\pm0.08$ & $0.21\pm0.02$ \\
397395014455786624 & UCAC4 679-007631 & $0.36\pm0.07$ & $0.16\pm0.01$ & $0.34\pm0.15$ & $0.16\pm0.03$ \\
2592234628262141312 & PG 0123+159 & $0.64\pm0.05$ & $0.17\pm0.01$ & $0.63\pm0.05$ & $0.17\pm0.01$ \\
344155939885410816 & ATO J029.0051+40.0561 & $0.51\pm0.02$ & $0.20\pm0.01$ & $0.47\pm0.02$ & $0.20\pm0.01$ \\
2514278566658235776 & PG 0209+017 & $0.44\pm0.06$ & $0.16\pm0.01$ & $0.40\pm0.09$ & $0.15\pm0.01$ \\
341195058150383104 & FBS 0229+439 & $0.55\pm0.06$ & $0.14\pm0.01$ & $0.55\pm0.05$ & $0.14\pm0.01$ \\
123108816565129344 & GALEX J030128.0+301537 & $0.19\pm0.05$ & $0.08\pm0.01$ & $0.19\pm0.05$ & $0.08\pm0.01$ \\
3266182479530739328 & PG 0313+005 & $0.16\pm0.03$ & $0.16\pm0.01$ & $0.19\pm0.05$ & $0.17\pm0.01$ \\
9132341717393792 & PG 0319+055 & $0.39\pm0.05$ & $0.13\pm0.01$ & $0.41\pm0.07$ & $0.13\pm0.01$ \\
64368713521485568 & UCAC4 561-007652 & $0.63\pm0.06$ & $0.22\pm0.01$ & $0.58\pm0.07$ & $0.21\pm0.01$ \\
3299455110137611776 & HZ 12 & $0.53\pm0.03$ & $0.31\pm0.01$ & $0.49\pm0.05$ & $0.30\pm0.01$ \\
\noalign{\smallskip}\hline
\end{tabular}
\end{center}
NOTE. This table is available in its entirety in machine-readable and Virtual Observatory (VO) forms in the online journal. A portion is shown here for guidance regarding its form and content.
\end{table*}

\section{Keplerian orbit fitting}
\label{orbit_fitting.sec}

\subsection{RV fitting}
\label{orbit.sec}

We first selected possible binaries for RV fitting. We quantified the RV variability of our sample using the following formulae \citep{2000MNRAS.319..305M,2017MNRAS.468.2910B}:
\begin{equation}
    \eta = -{\rm log}_{10}[P(\chi^{2} > \chi^{2}_{m})],
\end{equation}
where $\chi^{2}_{m}$ represents the $\chi^{2}$ value calculated under the null hypothesis that the RV of each HSD remains constant.
Some studies \citep{2024A&A...690A.368G,2025A&A...693A.121H} adopted $\eta>4$ as the criterion for distinguishing binary and single-star systems, while others \citep{2017MNRAS.468.2910B} used the threshold of $\eta>2.5$. 
In our sample, 73 systems have $\eta>2.5$ and 66 systems have $\eta>4$. 
Given that our sample has more RV observations, we adopted $\eta>2.5$ as the criterion for distinguishing binary and single-star systems.
%
%
The blue stars in Figure \ref{HR.fig} denote those systems.
Furthermore, only RV measurements with the uncertainty ($\sigma_{\rm RV}$) less than 10 km/s were used in RV fitting.

Close HSD binaries are thought to form through CE ejection. 
Although the CE phase is very short, it is highly efficient in circularizing the binary orbit. 
As a result, this process typically leads to systems with very small orbital eccentricities ($e<0.06$) \citep{2005A&A...442.1023E,2015A&A...576A..44K}.
%
%
Most systems in our sample exhibit short orbital periods (Table \ref{orbital_params_from_cir.tab}), suggesting that they have likely undergone sufficient circularization during the CE phase. 
Thus, we adopted a circular orbit fitting by fixing the eccentricity $e=0$ for our sample.

For a circular orbit, the RV curve of the HSD can be expressed by a sine function as follows:
\begin{equation}
    \mathrm{RV} = K \sin(\theta + \omega) + V_0,
    \label{eq:rv_curve}
\end{equation}
where $K$ is the semi-amplitude of RV of the HSD, $V_0$ is the systemic velocity, $\omega$ is the argument of periastron, and $\theta$ is the true anomaly. 
We initially estimated the orbital period ($P$) from the RV data using the LS method over a period range of 0.02 to 100 days. 
To assess the period uncertainty, we performed a uniform grid search by sampling 10000 period values within the range of 0.5$P$ to 1.5$P$. 
For each sampled period, we fitted the observed RV data using Equation \eqref{eq:rv_curve} and calculated corresponding $\chi^{2}$. 
The period uncertainty was then derived based on the $\Delta \chi^{2}$ profile.
However, for sources with sparse RV phase coverage or large RV uncertainties, we could not derive a reliable period uncertainty and only gave the best‑fit period value (Table \ref{orbital_params_from_cir.tab}).
%

%

%

After visually checking the fitting results, we successfully obtained orbital solutions for 17 systems using circular orbit fitting.
Taking those parameters, we calculated the minimum mass ($i=90^{\circ}$) of the companion using the mass function as follows:
\begin{equation}
    f(M_{2}) = \frac{M_{2} \, \textrm{sin}^3 i} {(1+q)^{2}} = \frac{P \, K_{1}^{3}}{2\pi G},
    \label{eq:mf}
\end{equation}
where $M_{2}$ is the mass of the companion in binary system, $q = M_{1}/M_{2}$ is the mass ratio, and $i$ is the orbital inclination angle.
%
%
%
The fitting results are listed in Table \ref{orbital_params_from_cir.tab}.
Figure \ref{periodograms.fig} presents the LS periodograms for systems with reliable period uncertainty estimates.
Figure \ref{circ_fitting.fig} shows the fitting results from the circular-orbit model.


\begin{table*}
\caption{Keplerian orbit for our binary sample under the assumption of circular orbits. For values of $\eta$ exceeding 300, we applied an upper truncation limit of 300. \label{orbital_params_from_cir.tab}}
\centering
\setlength{\tabcolsep}{2mm}
\begin{center}
\begin{tabular}{lcccccc}
\hline\noalign{\smallskip}
Gaia ID & $P$ & $K_{1}$ & $e$ & $v_{0}$ & $\eta$ & Ref.$^a$ \\
 & (days) & (km/s) &  & (km/s) \\
\hline\noalign{\smallskip}
3637481302758519040 & 0.10101582 & $84.83\pm8.74$ & 0 (fixed) & $2.42\pm0.10$ & 300.00 & (1) \\
2080063931448749824 & $0.11370455\pm0.00480500$ & $74.85\pm1.43$ & 0 (fixed) & $-22.12\pm1.04$ & 300.00 & (2) \\
1876904938896587008 & $0.16281736\pm0.04021991$ & $32.42\pm1.55$ & 0 (fixed) & $1.21\pm0.09$ & 57.16 & New \\
1914301803258669440 & $0.19877046\pm0.02470964$ & $86.18\pm0.71$ & 0 (fixed) & $-34.71\pm0.49$ & 300.00 & New \\
859683853719128192 & $0.32948825\pm0.05097693$ & $20.48\pm1.16$ & 0 (fixed) & $-37.39\pm0.95$ & 300.00 & (3) \\
1385361084513370368 & $0.33022568\pm0.14577619$ & $18.00\pm1.55$ & 0 (fixed) & $-2.80\pm2.08$ & 26.68 & New \\
3344334627867111168 & $0.39814865\pm0.04364146$ & $75.73\pm3.92$ & 0 (fixed) & $1.58\pm0.06$ & 300.00 & New \\
2551900379931546240 & $0.56990086\pm0.00048446$ & $104.82\pm0.62$ & 0 (fixed) & $6.93\pm0.48$ & 300.00 & (4) \\
697406932576439680 & 0.7561205 & $75.88\pm8.51$ & 0 (fixed) & $-3.52\pm0.26$ & 44.80 & New \\
4023163971460301952 & 1.01265504 & $81.64\pm6.28$ & 0 (fixed) & $-1.20\pm0.19$ & 300.00 & (5) \\
694109462844643072 & 1.19243697 & $42.59\pm1.02$ & 0 (fixed) & $-75.37\pm0.72$ & 15.17 & (6) \\
1427969118595251072 & 2.32429309 & $38.74\pm0.61$ & 0 (fixed) & $-0.67\pm0.02$ & 300.00 & New \\
3452480293768439296 & $6.22739615\pm0.00093420$ & $60.67\pm1.82$ & 0 (fixed) & $-4.18\pm1.41$ & 300.00 & New \\
3373814561832902272 & $6.25647547\pm0.00250284$ & $35.60\pm1.21$ & 0 (fixed) & $31.08\pm0.77$ & 300.00 & New \\
1595357428778377344 & $8.72207387\pm0.003925326$ & $53.42\pm1.47$ & 0 (fixed) & $2.12\pm0.02$ & 300.00 & New \\
1891098500140100352 & $21.64237096\pm0.00974004$ & $40.71\pm0.28$ & 0 (fixed) & $-73.32\pm0.21$ & 300.00 & New \\
611684298790301568 & $27.78048114\pm1.10160624$ & $28.12\pm1.66$ & 0 (fixed) & $31.87\pm1.31$ & 300.00 & (5) \\
\noalign{\smallskip}\hline
\end{tabular}
\end{center}
$^a$Ref. means whether the orbital parameters are first derived in this work or have been previously determined in the literature. (1) \cite{2007A&A...471..605V} and \cite{2008A&A...489..377C}, (2) \cite{2010MNRAS.408L..51O}, (3) \cite{2004A&A...422.1053O}, \cite{2006BaltA..15...61O} and \cite{2023A&A...673A..90S}, (4) \cite{2008A&A...477L..13G}, (5) \cite{2003MNRAS.338..752M} and \cite{2001MNRAS.326.1391M}, (6) \cite{2011MNRAS.415.1381C}.
\end{table*}

\subsection{Joint fitting of RV and LC}
\label{lc_fitting.sec}

\subsubsection{Light curves}
\label{lc.sec}

LCs can be used to distinguish the nature of the companions in most close HSD binaries, as different types of companion produce distinct LC morphologies.
In close systems with low-mass MS companions, a large temperature difference can lead to strong irradiation on the hemisphere facing the HSD. This produces a reflection effect, characterized by a sharp peak and a wide trough in the LCs.
%
%
%
%
Due to the high irradiating flux from the HSD, the temperature of the heated hemisphere can be significantly increased, giving rise to photometric amplitudes ranging from a few percent up to 20\% (see Figure \ref{irad_fitting.fig}). 
For example, for a companion with a intrinsic effective temperature of 3000 K, the irradiated side can reach temperatures of 10000--20000 K \citep{2000A&A...364..199K,2008ASPC..392..199V}.
When the companion reaches the superior conjunction (i.e., $\phi$ = 0 in Figure \ref{irad_fitting.fig}), the heated hemisphere is visible, resulting in the maximum brightness.
Given the small orbital separations, eclipses may occur in the binaries with a high inclination, producing obvious eclipse features in LC \citep{1993MNRAS.261..103W}.
Such systems are known as HW Vir binaries.

Compared to reflection systems, HSD+WD binaries typically do not exhibit prominent reflection features in their LCs, due to the smaller radius and high effective temperature of WDs. 
Instead, the LCs are primarily dominated by ellipsoidal modulation caused by tidal distortion.
Maximum brightness of HSD+WD binaries occurs at the phase of quadrature for the HSD ($\phi$ = 0.25 and 0.75), corresponding to the times of maximum redshift and blueshift in the RV curve, which is different with reflection effect.
In addition, the Doppler beaming may be detectable in HSD+WD systems, resulting in significant flux differences at quadrature \citep{2013A&A...554A..54G,2014A&A...570A.129T,2017ApJ...851...28K,2017ApJ...835..131K,2021NatAs...5.1052P,2025SCPMA..6869511L,2025A&A...693A.322Y}.

To obtain more accurate estimates of orbital periods, we first derived the orbital periods for each system in our binary sample using photometric data. 
Additionally, the LCs of these systems help constrain the orbital inclination, thereby enabling more precise determination of the companion masses. 
Thus, we compiled LCs from ASAS-SN, ZTF, and TESS for the binaries in our sample. 
For each LC, we applied the LS method to search for periodic signals within the range of 0.02 to 100 days. 
To ensure the photometric period corresponds to the orbital period, we require that it can fold the RV data well.
After visually inspecting the results of the LS method, we identified 18 binary systems whose LCs exhibit clear reflection features, and 8 systems whose LCs show shapes resembling ellipsoidal modulation, suggesting the possible presence of compact companions.

\subsubsection{Reflection}
\label{reflection.sec}

\begin{figure*}
    \includegraphics[width=0.24\textwidth]{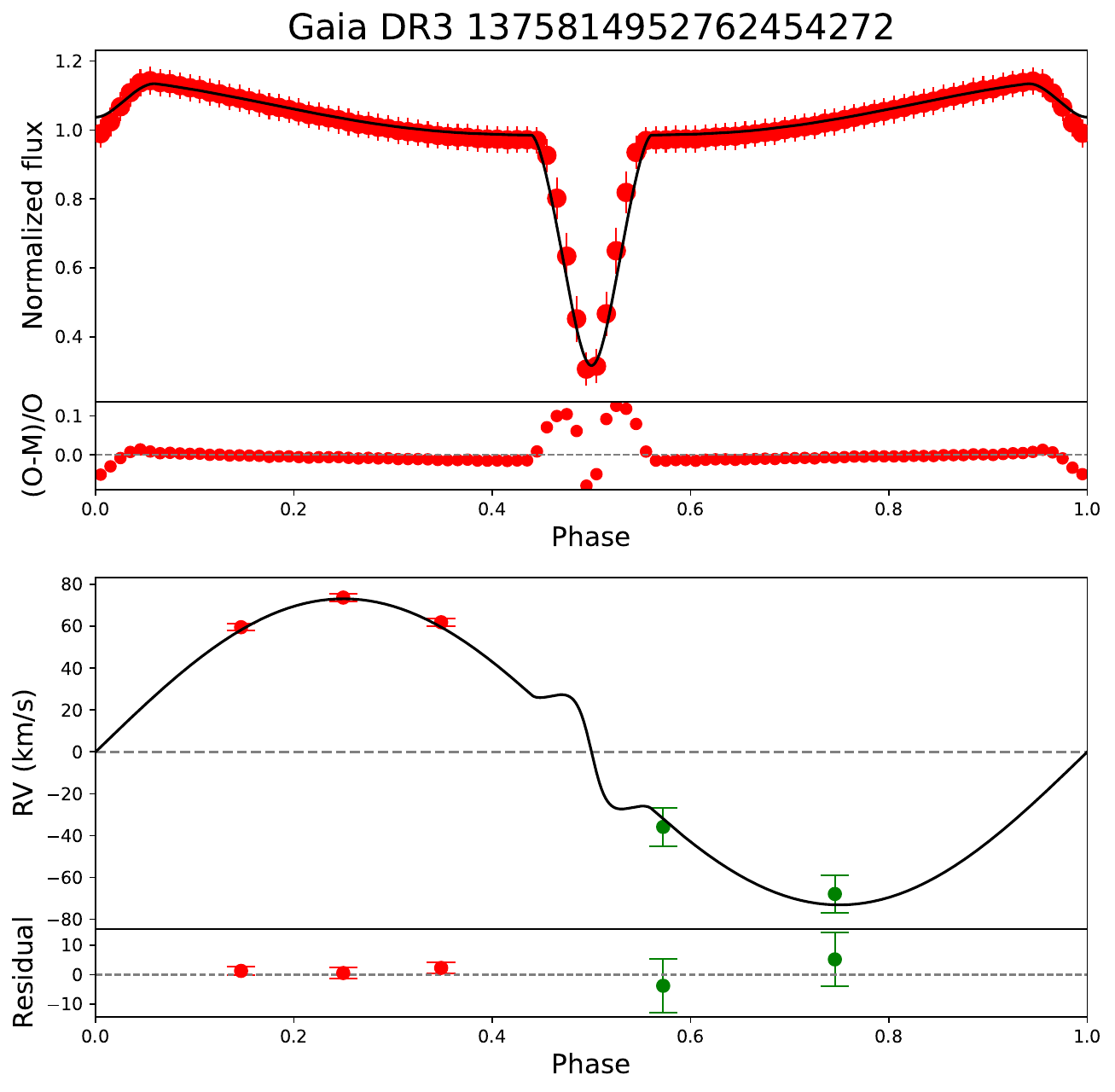}
    \includegraphics[width=0.24\textwidth]{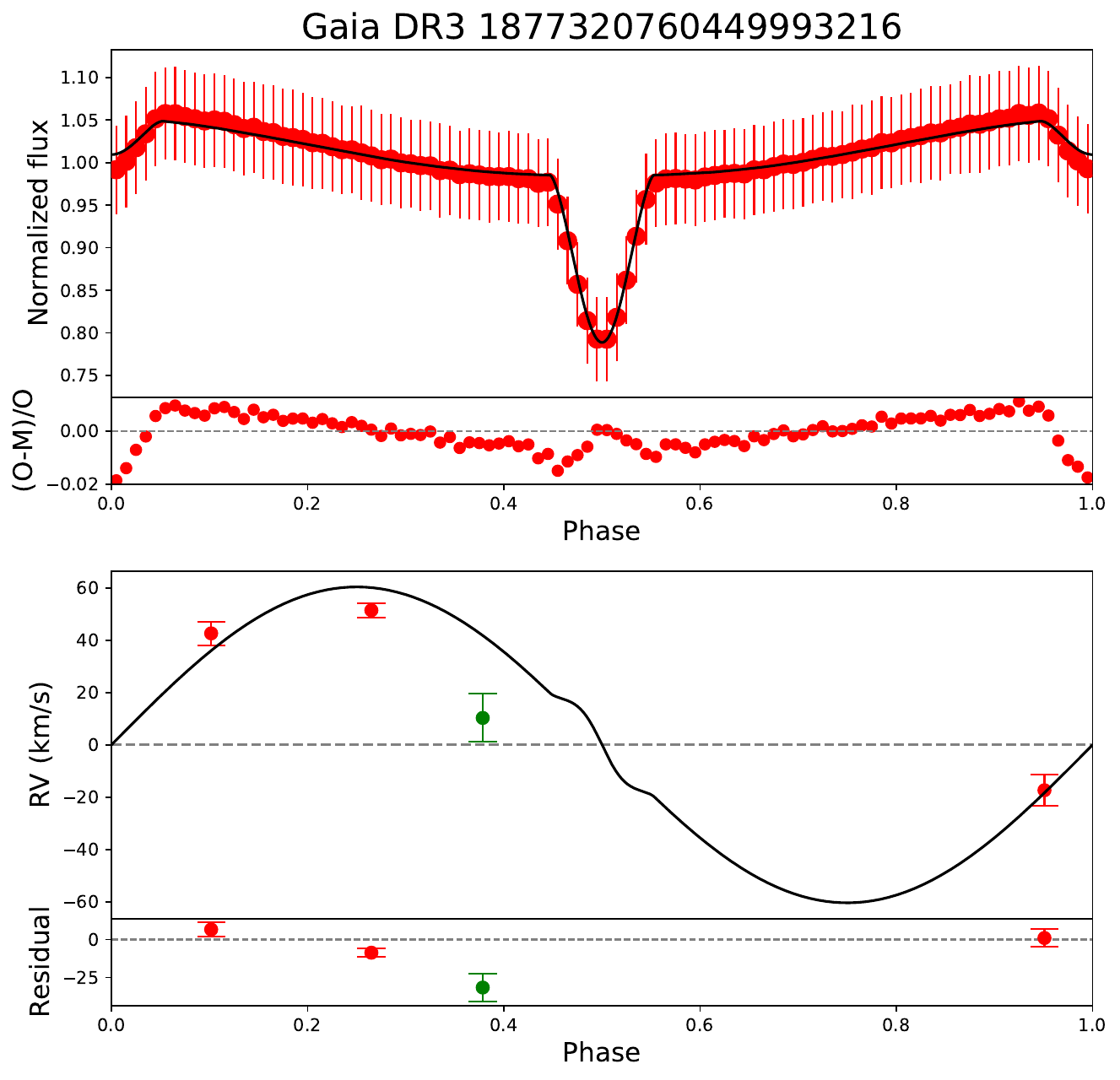}
    \includegraphics[width=0.24\textwidth]{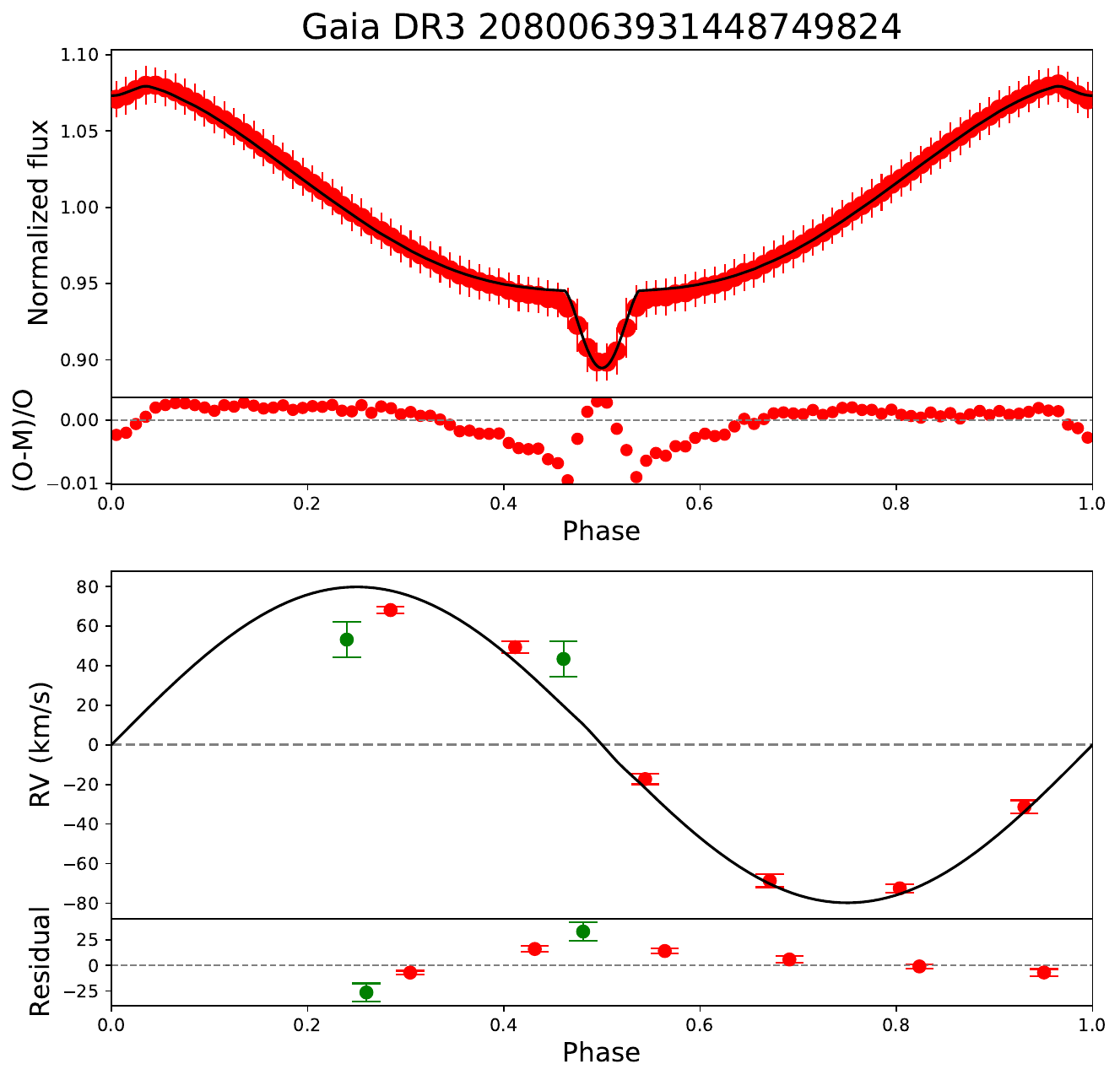}
    \includegraphics[width=0.24\textwidth]{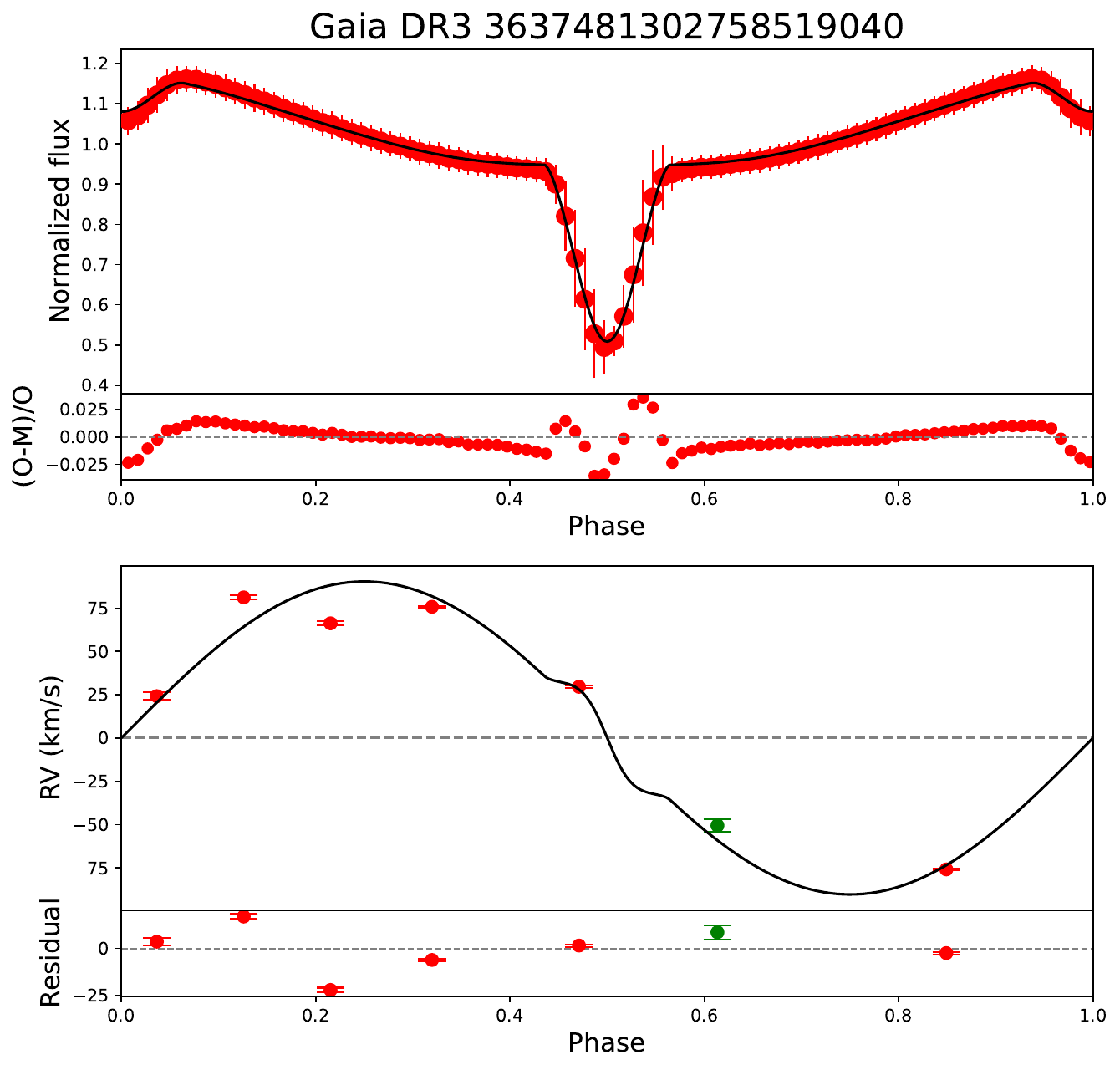}
    \vspace{1.5ex}
    \includegraphics[width=0.24\textwidth]{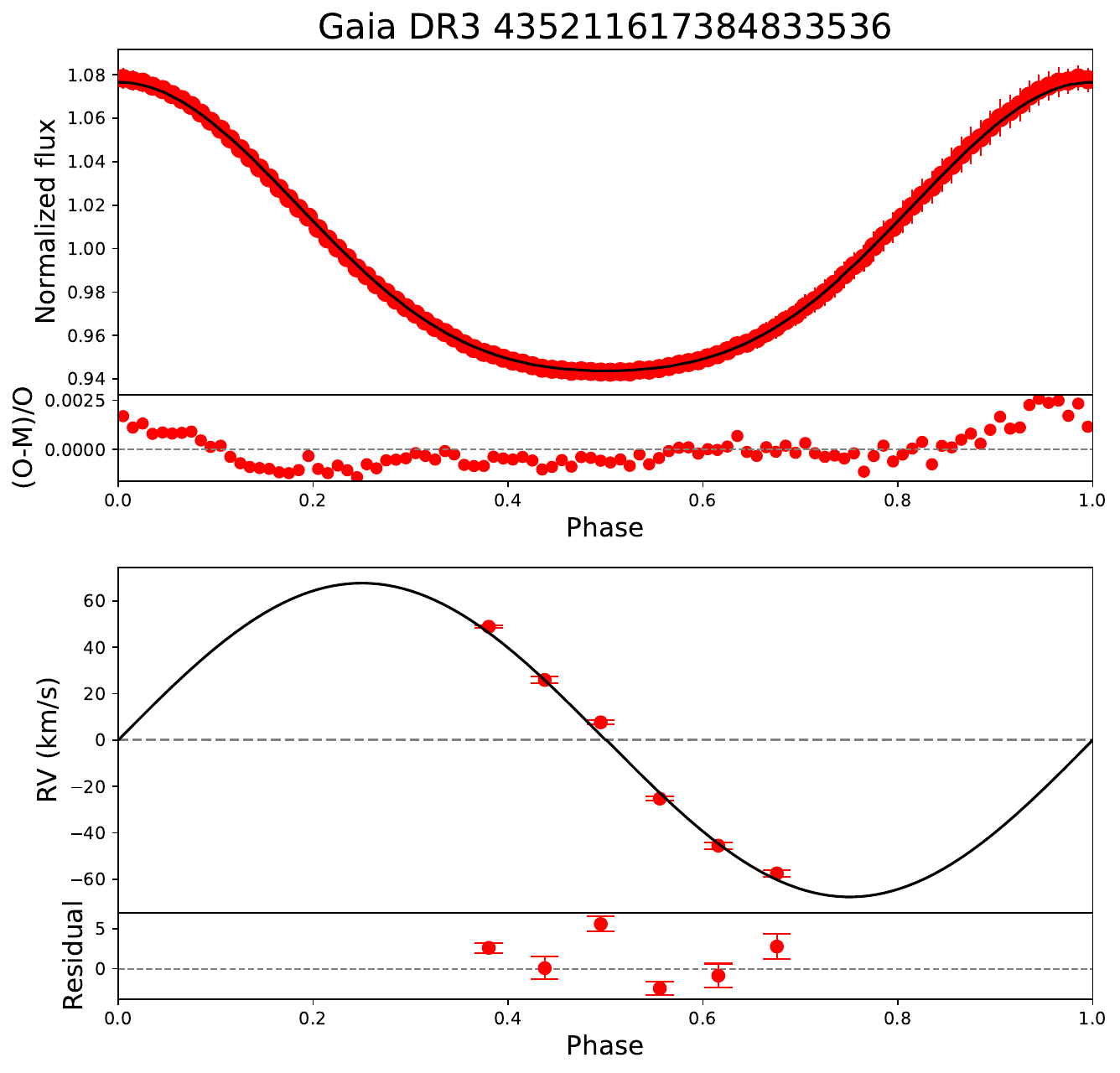}
    \includegraphics[width=0.24\textwidth]{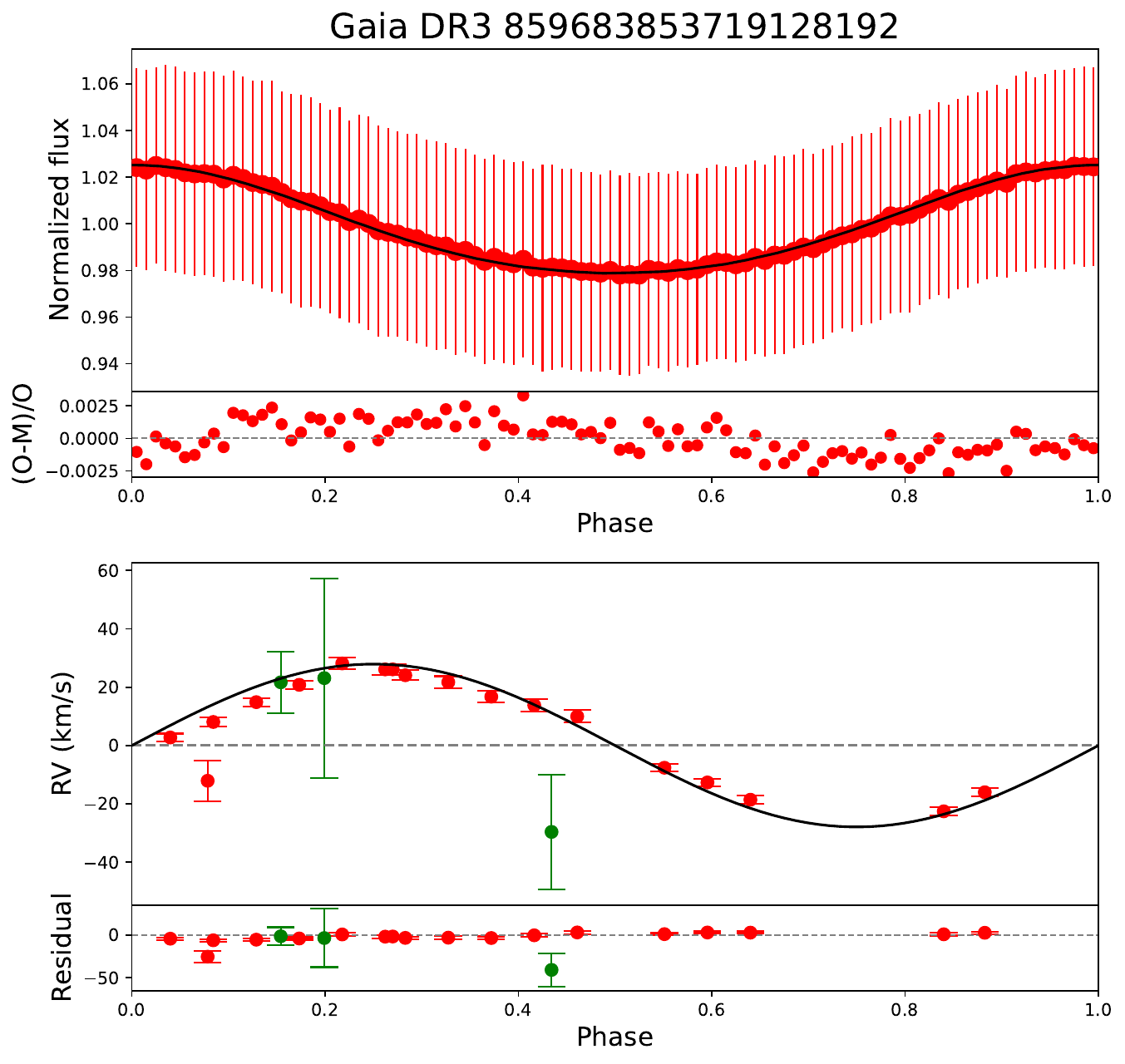}
    \includegraphics[width=0.24\textwidth]{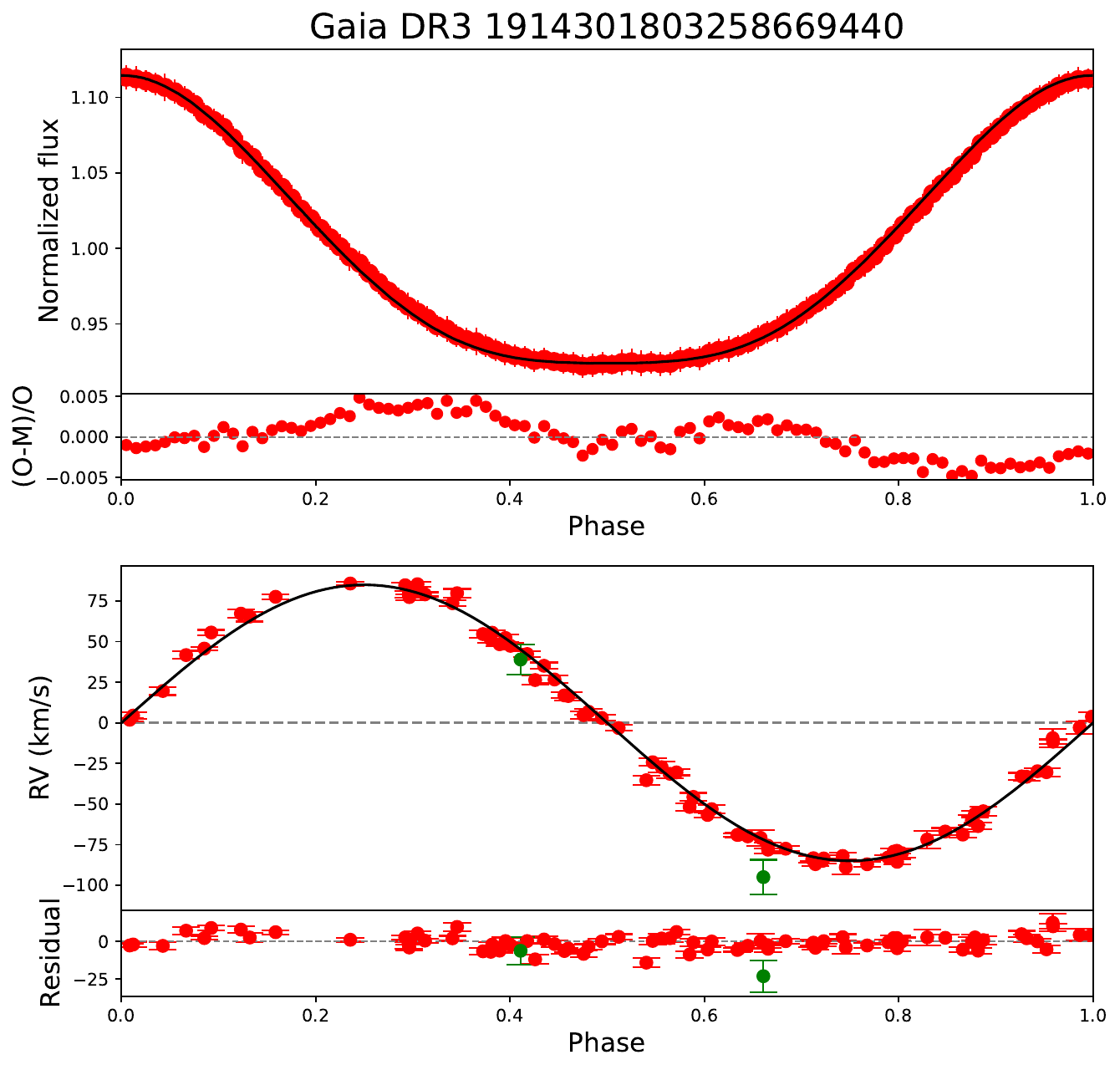}
    \includegraphics[width=0.24\textwidth]{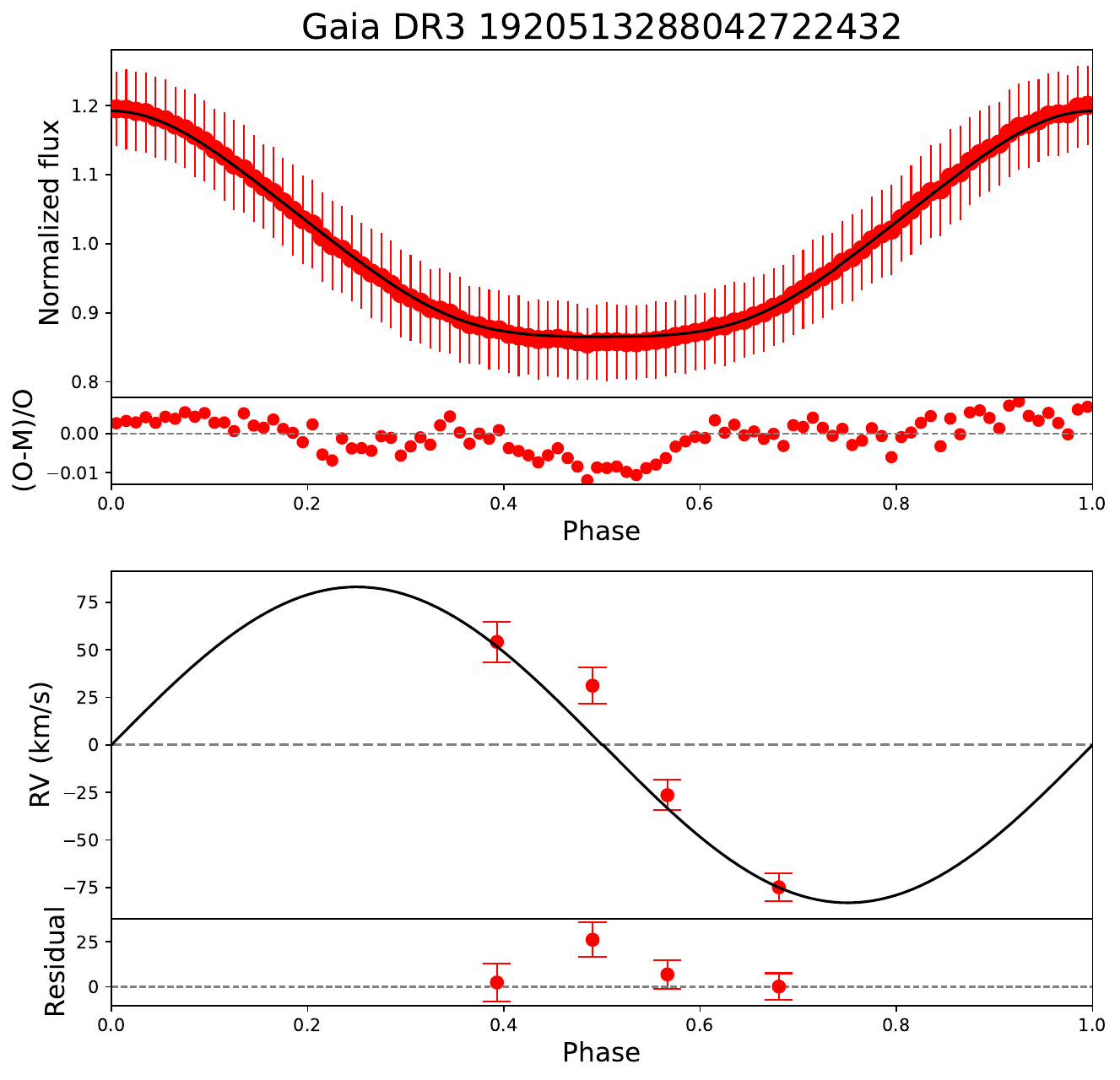}
    \vspace{1.5ex}
    \includegraphics[width=0.24\textwidth]{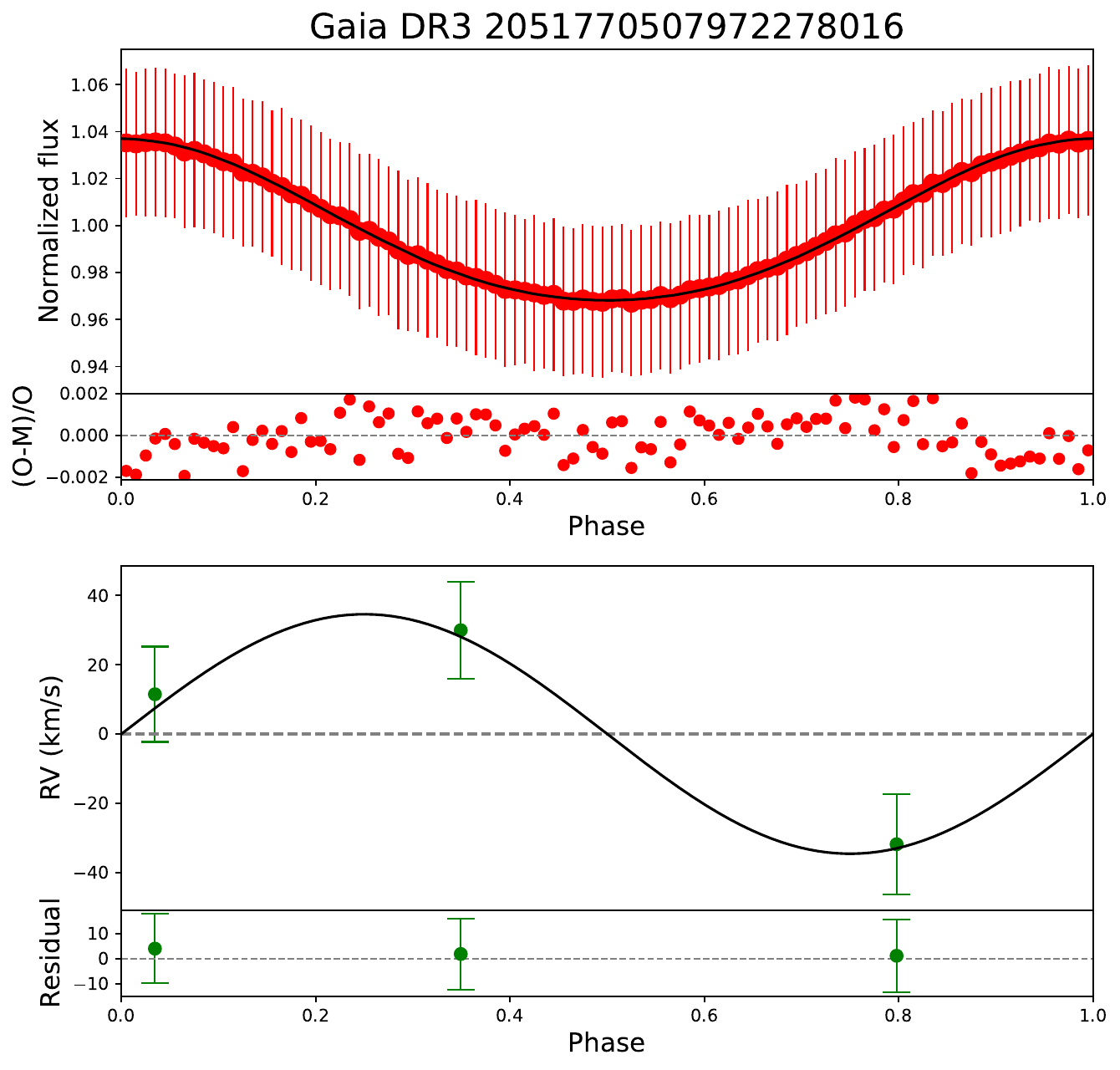}
    \includegraphics[width=0.24\textwidth]{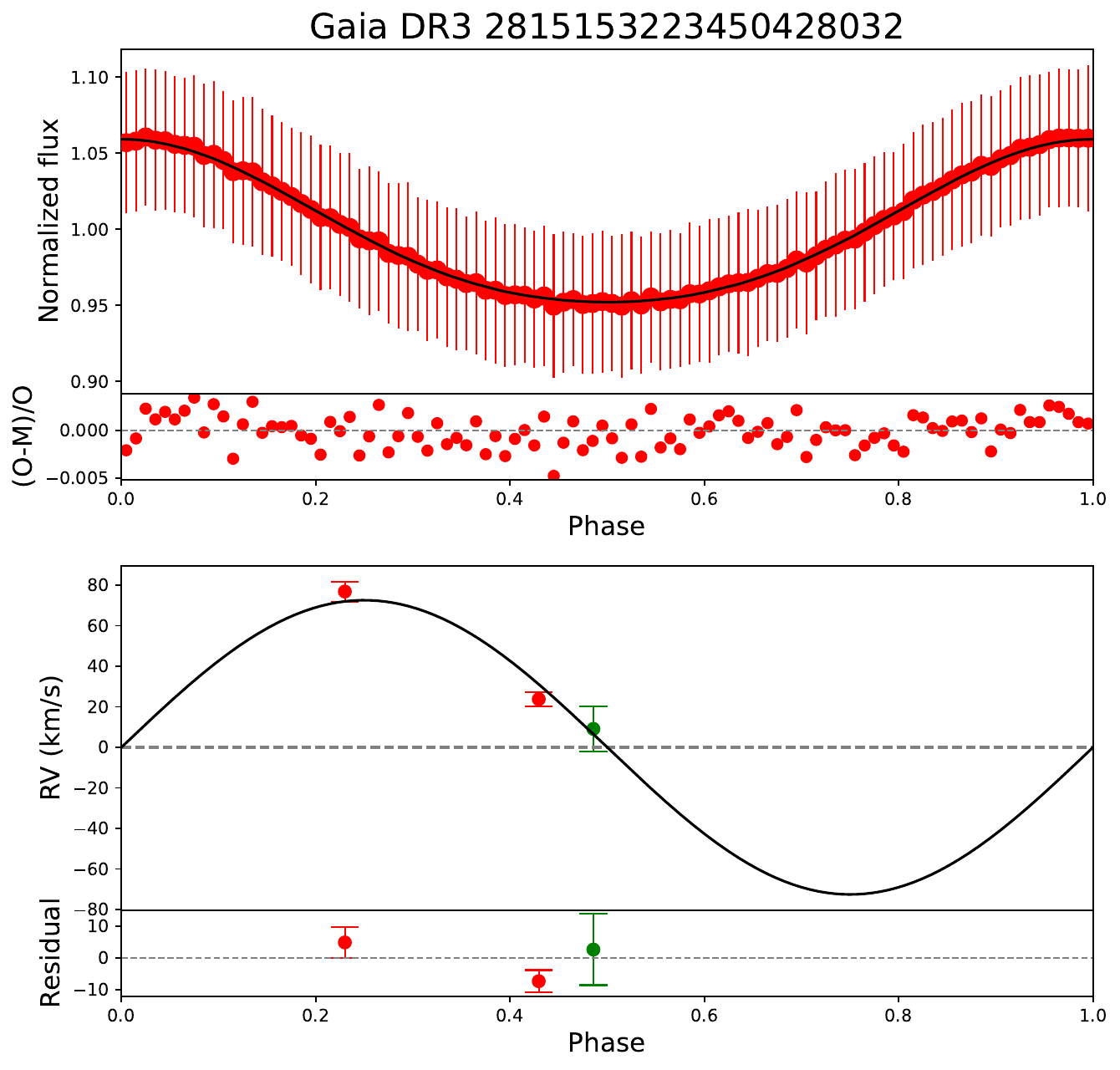}
    \caption{The joint LC and RV fitting for the 10 reflection binaries in our binary sample. The red dots represent TESS LC binned into 100 phase intervals. The red and green circles indicate the RV data from LAMOST MRS and LRS, respectively. The black lines are the best-fitting models derived from PHOEBE. Given that the LCs yield robust orbital periods, only RV data with uncertainties smaller than 15 km/s are used in the fitting, and RVs with larger errors are shown only for completeness.} 
    \label{irad_fitting.fig}
\end{figure*}

In close binaries composed of an HSD and a late-type companion, reflection from the HSD typically induces a characteristic LC with a single maximum and a single minimum, often exhibiting a large amplitude of variability (as shown in Figure \ref{irad_fitting.fig}).
To investigate the nature of the companions and the orbital parameters in such reflected systems, we jointly fitted the RV and LC data for those systems in our binary sample.
After excluding RV data with large uncertainties ($\sigma_{\rm RV}>$15 km/s), only systems whose RV phase coverage larger than 0.1 were retained for analysis.
Finally, 6 reflection binaries and 4 HW Vir binaries were successfully modeled (Figure \ref{irad_fitting.fig}).

The software Physics of eclipsing binaries \citep{2016ApJS..227...29P,2018ApJS..237...26H,2020ApJS..250...34C} (PHOEBE) was employed to generate a grid of synthetic LCs for the subsequent fitting process.
A detached binary configuration was assumed for the system modeling.
For the systematic parameters, the orbital period was fixed according to the observed LC, and the eccentricity was set to $e=0$.
%
For the HSDs, the effective temperature $T_{\rm eff}$ was fixed to the atmospheric value obtained in Section \ref{atmo.sec}, while the mass $M$ and radius $R$ were set according to the results presented in Section \ref{est_mass_rad.sec}.
For the companion, Gaussian priors on radius and mass were adopted based on stellar evolution models.
Additionally, uniform priors were applied for the effective temperature and orbital inclination ($i$) with ranges of [2500, 6000] K and [$0^{\circ}$, $90^{\circ}$], respectively.

The TESS LCs were used in the joint fitting due to their low dispersion.
In the fitting, an MCMC sampler was employed to perform the fit, using 20 walkers and 10000 iterations.
Figure \ref{irad_fitting.fig} presents the best-fit models for the 10 systems, and the corresponding fitting results are summarized in Table \ref{orbital_params_irad.tab}.
The companions in these systems are mostly M-type MS stars with small radii and low masses.

\begin{table*}
\caption{Best-fit orbital parameters and companion properties of the 10 reflection-dominated binaries in our sample. \label{orbital_params_irad.tab}}
\centering
\setlength{\tabcolsep}{1mm}
\begin{center}
\begin{tabular}{lcccccccc}
\hline\noalign{\smallskip}
Gaia ID & $P$ & $T_{\rm eff, 2}$ & $R_{\rm 2}$ & $M_{\rm 2}$ & $i$ & $v_{\rm 0}$ & Ref.$^a$ & CROWDSAP \\
 & (days) & (K) & ($R_{\odot}$) & ($M_{\odot}$) & ($^{\circ}$) & (km/s) & & \\
\hline\noalign{\smallskip}
1375814952762454272 & 0.16177 & $3105.85\pm131.35$ & $0.21\pm0.02$ & $0.13\pm0.01$ & $83.08\pm0.43$ & $-4.10\pm6.33$ & (1) & 0.99 \\
1877320760449993216 & 0.11059 & $2502.41\pm15.26$ & $0.12\pm0.01$ & $0.10\pm0.01$ & $77.65\pm0.73$ & $-64.28\pm4.64$ & (2) & 0.86 \\
2080063931448749824 & 0.12578 & $3105.30\pm6.36$ & $0.21\pm0.01$ & $0.17\pm0.01$ & $68.28\pm0.15$ & $-17.07\pm4.11$ & (3) & 0.86 \\
3637481302758519040 & 0.10101 & $3119.56\pm81.34$ & $0.20\pm0.01$ & $0.14\pm0.01$ & $76.98\pm0.04$ & $-19.82\pm1.21$ & (4) & 0.95 \\
435211617384833536 & 0.26586 & $3044.10\pm187.96$ & $0.29\pm0.01$ & $0.18\pm0.01$ & $45.04\pm1.97$ & $64.41\pm2.53$ & (5) & 0.85 \\
859683853719128192 & 0.34360 & $3094.46\pm30.14$ & $0.25\pm0.02$ & $0.18\pm0.03$ & $27.54\pm4.82$ & $-41.14\pm1.83$ & (6) & 0.99 \\
1914301803258669440 & 0.19877 & $3097.21\pm23.31$ & $0.30\pm0.01$ & $0.19\pm0.01$ & $65.25\pm2.62$ & $-34.66\pm2.38$ & New & 0.93 \\
1920513288042722432 & 0.17189 & $3213.64\pm23.82$ & $0.33\pm0.01$ & $0.21\pm0.01$ & $62.88\pm1.40$ & $-14.57\pm3.32$ & (5) & 0.78 \\
2051770507972278016 & 0.29237 & $3210.60\pm54.54$ & $0.33\pm0.03$ & $0.33\pm0.03$ & $19.37\pm4.14$ & $-7.72\pm6.11$ & (5) & 0.59 \\
2815153223450428032 & 0.16353 & $3110.93\pm36.01$ & $0.23\pm0.01$ & $0.28\pm0.03$ & $30.88\pm2.27$ & $-57.72\pm6.89$ & New & 0.94 \\
\noalign{\smallskip}\hline
\end{tabular}
\end{center}
$^a$Ref. means whether the orbital parameters are first derived in this work or have been previously determined in the literature. (1) \cite{2010ApJ...708..253F}, (2) \cite{2007ASPC..372..483O}, (3) \cite{2010MNRAS.408L..51O}, (4) \cite{2007A&A...471..605V} and \cite{2008A&A...489..377C}, (5) \cite{2023A&A...673A..90S}, (6) \cite{2004A&A...422.1053O} ,\cite{2006BaltA..15...61O} and \cite{2023A&A...673A..90S}.
\end{table*}

\subsubsection{Other types of light curves}
\label{other_types.sec}

In addition to systems displaying a clear reflection effect, we identified a group of systems exhibiting low‑amplitude photometric modulations, which can be attributed to a weak reflection effect, tidal distortion, or Doppler beaming (Table \ref{orbital_params_other.tab}).
%
%
However, the sparse RV data of these systems prevent us from constraining their nature. 
Further high-quality spectroscopic observations are required to determine their properties.

\begin{table}
\caption{Systems exhibiting low-amplitude photometric modulations without orbital solutions. \label{orbital_params_other.tab}}
\centering
\setlength{\tabcolsep}{3mm}
\begin{center}
\begin{tabular}{lc}
\hline\noalign{\smallskip}
Gaia ID &  LC Type \\
\hline\noalign{\smallskip}
312628749626419328 & Weak reflection? \\
619153556155078272 & Weak reflection? \\
781164326766404736 & Tidal distortion? \\
795959630108543232 &  Tidal distortion? \\
1639999937328089088 & Weak reflection? \\
1876904938896587008 &  Tidal distortion? \\
2160160432952810496 & Doppler beaming? \\
2846162921688127360 & Tidal distortion? \\
\noalign{\smallskip}\hline
\end{tabular}
\end{center}
\end{table}

\section{Discussion}
\label{dis.sec}

\subsection{Comparison with previous studies}
\label{mass2.sec}

\begin{figure*}
    \center
    \includegraphics[width=0.98\textwidth]{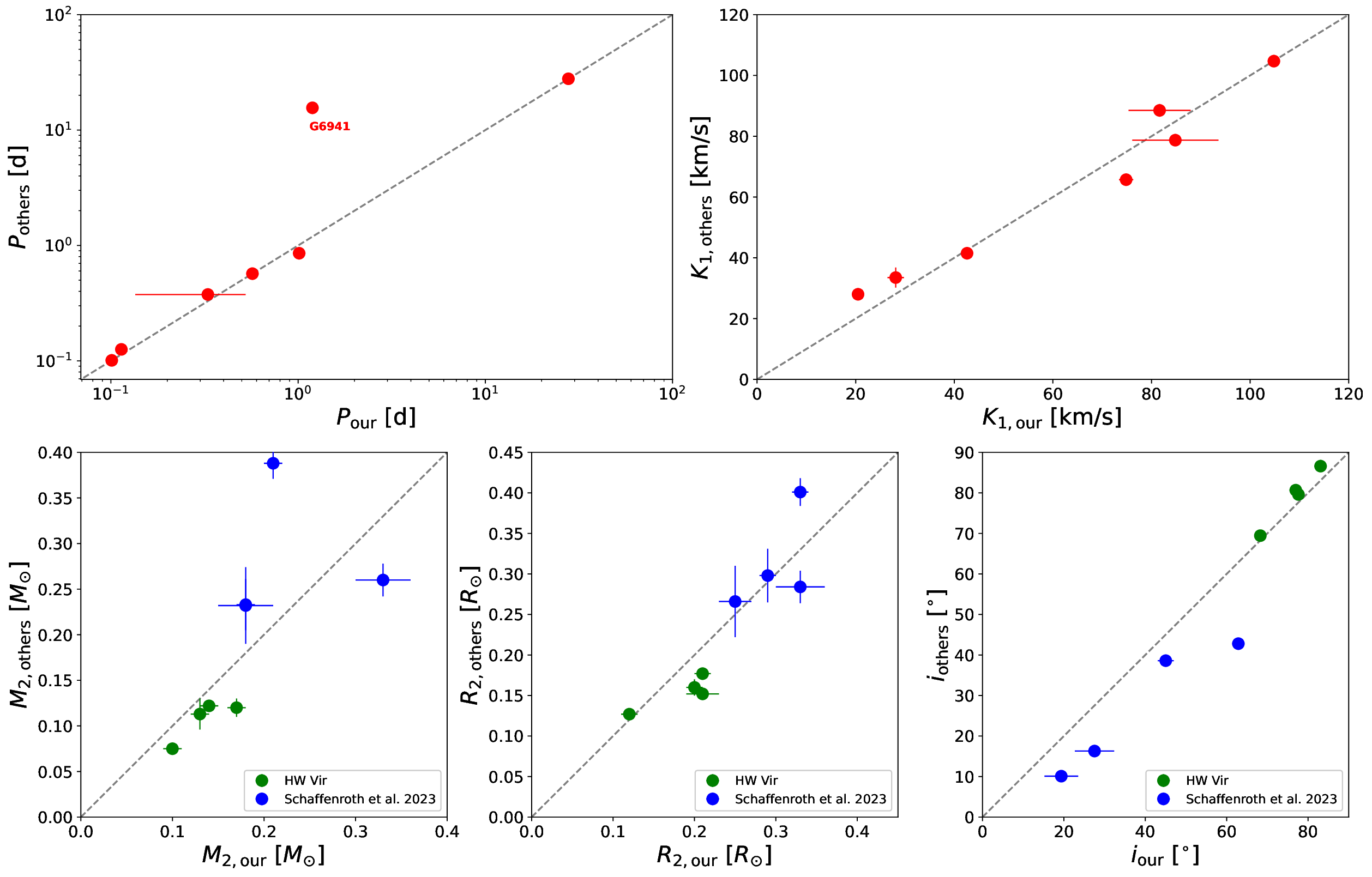}
    \caption{Top row: Comparison of orbital parameters for systems in Table \ref{orbital_params_from_cir.tab} with results from previous studies. Bottom row: Comparison of companion masses, radii, and inclinations for systems in Table \ref{orbital_params_irad.tab} with results from previous studies. Green dots represent the four HW Vir systems.}
    \label{comprios_with_previous_studies.fig}
\end{figure*}

To assess the reliability of our results, we cross-matched the systems in Table \ref{orbital_params_from_cir.tab} with the catalog provided by \cite{2015A&A...576A..44K}, which collected orbital parameters in previous studies.
Seven sources are common to both catalogs.
The RV semi-amplitudes of all systems and the periods of six systems are in good agreement with previous results (Figure \ref{comprios_with_previous_studies.fig}), supporting the credibility of our measurements.
However, the period of Gaia DR3 694109462844643072, (hereafter G6941) show significant discrepancies with previously reported values (Table \ref{campare_table3.tab}).
The period ($\sim$15.6 days) reported by \citep{2011MNRAS.415.1381C} was derived from more data points covering a larger fraction of the orbit, and is therefore likely more robust than our result. 

Additionally, several previous studies \citep{2007A&A...471..605V,2007ASPC..372..483O,2010ApJ...708..253F,2010MNRAS.408L..51O,2023A&A...673A..90S} performed the LC fitting to determine the nature of the companion in reflection-dominated binaries.
Eight of our reflection systems (including four HW Vir systems) have been studied in those works. 
Our results of HW Vir systems are in good agreement with previous studies \citep{2007A&A...471..605V, 2007ASPC..372..483O, 2010ApJ...708..253F, 2010MNRAS.408L..51O}.
However, compared to the population-based study \citep{2023A&A...673A..90S}, our derived companion masses and orbital inclinations show a clear discrepancy (the blue dots in Figure \ref{comprios_with_previous_studies.fig}).
This difference stems from a combination of differing input parameters between our work and \cite{2023A&A...673A..90S}, such as the temperature, mass, and radius of the HSD, as well as the temperature of the companion star. For example, \citet{2023A&A...673A..90S} fixed the HSD mass at 0.47 $M_{\odot}$ and the companion temperature at 3000 K. 
Apart from these settings, \citet{2023A&A...673A..90S} performed their fitting using only LC data, whereas we carried out a joint fit incorporating both LC and RV data.


\begin{table*}
\caption{Comparison of orbital parameters for systems in Table \ref{orbital_params_from_cir.tab} with results from previous studies. \label{campare_table3.tab}}
\centering
\setlength{\tabcolsep}{1mm}
\begin{center}
\begin{tabular}{lccccc}
\hline\noalign{\smallskip}
Gaia ID & $P_{\rm our}$ & $K_{\rm 1, our}$ & $P_{\rm others}$ & $K_{\rm 1, others}$ & Ref.$^a$ \\
 & (days) & (km/s) & (days) & (km/s) & \\
\hline\noalign{\smallskip}
3637481302758519040 & 0.10101582 & $84.83\pm8.74$ & $0.1010159990\pm00000001$ & $78.70\pm0.60$ & (1) \\ 
2080063931448749824 & 0.11370455 & $74.85\pm1.43$ & $0.12576530\pm0.000000021$ & $65.70\pm0.60$ & (2) \\
859683853719128192 & $0.32948825\pm0.19460401$ & $20.48\pm1.16$ & $0.376\pm0.003$ & $28.00\pm0.2$ & (3) \\
2551900379931546240 & $0.56990086\pm0.00000001$ & $104.82\pm0.62$ & $0.569899\pm0.000001$ & $104.70\pm0.4$ & (4) \\
4023163971460301952 & 1.01265504 & $81.64\pm6.28$ & $0.856210\pm0.000003$ & $88.50\pm2.1$ & (5) \\
694109462844643072 & 1.19243697 & $42.59\pm1.02$ & $15.583\pm0.001$ & $41.50\pm0.8$ & (6) \\
611684298790301568 & 27.78048114 & $28.12\pm1.66$ & $27.815\pm0.005$ & $33.50\pm3.3$ & (5) \\
\noalign{\smallskip}\hline
\end{tabular}
\end{center}
$^a$Ref. means whether the orbital parameters are first derived in this work or have been previously determined in the literature. (1) \cite{2007A&A...471..605V} and \cite{2008A&A...489..377C}, (2) \cite{2010MNRAS.408L..51O}, (3) \cite{2004A&A...422.1053O} ,\cite{2006BaltA..15...61O} and \cite{2023A&A...673A..90S}, (4) \cite{2008A&A...477L..13G}, (5) \cite{2003MNRAS.338..752M} and \cite{2001MNRAS.326.1391M}, (6) \cite{2011MNRAS.415.1381C}.
\end{table*}

\subsection{Mass-radius distribution of the companion}
\label{mass_rad.sec}

\begin{figure}
    \center
    \includegraphics[width=0.49\textwidth]{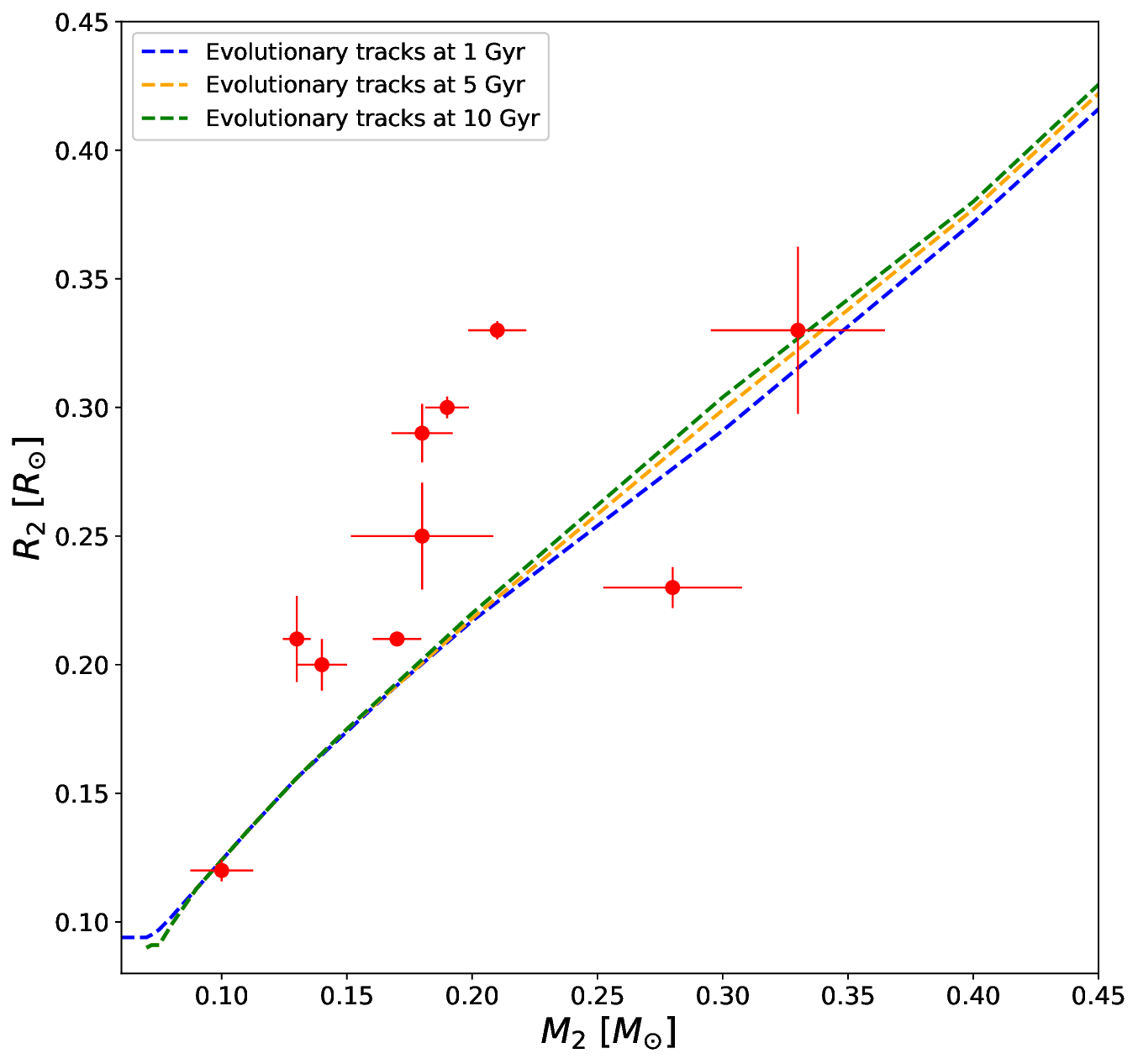}
    \caption{Mass and radius distribution of the companions in the 10 reflection binaries within our binary sample. The blue, yellow, and green dashed lines represent the evolutionary tracks of low-mass MS stars from \cite{2015A&A...577A..42B} at 1 Gyr, 5 Gyr, and 10 Gyr, respectively.} 
    \label{iso.fig}
\end{figure}

To assess the plausibility of our measurements, we compared the mass-radius relation of the companions in the 10 reflection systems (Section \ref{reflection.sec}) with theoretical evolutionary models derived from \cite{2015A&A...577A..42B}. 
Figure \ref{iso.fig} shows the observed mass-radius distribution and the theoretical evolutionary tracks for low-mass MS stars. 
The overall agreement between our results and the models indicates that the derived parameters of companion are physically self-consistent, lending credibility to our analysis method. 
However, some scatter is still present, in agreement with previous analyses \citep{2018MNRAS.481.1083P,2023A&A...673A..90S}. 
This scatter may arise from several factors. 
First, degeneracy between the radius of companion and orbital inclination in the LC fitting can introduce uncertainties in the derived radius. 
Second, limited RV data or relatively large RV uncertainties may contribute to the scatter in the fitting results.
However, the dispersion of radius caused by these two effects should be symmetrically distributed on both sides of the evolutionary track, inconsistent with the pattern shown in Figure \ref{iso.fig}.
Rotational effects in the companion may be a key factor behind the inflation of its radius.
In rapidly rotating low-mass MS stars, enhanced magnetic activity can suppress convective efficiency, disrupting energy transport and leading to radius inflation \citep{2018MNRAS.481.1083P}. 
\cite{2011ApJ...728...48K} reported that M dwarfs in close binaries (e.g., $P<1$ day) tend to be more inflated than those in wide binaries. 
Based on a sample of 23 M dwarfs with precise radius measurements, \citet{2018MNRAS.481.1083P} found that low-mass stars with rotation periods shorter than 5 days also exhibit significant radius inflation.
All 10 reflection systems in our sample have short orbital periods ($P<1$ day), suggesting the systematic offset toward larger radii in our fits, compared to prediction of theoretical models, likely arises from the rapid rotation of the M dwarfs.

\subsection{Mass-period distribution of companion}
\label{mass_per.sec}

\begin{figure}
    \center
    \includegraphics[width=0.49\textwidth]{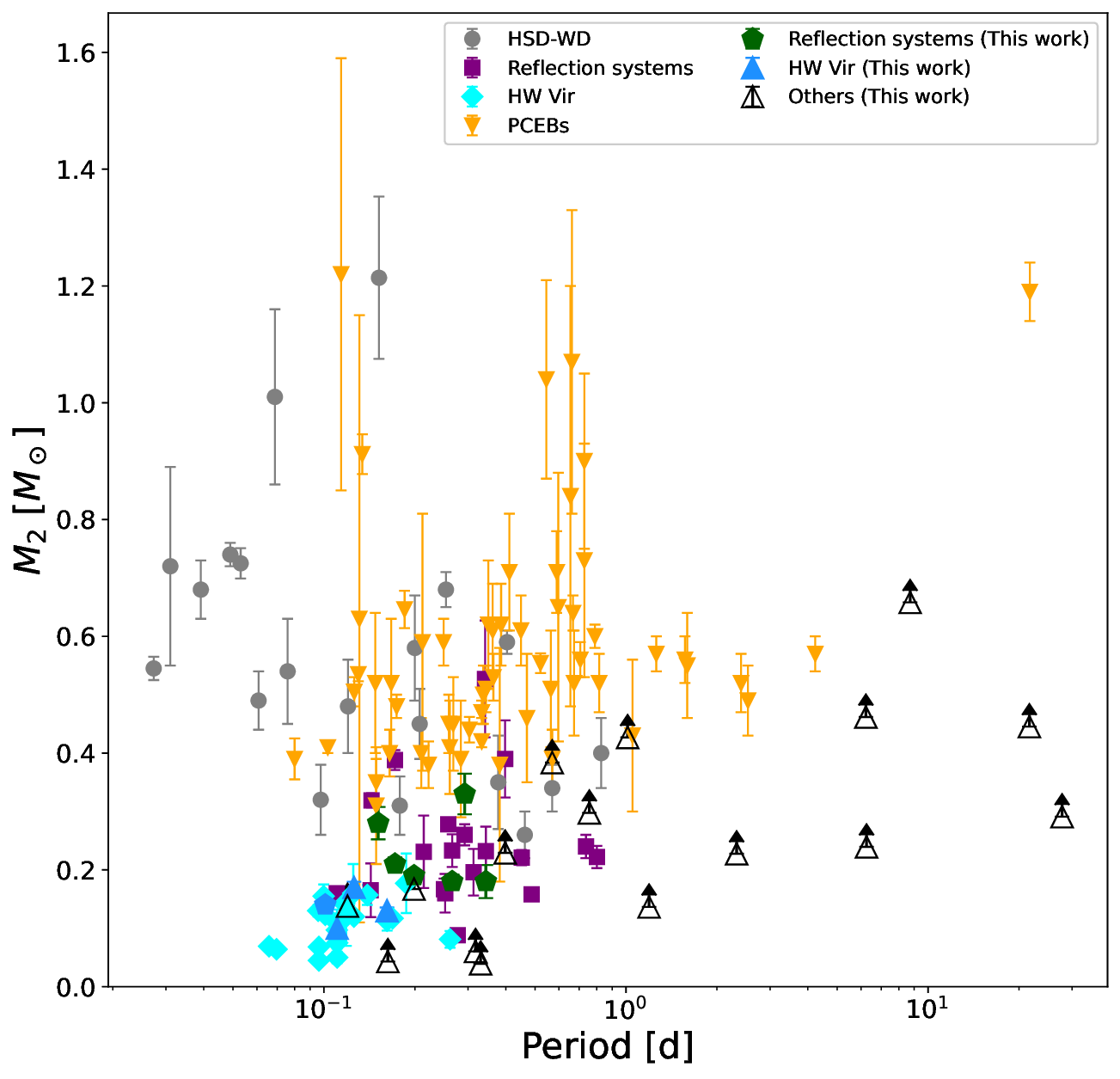}
    \caption{A comparison of orbital periods and the mass of the companion for different binaries in this work and literature.} 
    \label{per_m2.fig}
\end{figure}

We investigated the distribution of orbital period and companion mass for the 10 HSD+M binaries (Section \ref{reflection.sec}). 
For comparison, we collected different binaries for HSD+WD systems \citep{2011MNRAS.410.1787B,2013A&A...554A..54G,2017ApJ...835..131K,2017ApJ...851...28K,2019ApJ...883...51R,2020ApJ...902...92R,2020ApJ...898L..25K,2020ApJ...891...45K,2021NatAs...5.1052P,2022ApJ...925L..12K,2022MNRAS.513.2215R,2025A&A...693A.322Y,2025SCPMA..6869511L}, reflection systems \citep{2023A&A...673A..90S}, HW Vir-type binaries \citep{2001A&A...379..893D,2007ASPC..372..487P,2007A&A...471..605V,2007ASPC..372..483O,2008ASPC..392..221O,2010MNRAS.408L..51O,2010ApJ...708..253F,2011ApJ...731L..22G,2012A&A...543A.138B,2012MNRAS.423..478A,2013A&A...553A..18S,2013MNRAS.430...22B,2013A&A...553A..97V,2014ASPC..481..259V,2014A&A...564A..98S,2015ApJ...808..179D,2015A&A...578A.125H,2015A&A...576A.123S,2015A&A...580A.117S,2015A&A...576A.123S,2018A&A...614A..77S}, and post common envelope binaries (PCEBs) \citep{2010A&A...520A..86Z}.
Figure \ref{per_m2.fig} plots the distribution of period and the mass of companion for all systems.

Among the reflection-dominated systems, the majority have orbital periods exceeding 0.1 days. Their companions are predominantly low‑mass main‑sequence stars, with typical masses around 0.2 $M_{\odot}$.  
In contrast, most HSD+WD binaries exhibit very short orbital periods (e.g., $P < 0.2$ days) and more massive WD companions.  

The observed HW Vir‑type binaries also exhibit short orbital periods (typically $P < 0.2$ days), yet they host companions of lower mass compared to the reflection‑dominated systems.

\subsection{Possible formation mechanisms}
\label{formation.sec}

\begin{figure}
    \center
    \includegraphics[width=0.49\textwidth]{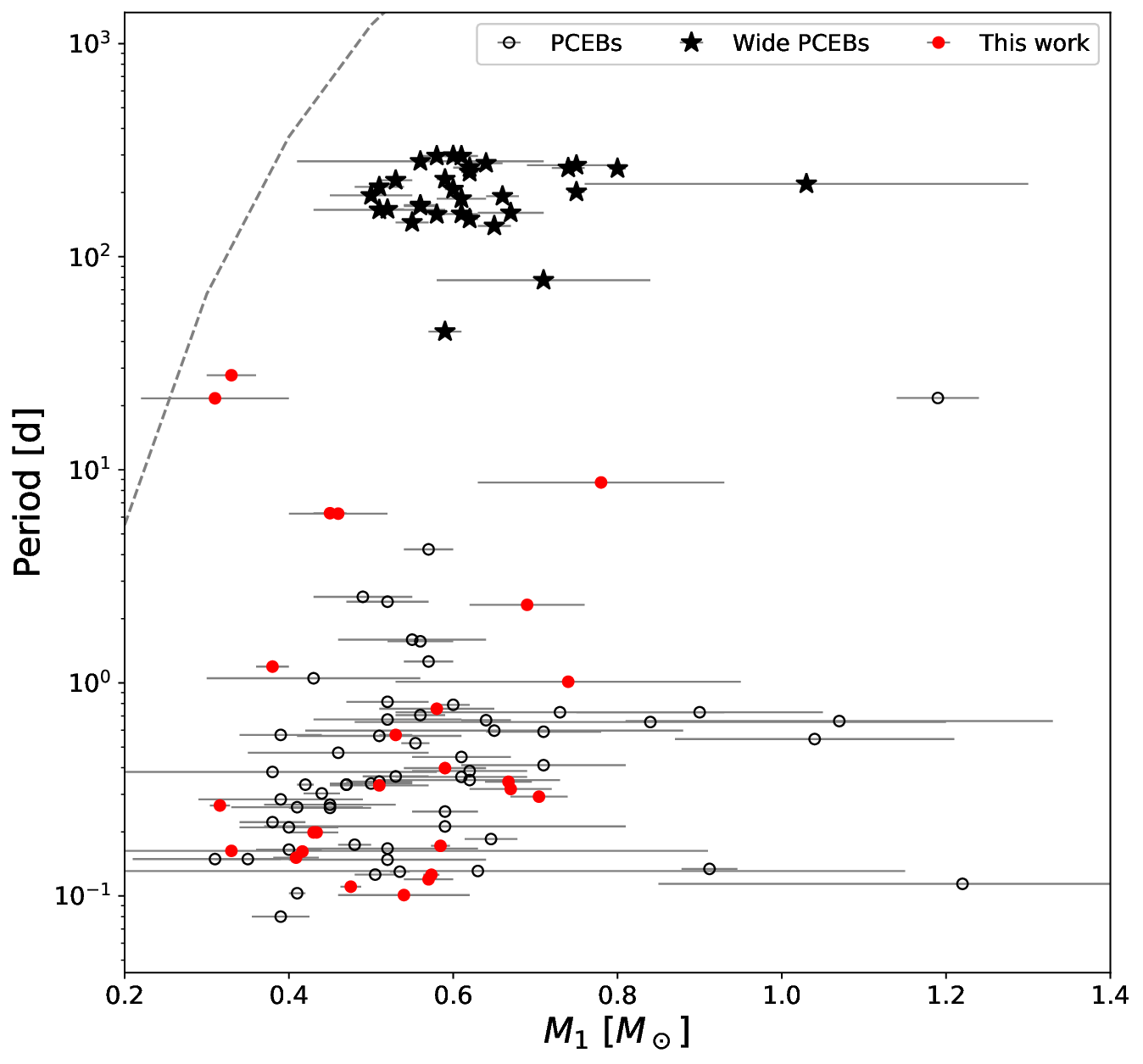}
    \caption{Comparison of our sample with PCEBs in $M_{1}$--Period diagram. The dashed line represents the evolutionary track predicted for binaries undergoing stable mass transfer \citep{1995MNRAS.273..731R}.} 
    \label{per_m1.fig}
\end{figure}

Our sample exhibits nearly circular orbits, suggesting these systems likely underwent a CE phase.  
To investigate whether they share the same evolutionary pathway as PCEBs, we compared our sample with published PCEB samples \citep{2010A&A...520A..86Z, 2024PASP..136h4202Y} in the $M_{1}$--Period diagram.
As is shown in Figure \ref{per_m1.fig}, PCEBs can be broadly divided into two subclasses based on orbital period: close binaries ($P_{\rm orb} \lesssim 10$\,days) and wide binaries ($P_{\rm orb} \gtrsim 10$\,days).  
Compared to close PCEBs, wide PCEBs typically host more massive WDs, and their longer orbital periods imply that additional energy sources beyond orbital energy may be required to successfully eject the envelope during the CE phase \citep{2010MNRAS.403..179D,2010A&A...520A..86Z,2016MNRAS.463.2125P,2023MNRAS.518.3966S,2024MNRAS.52711719Y}.
Our sample shares the same region as close PCEBs.
In addition, the gray dashed line represents the theoretical $M_{1}$--Period relation \citep{1995MNRAS.273..731R} for binaries evolving through stable Roche lobe overflow.
Most systems in our sample lie significantly below this line, further indicating they have undergone CE evolution.

\section{Summary}
\label{summary.sec}

We conducted a systematic parameter measurement of HSD binaries selected from the Gaia EDR3 catalog \citep{2022A&A...662A..40C}, utilizing spectroscopic data from LAMOST, BOSS, SEGUE, APOGEE, and GALAH. 
For LAMOST LRS and MRS spectra, we remeasured RVs by computing the Doppler shifts of the $H_{\alpha}$ line. 
For systems with LAMOST LRS data, we employed a CNN approach to derive atmospheric parameters of the HSDs, including effective temperature, surface gravity, and helium abundance. 
Using these atmospheric parameters, the HSDs in our sample have a median mass of approximately $0.45^{+0.19}_{-0.17} M_{\odot}$ and a typical radius of about $0.18^{+0.04}_{-0.05} R_{\odot}$, consistent with results from previous studies \citep{2023ApJ...953..122L}.

We employed a circular orbital model to derive the orbital parameters from RV data for our sample. 
We derived orbital parameters for 17 systems, among which 10 are newly solved in this work.
Furthermore, we determined orbital inclinations ($i$) and companion properties (e.g., mass, radius, and effective temperature) by jointly fitting RV and LCs from TESS.
Among these systems, 10 are classified as reflection-dominated (including four HW Vir binaries), two of which are newly solved.
Our orbital solutions (e.g., $i$, $P$) are generally consistent with previous studies.
A discrepancy is observed in the companion mass, which likely arises from the use of a typical but inaccurate HSD mass in earlier population-based studies.

%

The majority of the reflection-dominated systems have orbital periods longer 0.1 days, and their companions are predominantly low‑mass main‑sequence stars, with typical masses around 0.2 $M_{\odot}$.  
The measured radii of these (mostly) M‑type companions are systematically larger than stellar‑model predictions, indicating radius inflation likely driven by rapid rotation.
Most HSD+WD binaries exhibit very short orbital periods ($P < 0.2$ days), and those systems with the shortest periods tend to host more massive WD companions.
HW Vir‑type binaries also exhibit short orbital periods (typically $P < 0.2$ days), but harbor low‑mass companions.
Most of the studied systems occupy the same region in the mass–-period diagram as PCEBs, strongly suggesting that they have undergone CE evolution.

\begin{acknowledgements}

We thank the anonymous referee for very helpful comments and suggestions that significantly improved paper. 
This work made use of the data from LAMOST (Large Sky Area Multi-Object Fiber Spectroscopic Telescope, also known as the Guoshoujing Telescope) (https://cstr.cn/31118.02.LAMOST). LAMOST is a Chinese national mega-science facility, operated by National Astronomical Observatories, Chinese Academy of Sciences.
SDSS telescopes are located at Apache Point Observatory, funded by the Astrophysical Research Consortium and operated by New Mexico State University, and at Las Campanas Observatory, operated by the Carnegie Institution for Science. 
This work made use of the Fourth Data Release of the GALAH Survey. The GALAH Survey is based on data acquired through the Australian Astronomical Observatory. 
This work presents results from the European Space Agency (ESA) space mission {\it Gaia}. {\it Gaia} data are being processed by the {\it Gaia} Data Processing and Analysis Consortium (DPAC). 

This work was funded by the Strategic Priority Program of the Chinese Academy of Sciences under grant number XDB1160302 (Song Wang), the National Key Research and Development Program of China under grant number 2023YFA1607901 (Song Wang), the National Natural Science Foundation of China (NSFC) under grant number 12588202 (Jifeng Liu), the NSFC under grant number 12273057 (Song Wang), science research grants from the China Manned Space Project (Song Wang), the New Cornerstone Science Foundation through the New Cornerstone Investigator Program and the XPLORER PRIZE (Jifeng Liu), Chongqing Natural Science Foundation under grant number CSTB2023NSCQ-MSX0093 (Xiaohong Yang), the NSFC under grand number 12547101 (Xinlin Zhao), and the Postdoctoral Fellowship Program of China Postdoctoral Science (CPSF) under grant number GZB20250739 (Xinlin Zhao).

\end{acknowledgements}

\bibliographystyle{aasjournal}
\bibliography{main.bib}{}   

\clearpage
\appendix
\renewcommand*\thetable{\Alph{section}.\arabic{table}}
\renewcommand*\thefigure{\Alph{section}\arabic{figure}}

\section{RV measurements}
\label{rvdata.sec}

\setcounter{figure}{0}
\setcounter{table}{0}

\begin{figure*} 
    \center
    \includegraphics[width=0.48\textwidth]{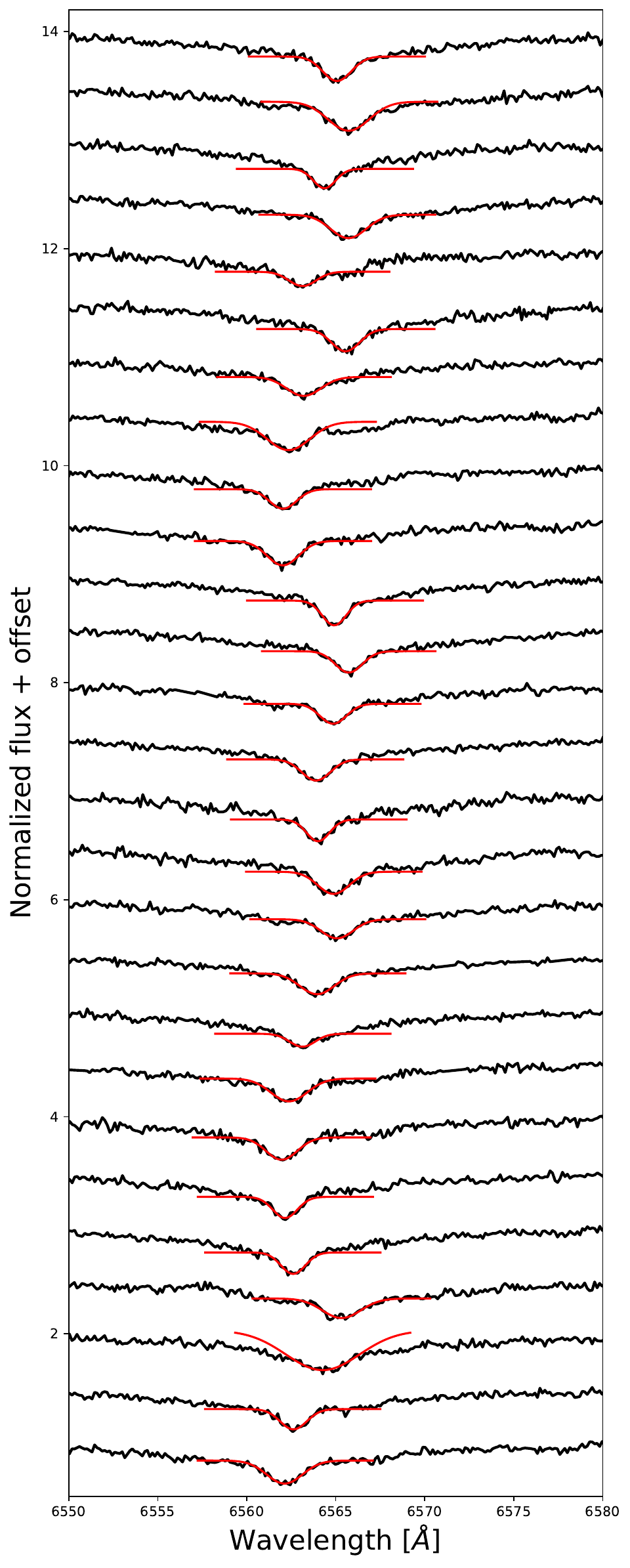}
    \caption{Schematic diagram of the $H_{\alpha}$ line fitting for Gaia DR3 1914301803258669440. The black lines denote the MRS spectra, while the red lines represent the best fitting.}
    \label{Ha_fitting_1914.fig}
\end{figure*}

Here we list the RV measurements from multiple spectroscopic surveys for our sample. 
An example of the $H_{\alpha}$ line fitting for Gaia DR3 1914301803258669440 using LAMOST MRS is shown in Figure \ref{Ha_fitting_1914.fig}.

\begin{table*}
\caption{The RV measurements for our sample. \label{rvdata.tab}}
\centering
\setlength{\tabcolsep}{1mm}
\begin{center}
\begin{tabular}{lcccc}
\hline\noalign{\smallskip}
Gaia ID & BJD & RV & SNR & Database \\
 &  & (km/s) &  &  \\
\hline\noalign{\smallskip}
2767874292175410560 & 2459565.94426 & $-4.58\pm1.26$ & 36.91 & MRS \\
2767874292175410560 & 2459565.95884 & $-2.90\pm0.79$ & 39.65 & MRS \\
2767874292175410560 & 2459565.97481 & $-4.71\pm0.81$ & 39.57 & MRS \\
2767874292175410560 & 2458108.93912 & $-12.39\pm11.06$ & 170.79 & LRS \\
2850670743266825600 & 2458434.05369 & $-72.28\pm16.30$ & 30.50 & LRS \\
2581810261598516096 & 2458423.08473 & $-15.19\pm11.56$ & 63.93 & LRS \\
2581810261598516096 & 2459142.45360 & $-19.01\pm11.44$ & 91.55 & LRS \\
2581810261598516096 & 2459564.98821 & $-23.72\pm2.98$ & 8.19 & MRS \\
2581810261598516096 & 2459565.01460 & $-21.73\pm5.72$ & 5.97 & MRS \\
2581810261598516096 & 2459564.97294 & $-24.35\pm5.58$ & 7.93 & MRS \\
2581810261598516096 & 2459566.98177 & $-29.95\pm5.19$ & 9.68 & MRS \\
2581810261598516096 & 2459569.93635 & $-13.96\pm11.22$ & 13.16 & MRS \\
2581810261598516096 & 2459569.95163 & $-23.40\pm7.50$ & 11.03 & MRS \\
2581810261598516096 & 2459569.96690 & $-20.69\pm2.68$ & 13.51 & MRS \\
2581810261598516096 & 2459566.96580 & $-20.29\pm7.55$ & 10.16 & MRS \\
312628749626419328 & 2456286.98227 & $64.21\pm8.94$ & 71.81 & LRS \\
312628749626419328 & 2456287.01011 & $54.62\pm8.71$ & 69.63 & LRS \\
312628749626419328 & 2456965.07859 & $54.56\pm8.67$ & 135.97 & LRS \\
2551900379931546240 & 2456199.21690 & $-19.91\pm9.69$ & 82.02 & LRS \\
2551900379931546240 & 2459139.48739 & $-81.77\pm8.62$ & 495.03 & LRS \\
\noalign{\smallskip}\hline
\end{tabular}
\end{center}
NOTE. This table is available in its entirety in machine-readable and Virtual Observatory (VO) forms in the online journal. A portion is shown here for guidance regarding its form and content.
\end{table*}


\section{Keplerian solutions derived from circular orbit}
\label{rvfitting.sec}

\setcounter{table}{0}
\setcounter{figure}{0}

\begin{figure}
    \center
    \includegraphics[width=0.32\textwidth]{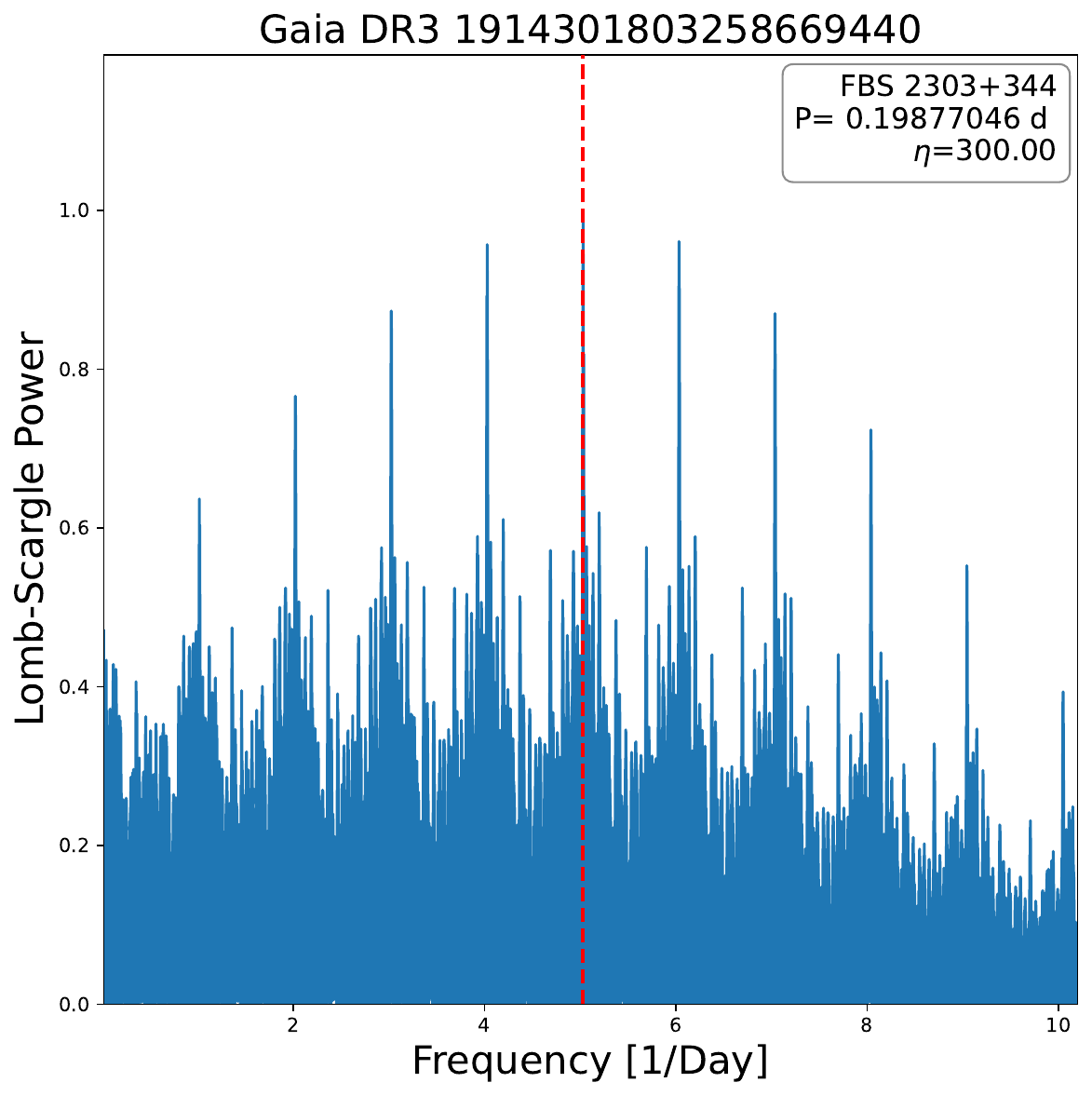}
    \includegraphics[width=0.32\textwidth]{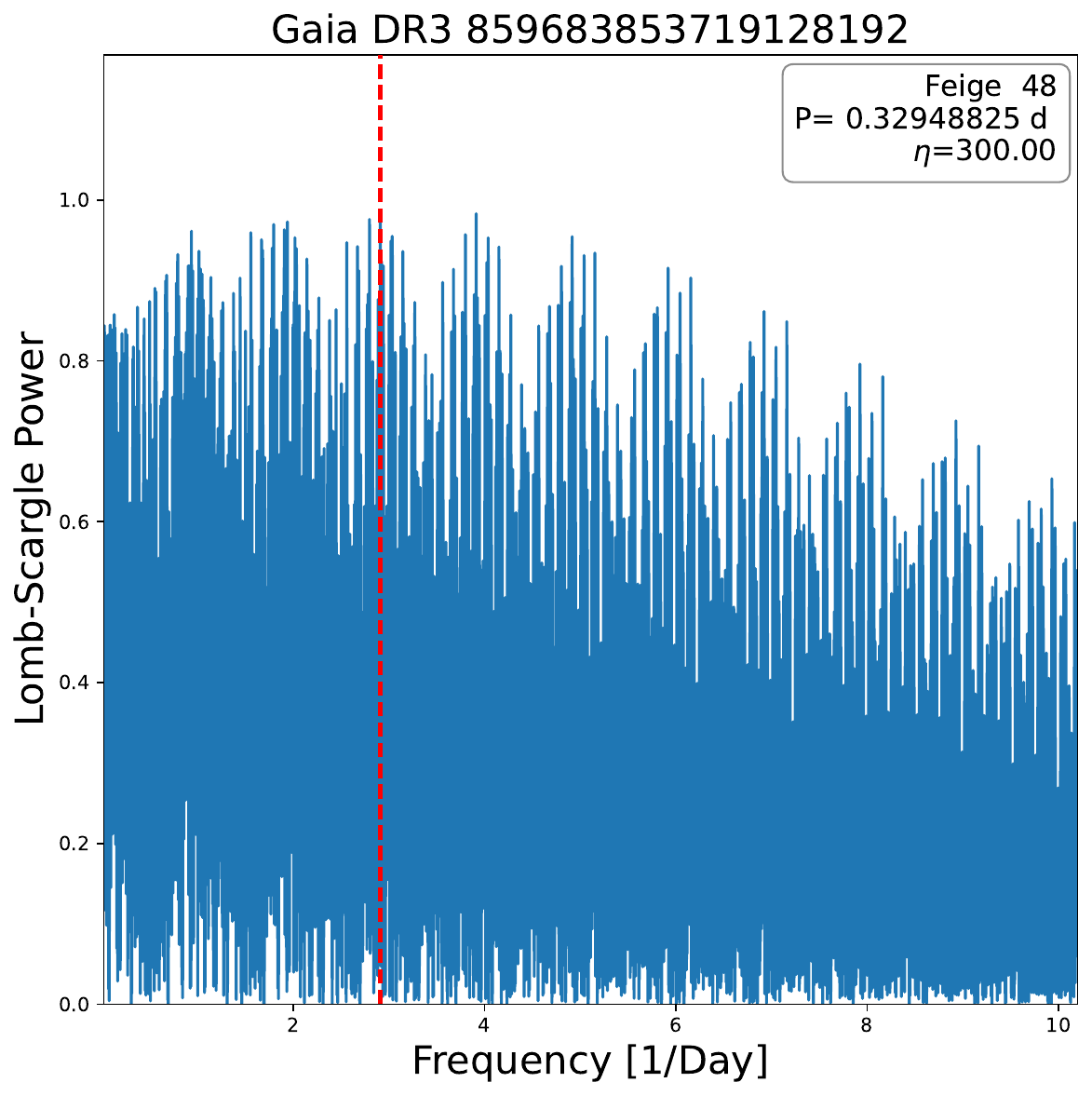}
    \includegraphics[width=0.32\textwidth]{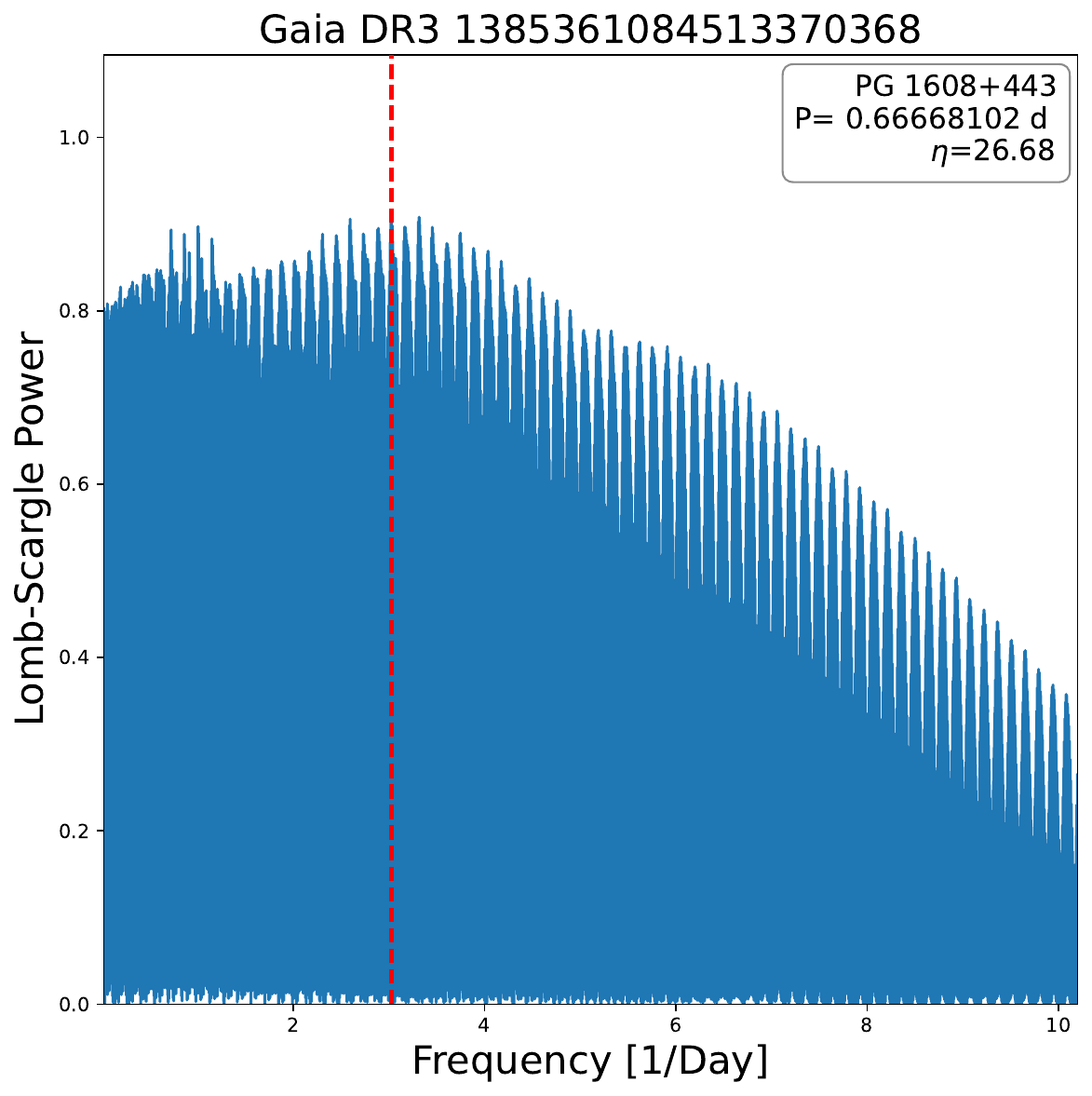}
    \vspace{1.2ex}
    \includegraphics[width=0.32\textwidth]{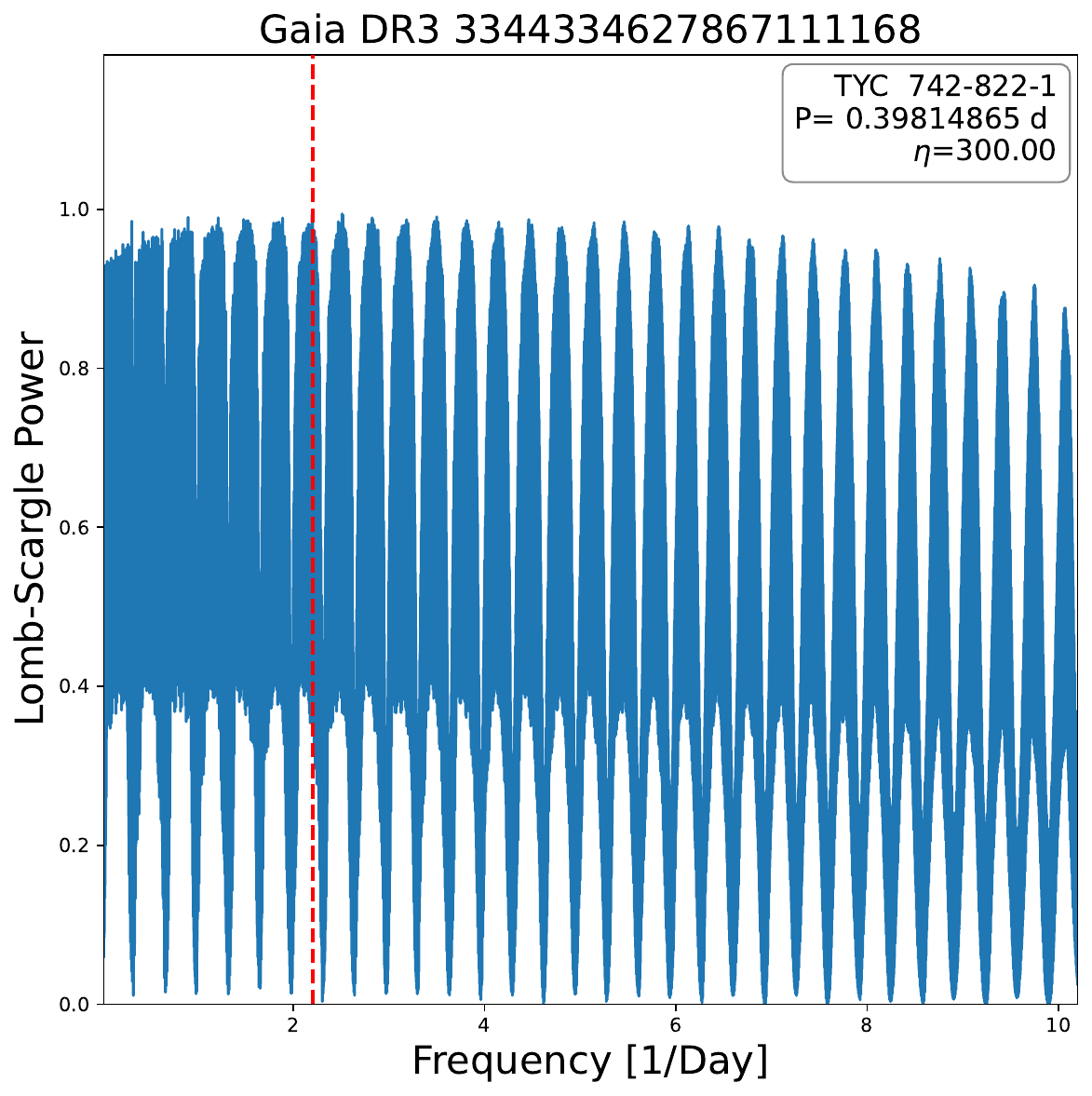}
    \includegraphics[width=0.32\textwidth]{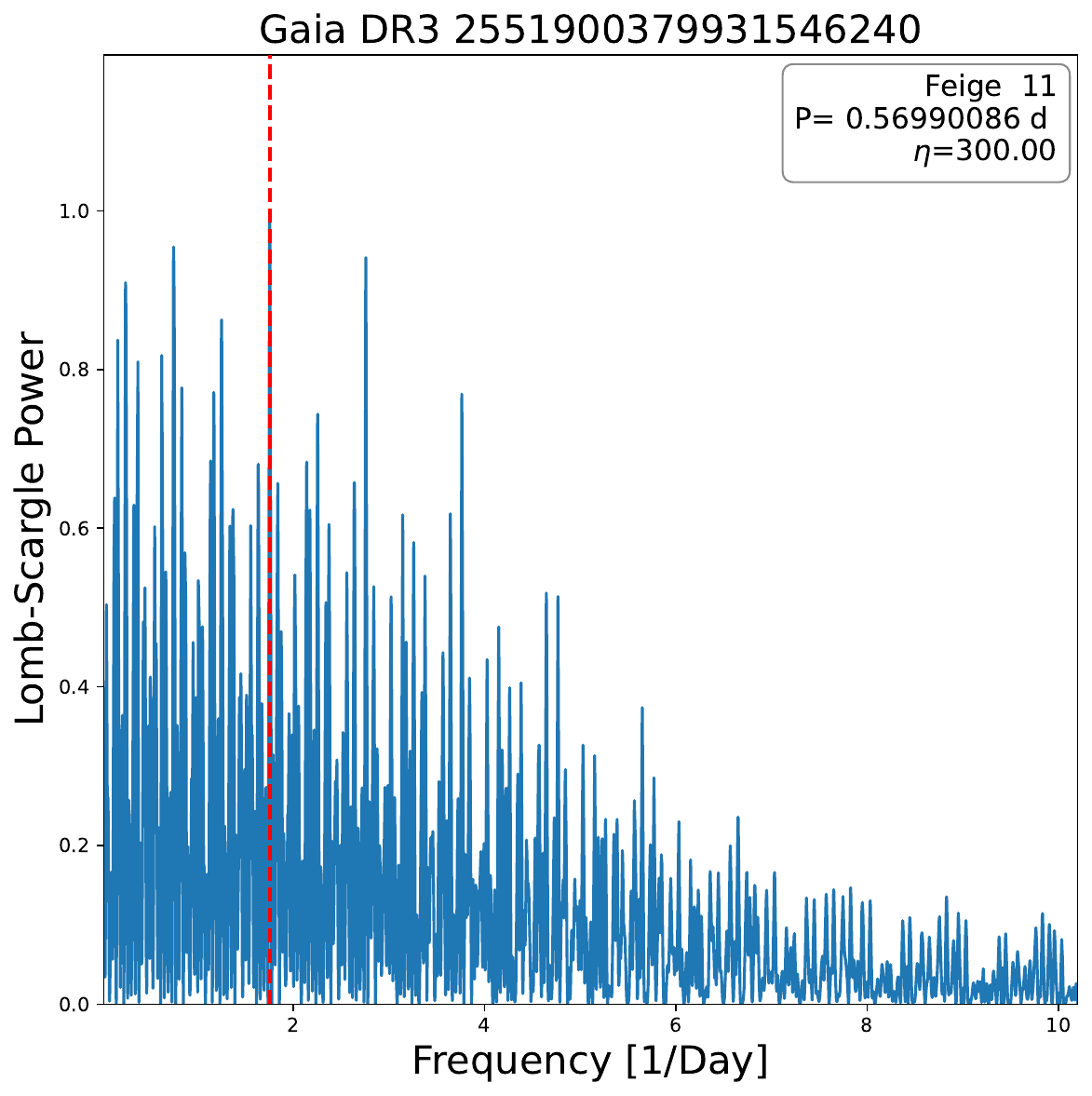}
    \includegraphics[width=0.32\textwidth]{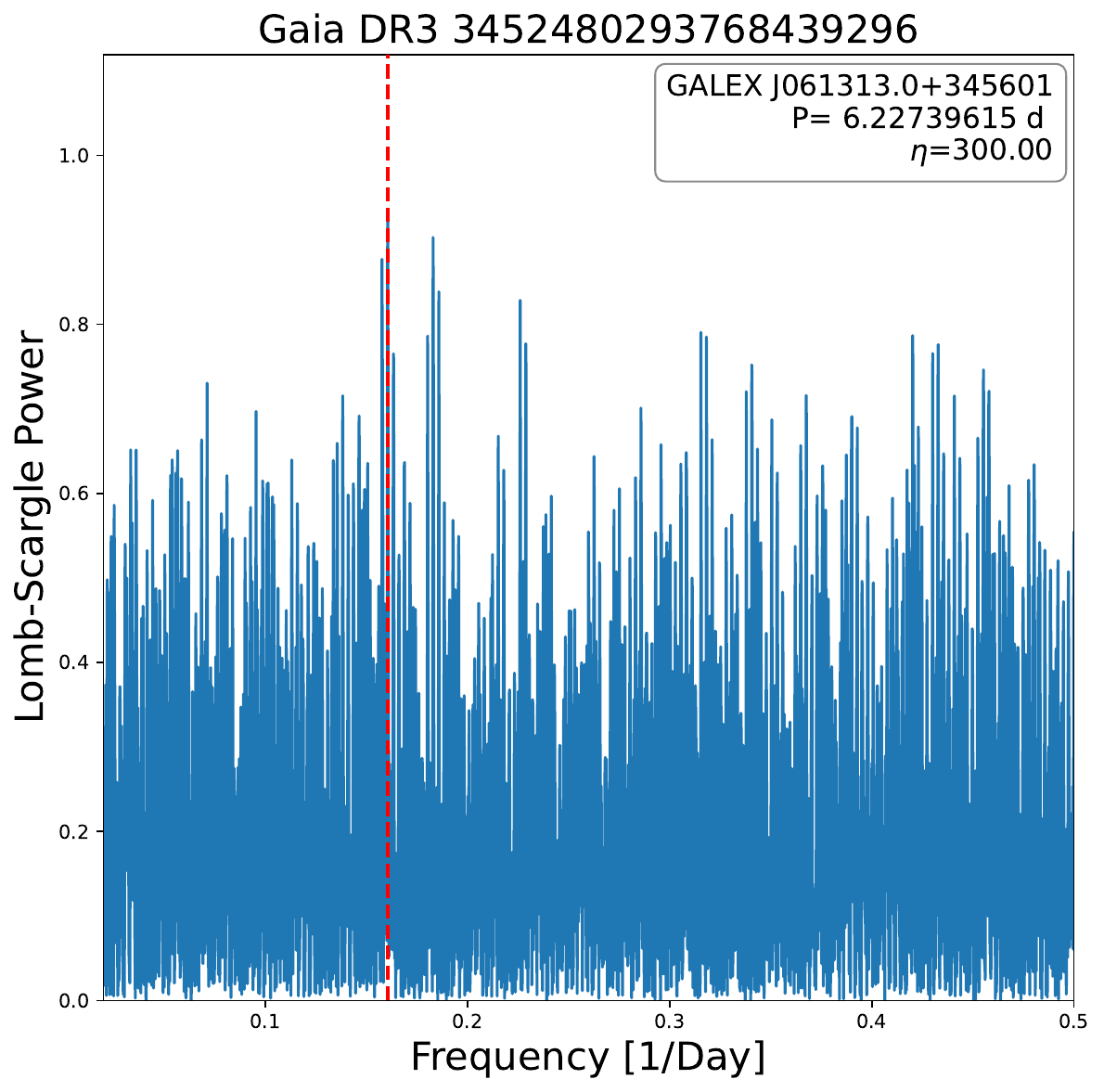}
    \vspace{1.2ex}
    \includegraphics[width=0.32\textwidth]{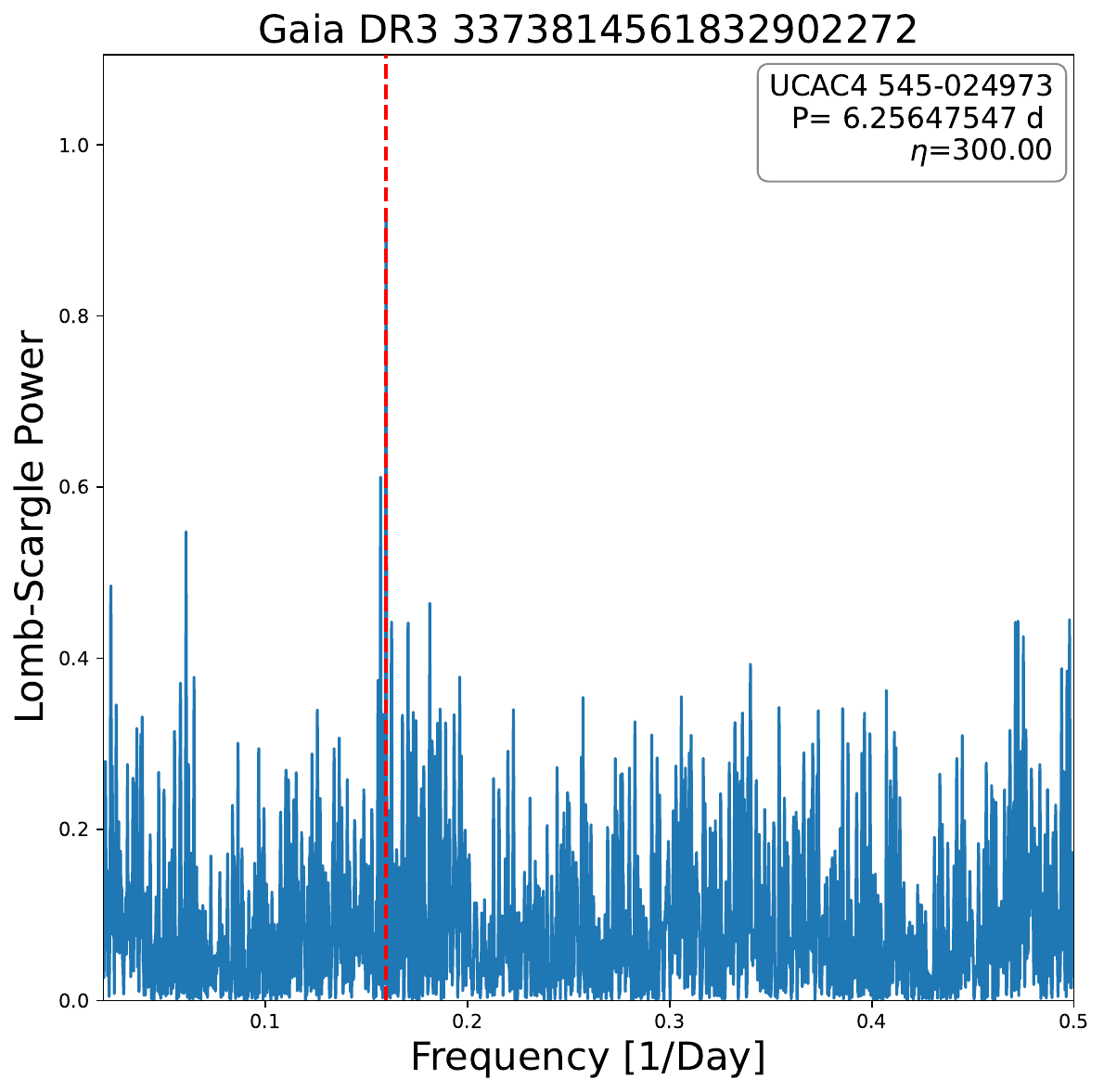}
    \includegraphics[width=0.32\textwidth]{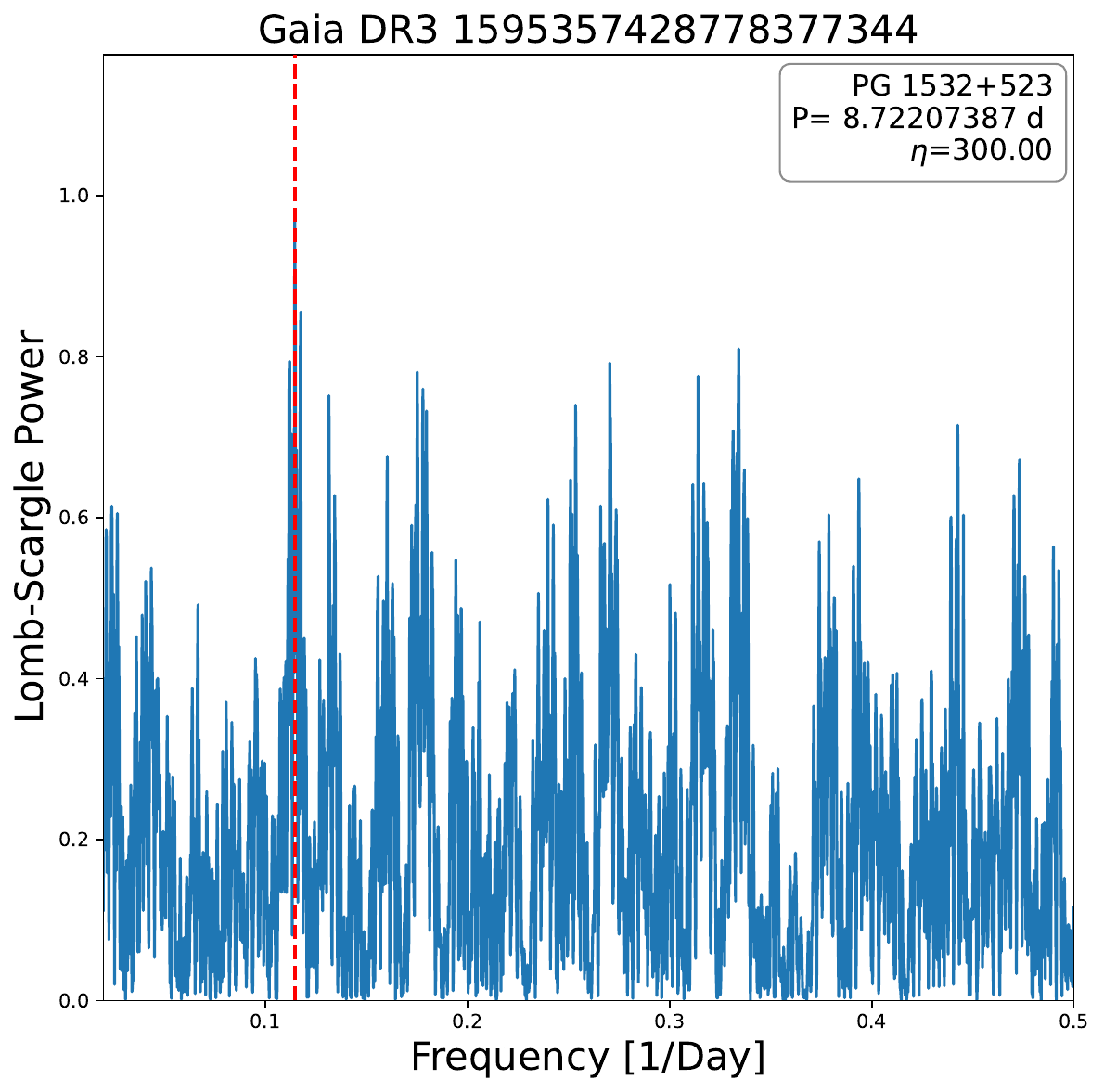}
    \includegraphics[width=0.32\textwidth]{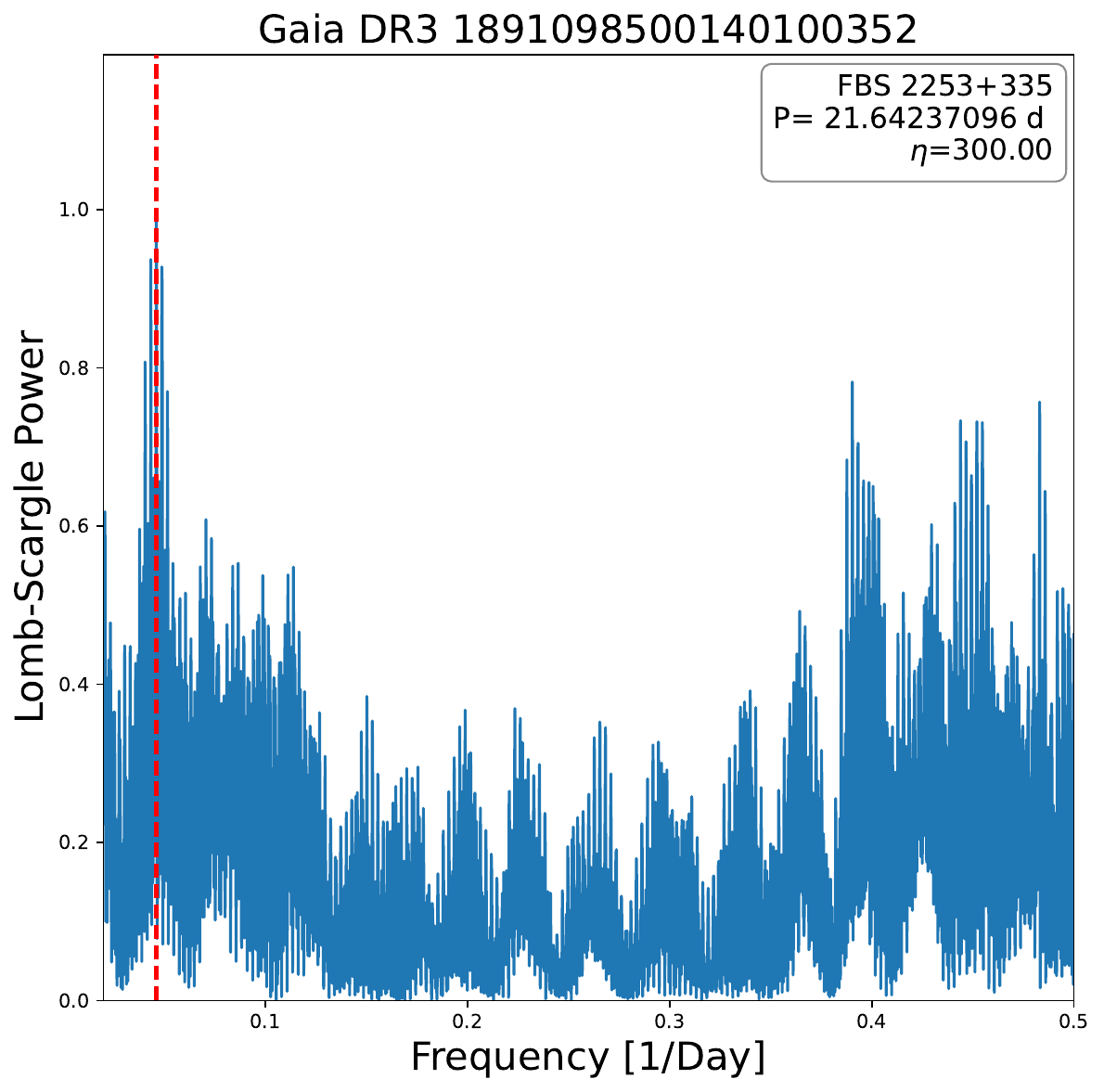}
    \caption{The LS periodograms for the systems with a reliable period uncertainty estimates. The $y$ axis is the normalized power given by the $astropy$ package. For values of $\eta$ exceeding 300, we applied an upper truncation limit of 300.}
    \label{periodograms.fig}
\end{figure}

Figure \ref{periodograms.fig} shows the periodograms of the systems listed in Table \ref{orbital_params_from_cir.tab} that have reliable period uncertainty estimates.
Figures \ref{circ_fitting.fig} presents the Keplerian solutions derived from a circular orbit fits ($e=0$). 
%
%
%

\begin{figure*}
    \center
    \includegraphics[width=0.32\textwidth]{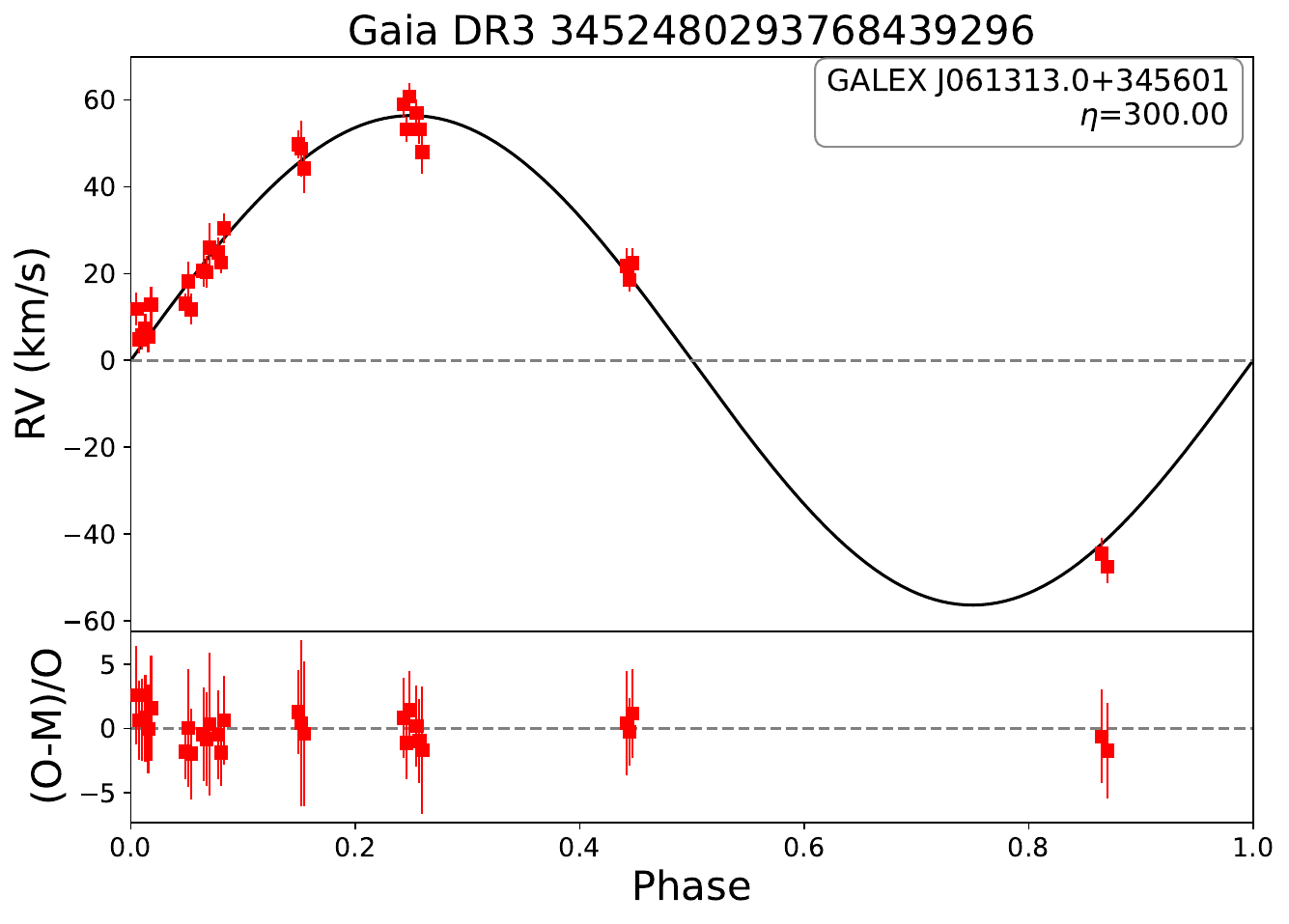}
    \includegraphics[width=0.32\textwidth]{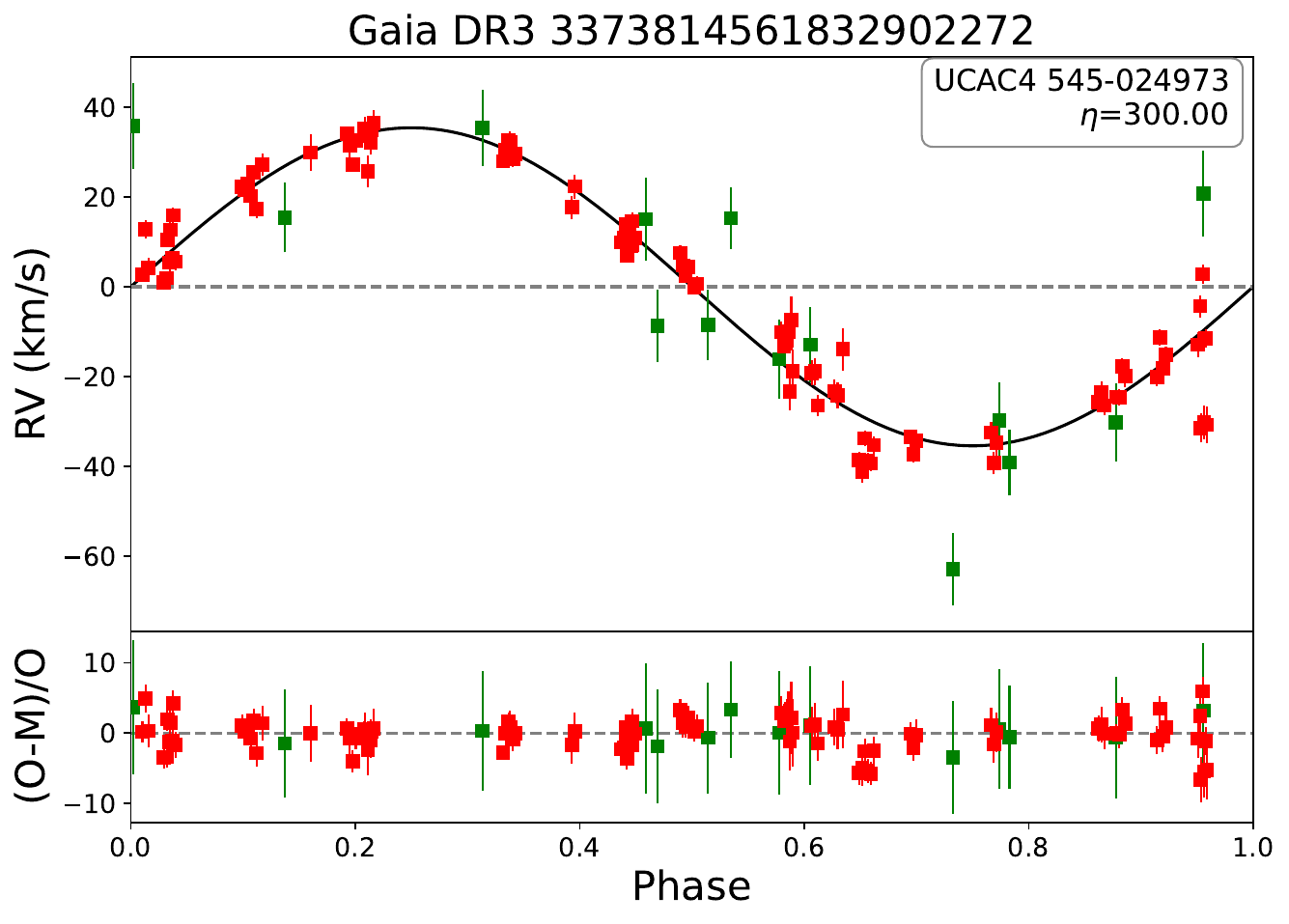}
    \includegraphics[width=0.32\textwidth]{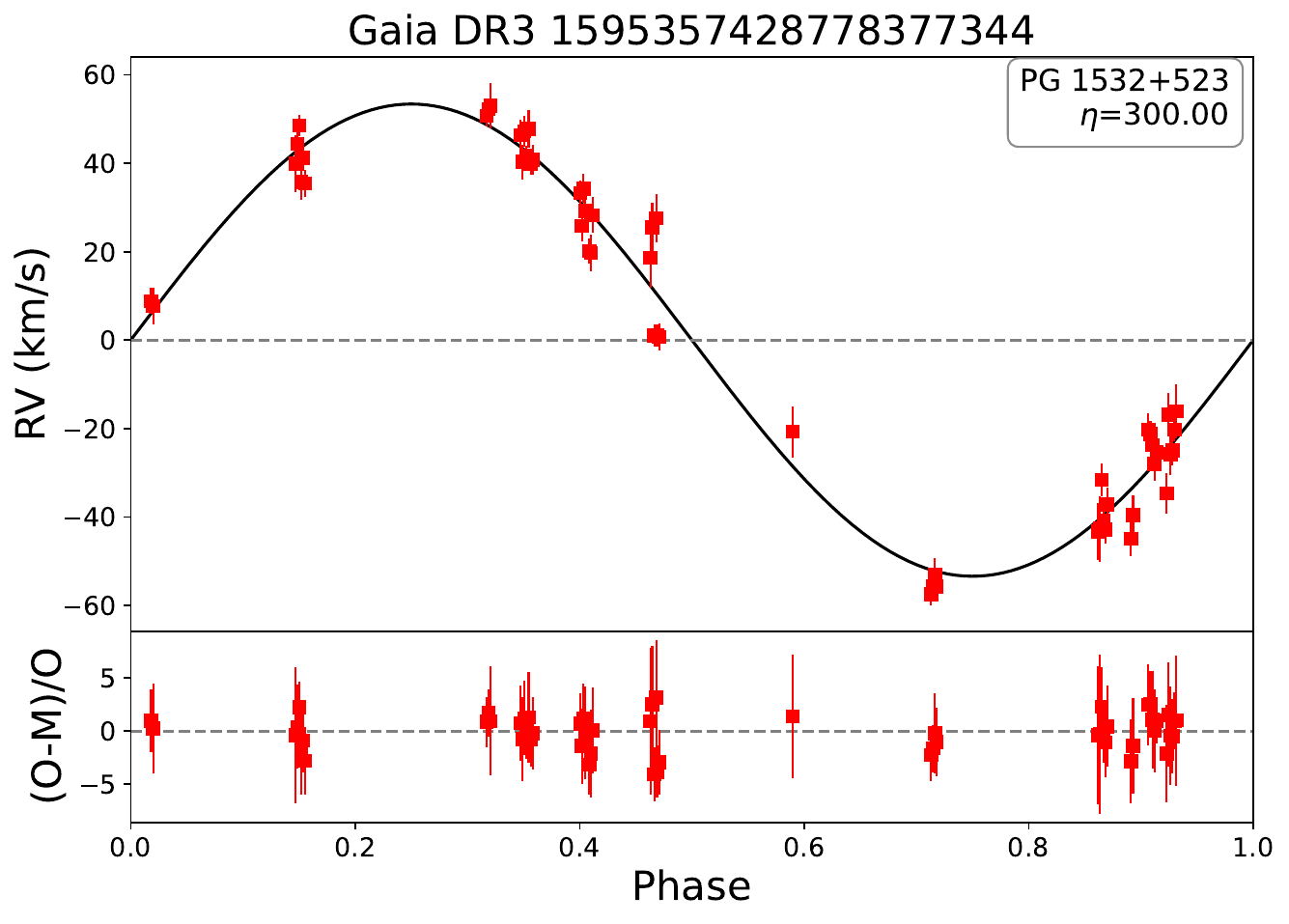}
    \vspace{1.5ex}
    \includegraphics[width=0.32\textwidth]{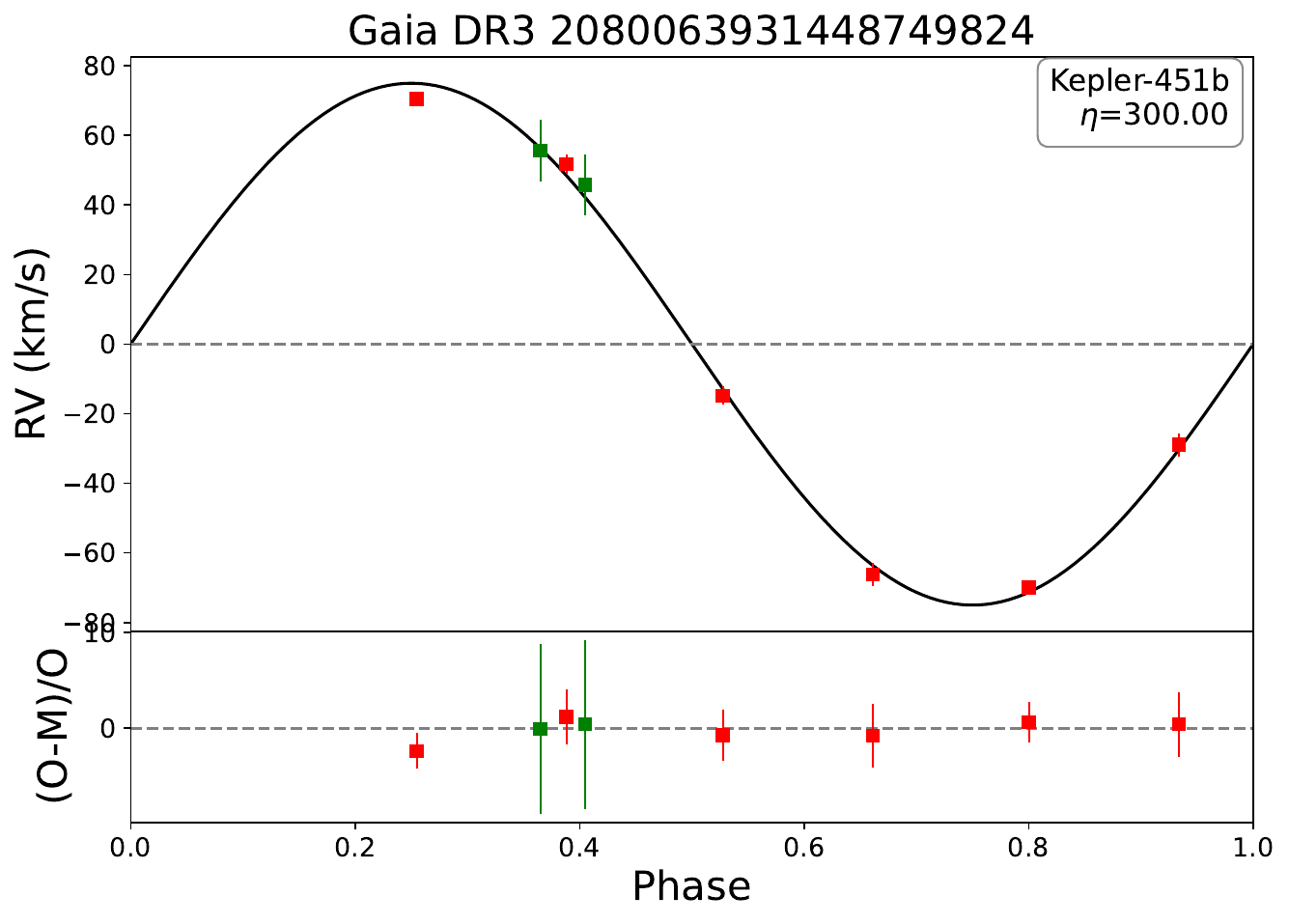}
    \includegraphics[width=0.32\textwidth]{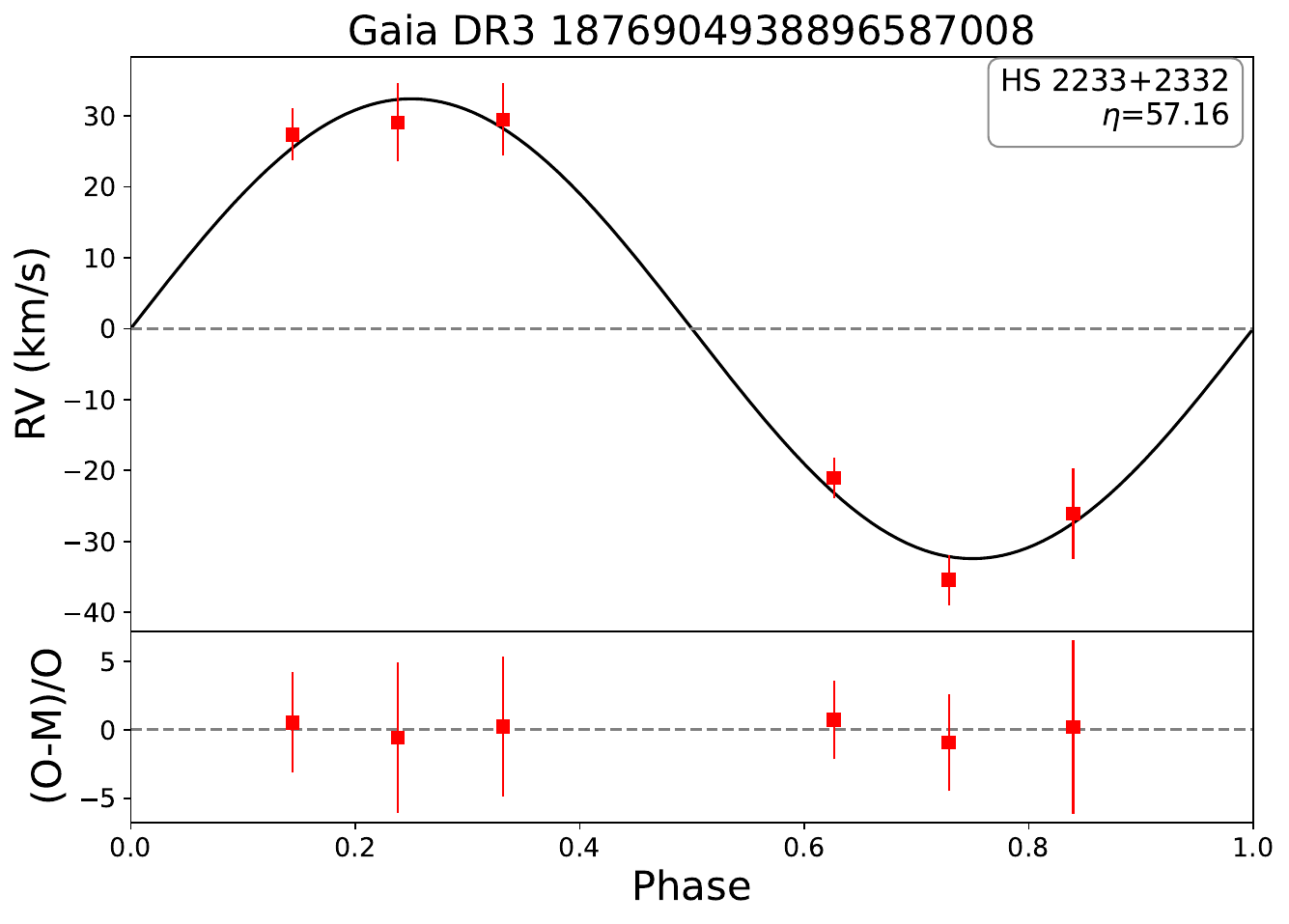}
    \includegraphics[width=0.32\textwidth]{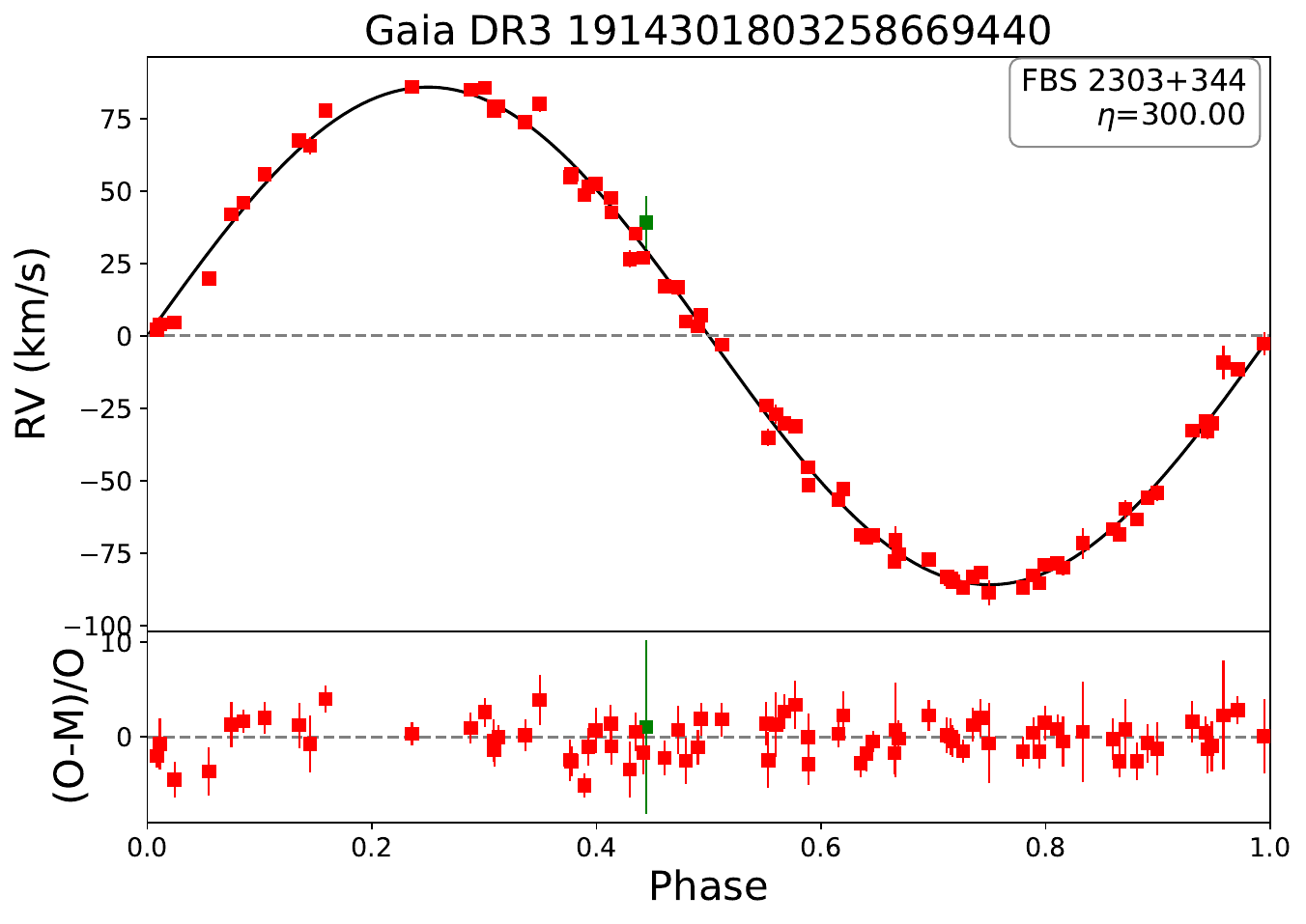}
    \vspace{1.5ex}
    \includegraphics[width=0.32\textwidth]{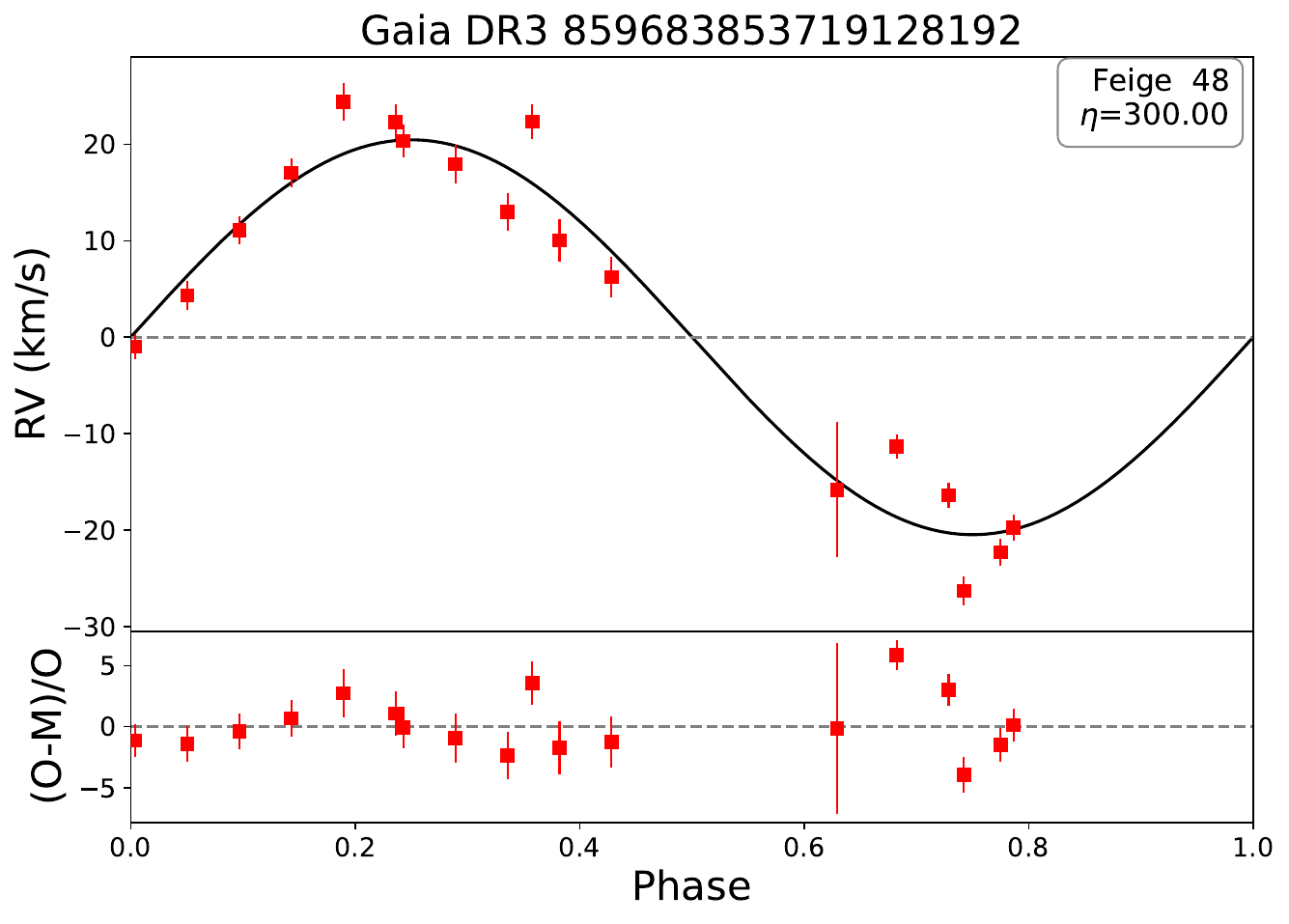}
    \includegraphics[width=0.32\textwidth]{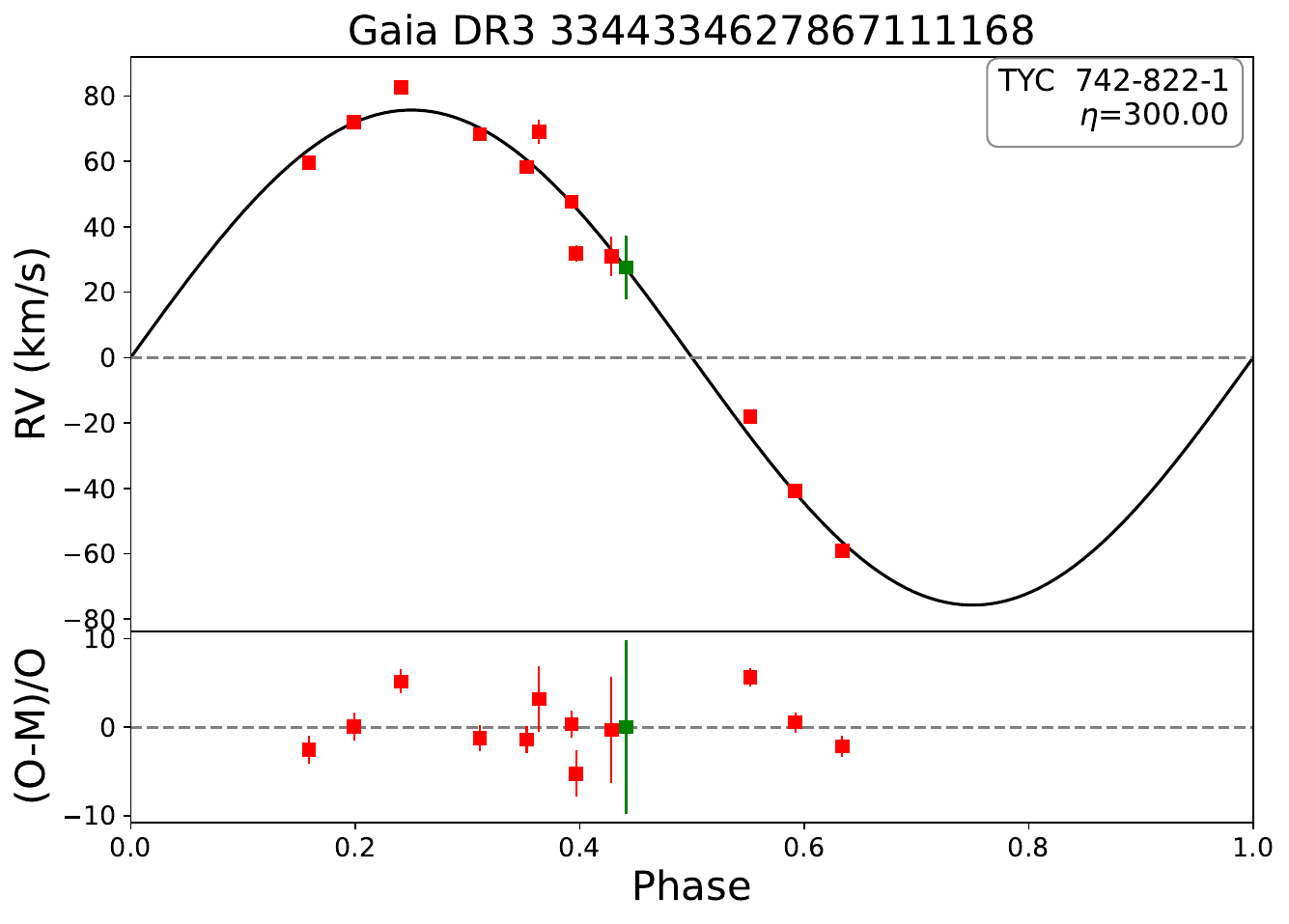}
    \includegraphics[width=0.32\textwidth]{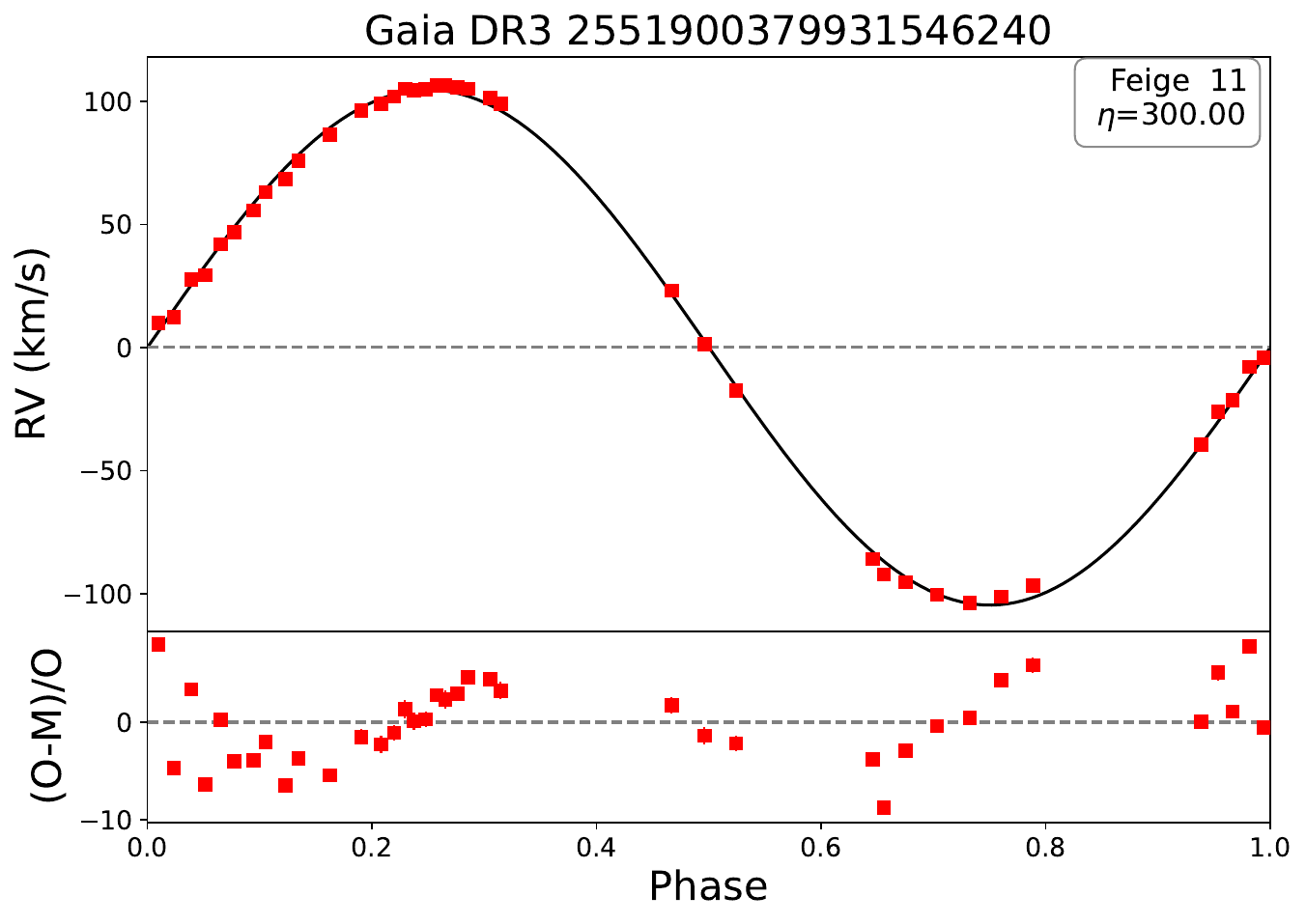}
    \vspace{1.5ex}
    \includegraphics[width=0.32\textwidth]{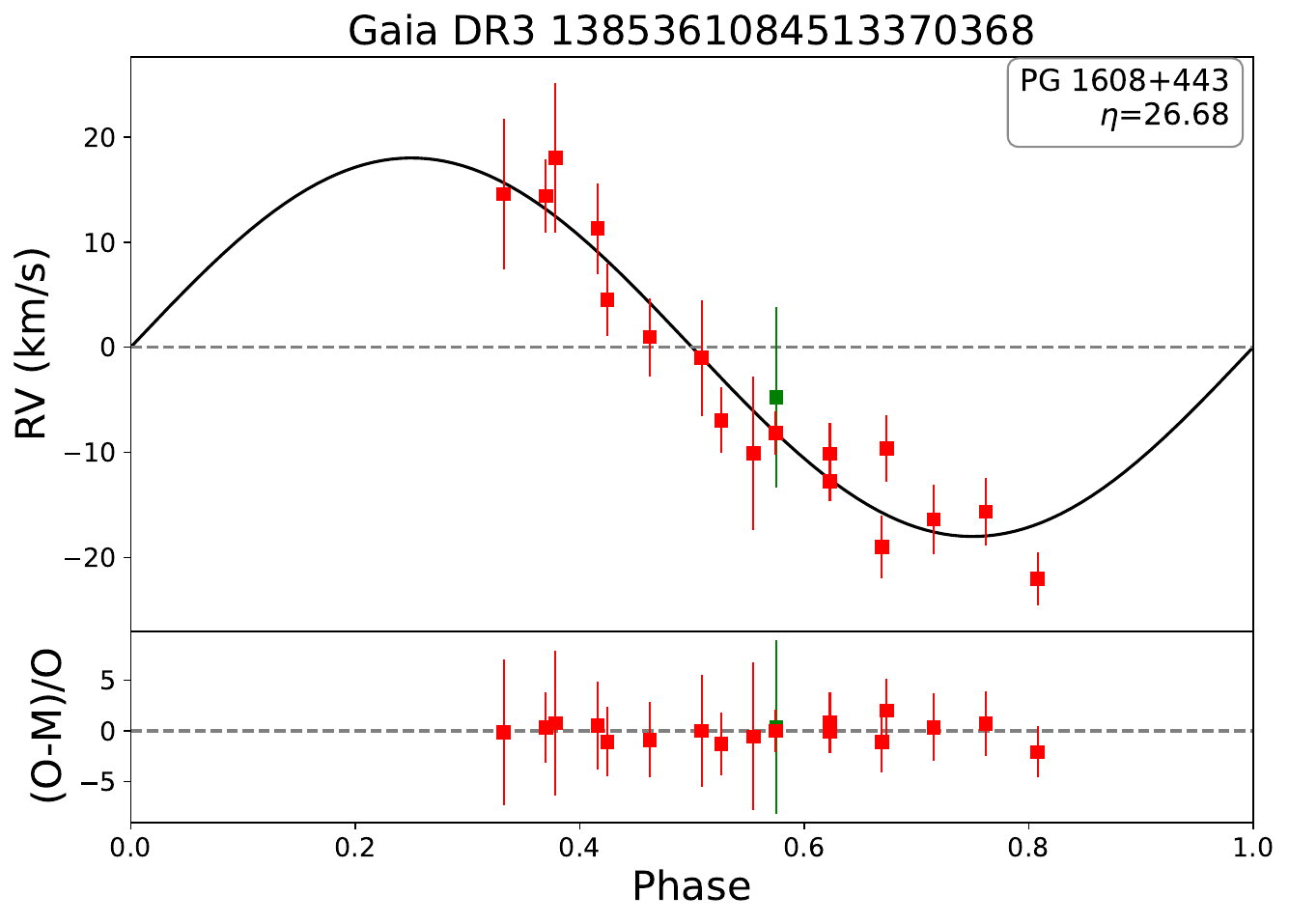}
    \includegraphics[width=0.32\textwidth]{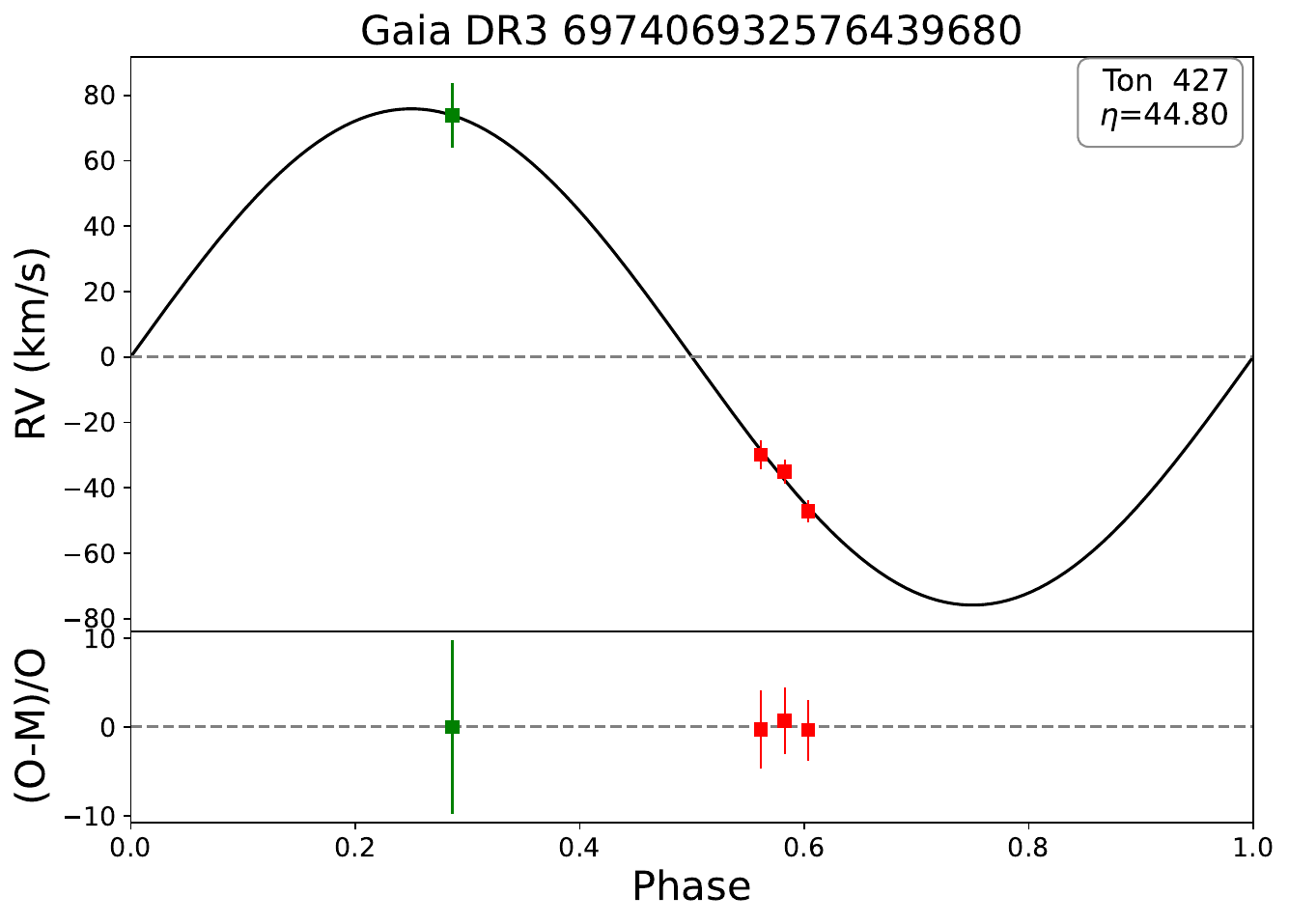}
    \includegraphics[width=0.32\textwidth]{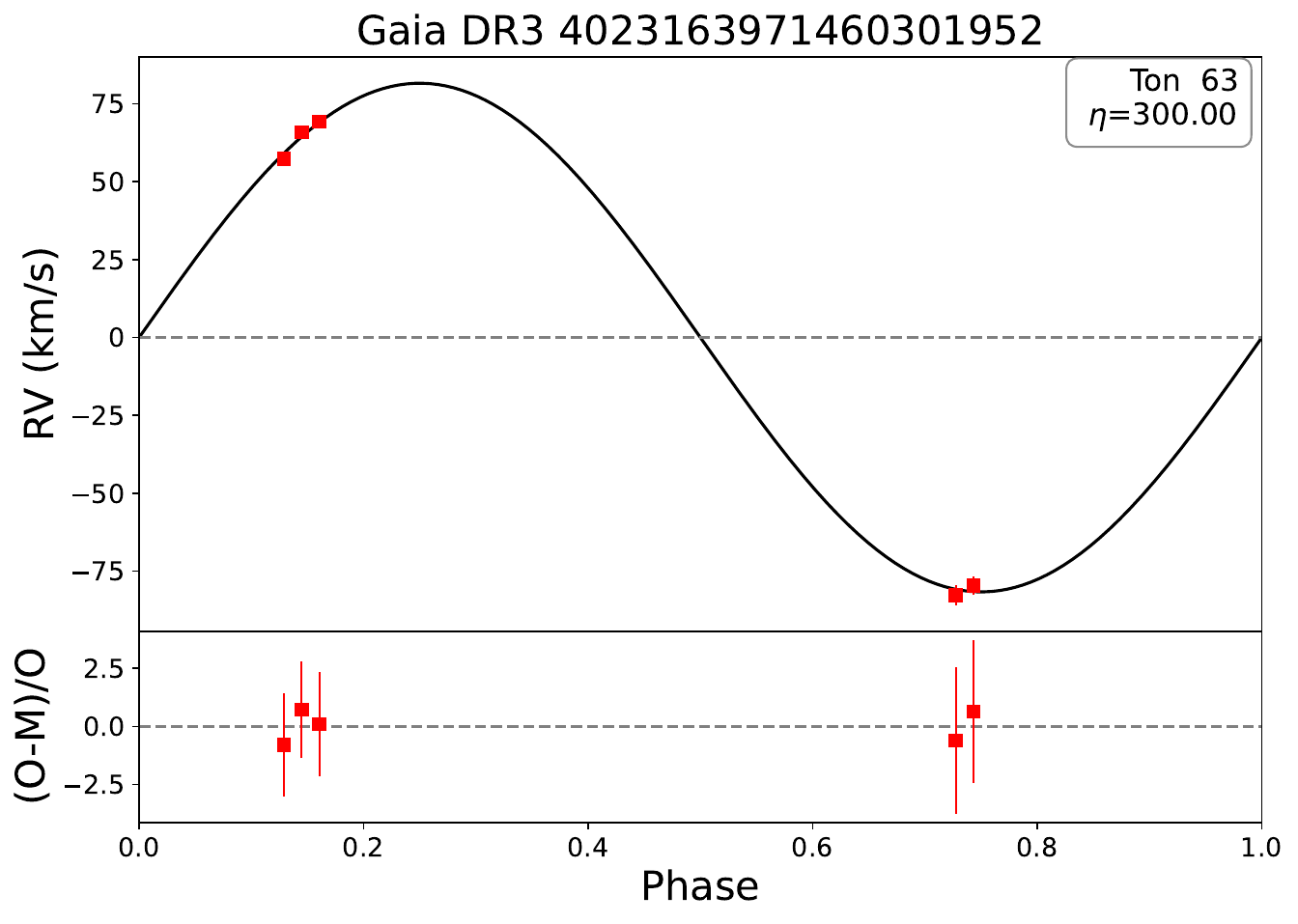}
    \vspace{1.5ex}
    \includegraphics[width=0.32\textwidth]{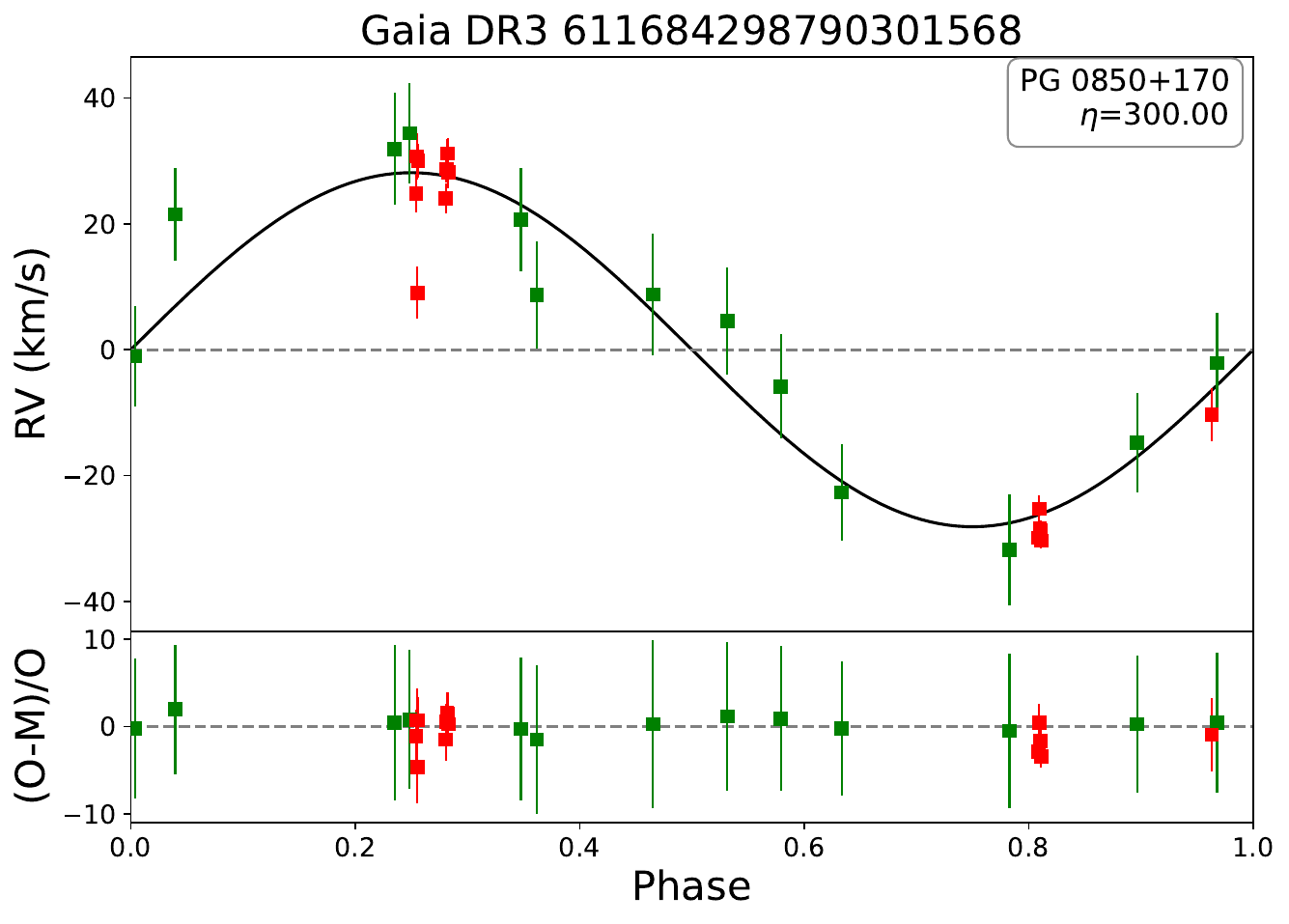}
    \includegraphics[width=0.32\textwidth]{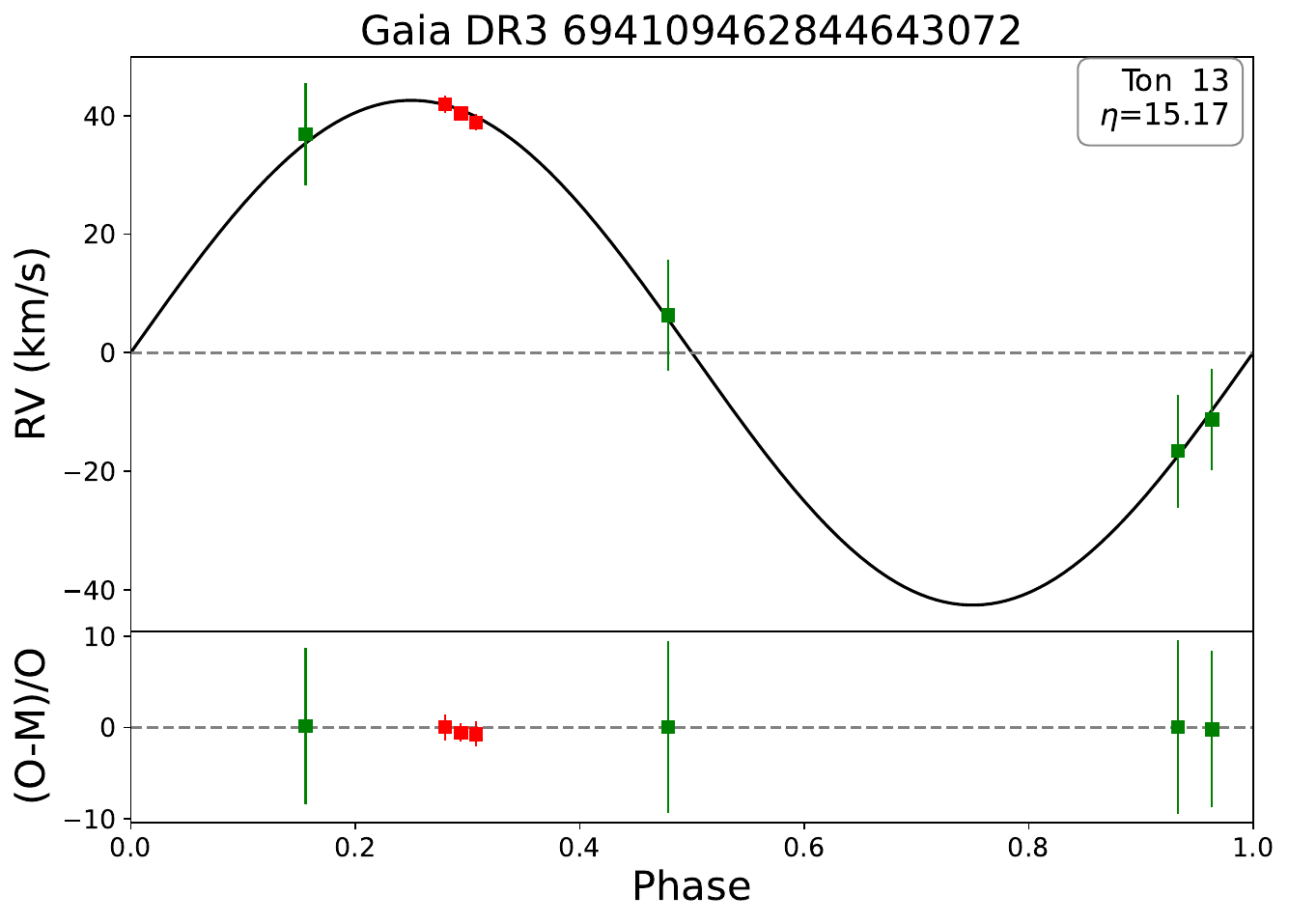}
    \includegraphics[width=0.32\textwidth]{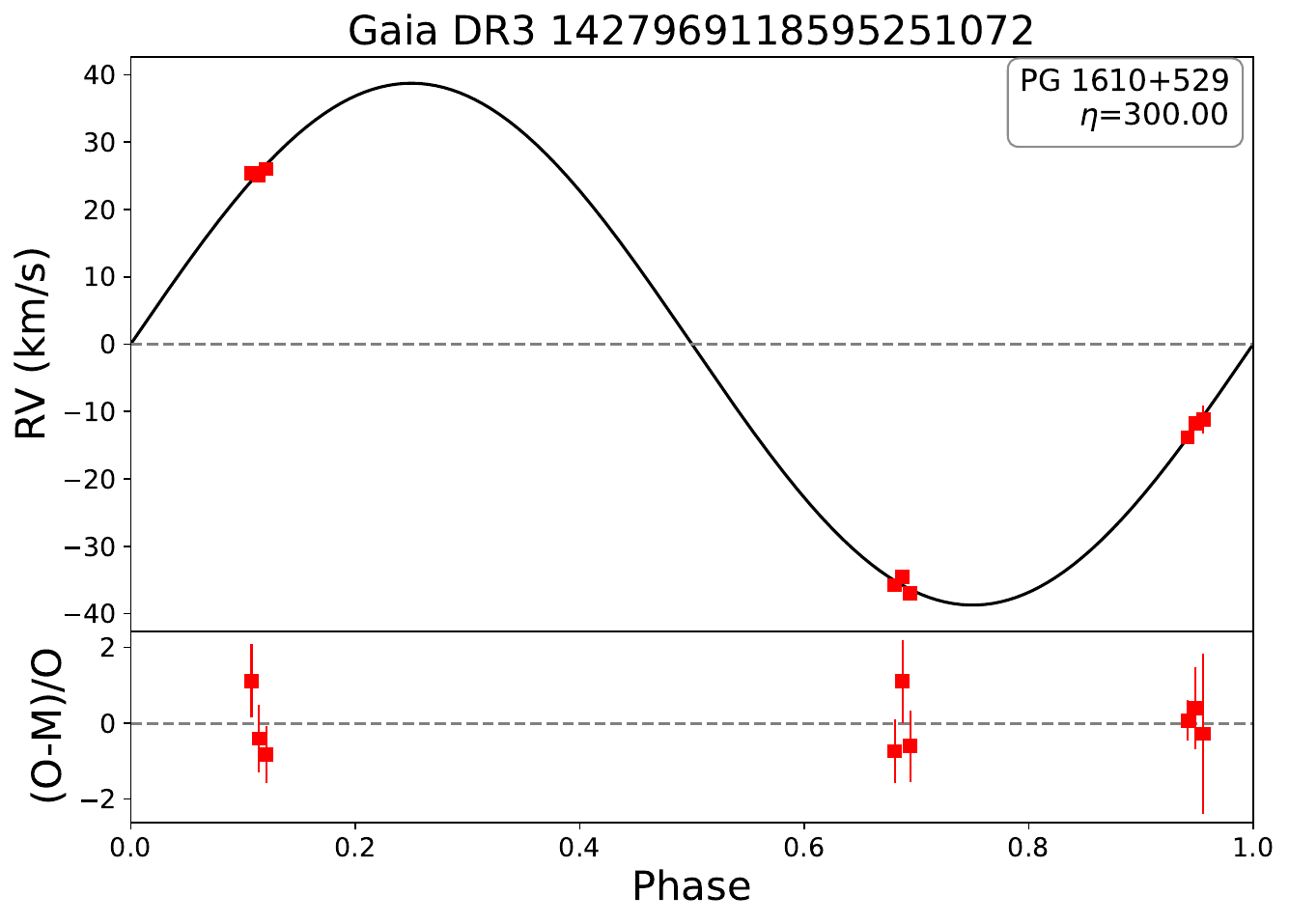}
    \caption{RV fitting of the systems in Table \ref{orbital_params_from_cir.tab} under the assumption of circular orbits. The red and green dots represent the RV data from LAMOST MRS and LRS, respectively. For values of $\eta$ exceeding 300, we applied an upper truncation limit of 300.}
\end{figure*}

\setcounter{figure}{1}
\begin{figure*} 
    \center
    \includegraphics[width=0.32\textwidth]{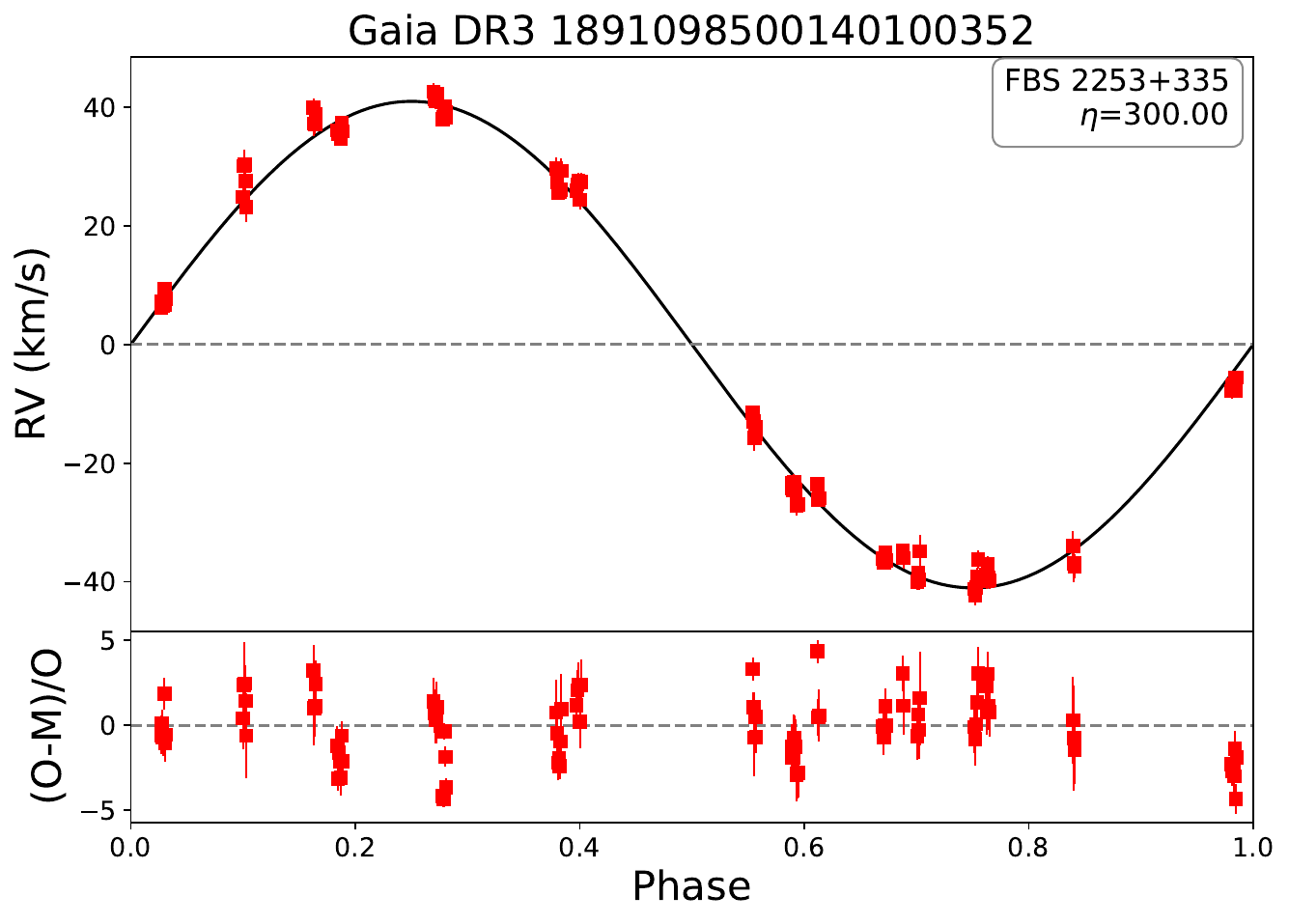}
    \includegraphics[width=0.32\textwidth]{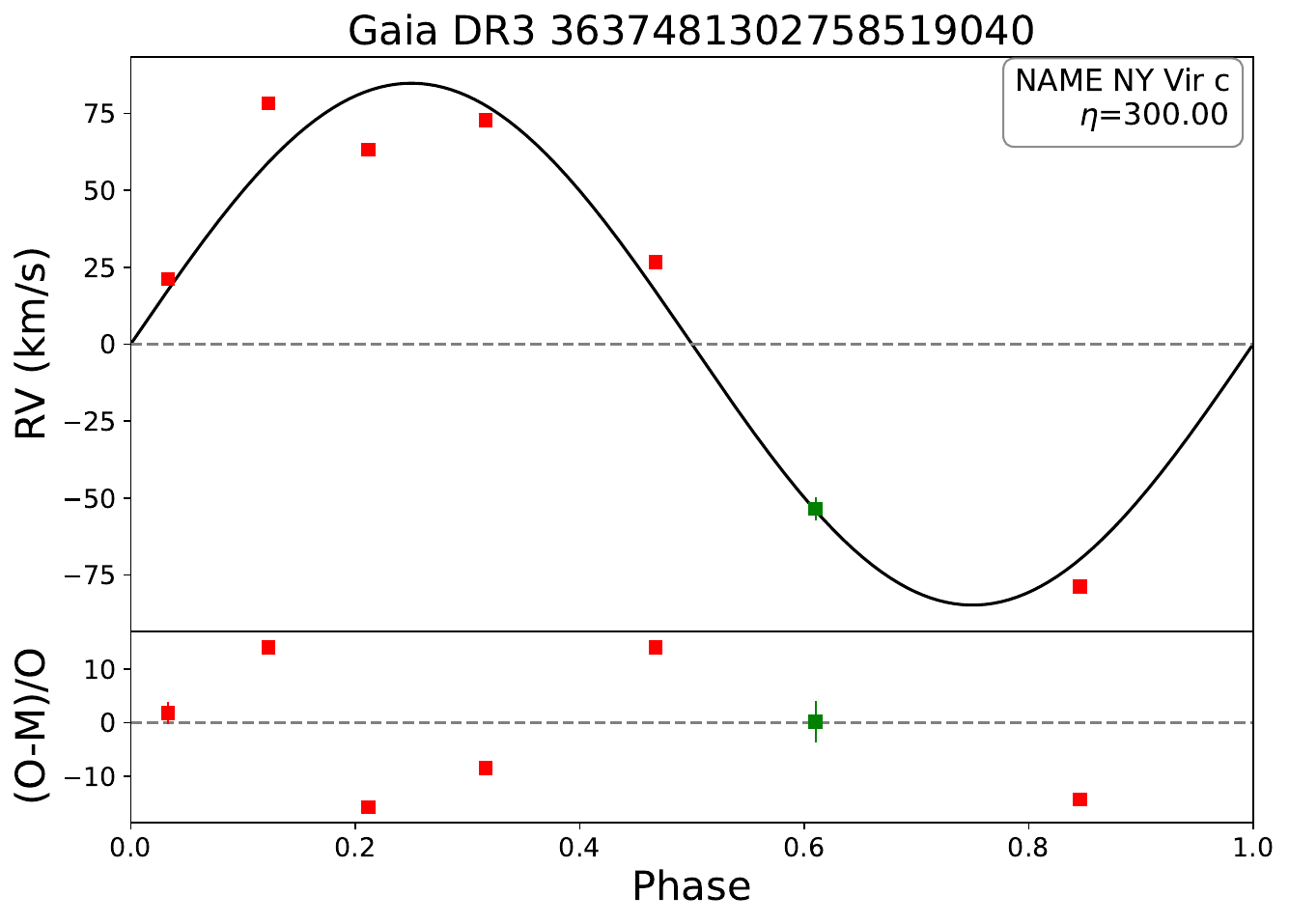}
    \caption{Continued. RV fitting of the systems in Table \ref{orbital_params_from_cir.tab} under the assumption of circular orbits.} 
    \label{circ_fitting.fig}
\end{figure*}

\end{document}